\PassOptionsToPackage{pdfpagelabels=false}{hyperref}
\documentclass[desactivate]{aa}
\usepackage{lineno}
\AtBeginDocument{\nolinenumbers}

\usepackage{graphicx}
\usepackage{amsmath}
\usepackage{amssymb}
\usepackage{placeins}
\usepackage[normalem]{ulem}

\usepackage{txfonts}
\usepackage{color}
\usepackage[colorlinks,linkcolor=magenta,anchorcolor=blue,citecolor=blue]{hyperref}

\begin{document}

\title{XGAP with uGMRT I: Old AGN plasma in merging galaxy groups}

\author{R. Santra
          \inst{1}\fnmsep\thanks{ramananda1999@gmail.com}
          \and
          R. Kale\inst{1}
          \and 
          K. Kolokythas \inst{2,3}
          \and
          M. Brienza \inst{4}
          \and
          E. O'Sullivan \inst{5}
          \and 
          D. Eckert \inst{6}
          \and
          F. de. Gasperin \inst{4}
          \and
          T. Pasini \inst{4}
          \and
          F. Gastaldello \inst{7}
          \and
          A. Finoguenov \inst{8}
          \and
          M. Sun \inst{9}
          \and
          G.  Gozaliasl \inst{8, 10}
          \and
          M. Bourne \inst{11, 12}
          }
\institute{National Centre for Radio Astrophysics, Tata Institute of Fundamental Research, S. P. Pune University, Ganeshkhind, Pune 411007, India
\and
Centre for Radio Astronomy Techniques and Technologies, Department of Physics and Electronics, Rhodes University, P.O. Box 94,
Makhanda 6140, South Africa
\and 
South African Radio Astronomy Observatory, Black River Park North, 2 Fir St, Cape Town, 7925, South Africa
\and
Istituto Nazionale di Astrofisica (INAF) - Istituto di Radioastronomia (IRA), via Gobetti 101, 40129, Bologna, Italy
\and
Center for Astrophysics | Harvard \& Smithsonian, 60 Garden Street, Cambridge, MA 02138, USA
\and
Department of Astronomy, University of Geneva, Ch. d’Ecogia 16, CH-1290 Versoix, Switzerland
\and
INAF - IASF Milano, via A. Corti 12, 20133 Milano, Italy
\and
Department of Physics, University of Helsinki, P. O. Box 64, FI00014 Helsinki, Finland
\and
Department of Physics and Astronomy, University of Alabama in Huntsville, 301 Sparkman Drive, Huntsville, AL 35899, USA
\and
Department of Computer Science, Aalto University, PO Box 15400,
Espoo, FI-00076, Finland
\and
Centre for Astrophysics Research, Department of Physics, Astronomy and Mathematics, University of Hertfordshire, College Lane, Hatfield AL10 9AB, UK
\and
Kavli Institute for Cosmology, University of Cambridge, Madingley Road, Cambridge CB3 0HA, UK
}



  \abstract
  {Galaxy groups are significantly affected by outflows from central Active Galactic Nuclei (AGN), especially due to the shallower gravitational potential compared to galaxy clusters. The group binding energy is comparable to the energy output from AGN, making it an important factor in mutual evolution.}
  {We present a multi-wavelength analysis of three dynamically active groups—SDSSTG8102, SDSSTG16393, and SDSSTG28674—which are part of the \textit{XMM-Newton} Group AGN Project X-GAP sample, a statistically complete sample of 49 galaxy groups, to understand the central AGN evolution.} 
  {We combine proprietary uGMRT 400 MHz observations, with 144 MHz LOFAR, and \textit{XMM-Newton} observations to study the  radio sources associated with the respective Brightest Group Galaxies (BGGs).}
  {The BGGs in SDSSTG8102 and SDSSTG16393 have extended radio emission with asymmetric distortions in their morphologies. SDSSTG28674 has a compact flat-spectrum radio source associated with the BGG and an extended lobe on one side, which is connected to it with a faint bridge detected with LOFAR. Integrated spectral indices of the three BGGs are $-0.96\pm0.09$ (SDSSTG8102), $-1.35 \pm 0.09$ (SDSSTG16393), and $-1.6\pm0.02$ (SDSSTG28674). X-ray images reveal elongated morphologies in all three groups, SDSSTG28674 showing evidence of a binary merger, while thermodynamical maps highlight temperature variations.}
   {In SDSSTG8102, lobes are bent and displaced by IGrM flows, while SDSSTG16393 hosts steep-spectrum relic-like plasma coinciding with X-ray emission. SDSSTG28674, with its ultra-steep spectrum lobe and disturbed morphology, likely traces merger-driven activity, consistent with a remnant or revived radio phoenix. The spectral diversity across the systems reflects different stages of AGN fading governed by duty cycle, source age, and confinement by the hot IGrM. The presence of bright group-scale X-ray halos ($\sim$300 kpc) and $>$50 kpc radio emission, combined with disturbed morphologies, underscores the central role of IGrM confinement and merger-driven gas motions in sustaining extended diffuse structures. The multi-band radio follow-up of the entire X-GAP sample will allow further insights.}


   \maketitle


\section{Introduction} \label{sec:intro}

Galaxy groups are gravitationally bound systems that house the majority of galaxies and stars in the Universe \citep[e.g.,][]{eke04, robotham11}. These groups encompass more than half of all galaxies, compared to only about 2\% found in galaxy clusters, making them the most common type of environment in the local Universe \citep[e.g.,][]{geller83}. Although galaxy groups are less massive than clusters, typically having M$_{500} < 10^{14} M_{\odot}$, and contain fewer members, they play a pivotal role in the formation of large-scale structures \citep[e.g.,][]{grogin00, lovisari21}. In the hierarchical model of structure formation, galaxy clusters are thought to form through the merging of smaller groups \citep[e.g.,][]{vandenbosch14, haines18, nelson24}. As a result, the early evolution of galaxies in rich clusters \citep[e.g.,][]{beck1999} is closely connected to the processes that shape galaxies within groups. Galaxy groups should not be seen simply as the scaled-down versions of galaxy clusters \citep[e.g.,][]{ponman99, ponman03, sun12}. They are characterised by shallow gravitational potentials and low velocity dispersions, which create environments conducive to galaxy mergers and tidal interactions, in turn driving rapid galaxy evolution \citep[e.g.,][]{msintosh08, alonso12, solanes18}. Therefore, studying galaxy groups is essential for developing a more complete understanding of galaxy formation and evolution, as well as the mechanisms behind these processes \citep[e.g.,][]{forbes06, kolokythas19}.

Unlike virialised clusters, where the hot gas permeating the system volume (called the intracluster medium) is consistently the dominant baryonic component, group gas content varies significantly \citep[e.g.,][]{sun09, eckert12, lovisari15}. It is now known that the more massive and evolved galaxy groups are dominated by early-type galaxies that present little cold gas (H\textsc{I}) but often host an extended hot X-ray emitting Intragroup medium (IGrM) \citep[e.g.][]{Osullivan17}. The shallow potential wells of galaxy groups bring galaxies into close proximity, fostering interactions like NGC~1550, NGC~5903 \citep[e.g.,][]{kolokythas20, Osullivan18} or mergers between galaxies \citep[e.g.,][]{pearson24}. The group scaling relations are known to deviate from self-similarity, suggesting a strong impact of non-gravitational processes such as Active Galactic Nuclei (AGN) feedback in their evolution \citep[e.g.][]{ponman99, finoguenov03, lovisari21}. 

Galaxy groups provide an important environment where feedback may exert the most significant influence on galaxy formation and evolution \citep[see e.g.,][for reviews]{eckert21,eckert24}. X-ray bright groups are a key environment for the study of AGN feedback, having almost universally short central cooling times and low entropies \citep{Osullivan17, osullivan24}. It is known that AGN feedback operates differently in groups \citep{giodini09,sun12}, in some cases as a near-continuous `bubbling' mode with moderate thermal regulation \citep{birzan12, panagoulia14, brienza21}, in other cases manifesting instead as powerful AGN outbursts (such as the ones sometimes observed in massive clusters), where the group environment is disrupted as in NGC~4261, NGC~193 \citep[e.g.,][]{Osullivan11,kolokythas15,kolokythas18} or gas is stripped \citep{morganti13}, at least in the central region due to their shallow gravitational potential. Hence, the group environment represents an excellent opportunity to study a rich array of processes that include galaxy interactions, triggering and quenching of star formation, and the feeding and feedback of the AGN \citep[e.g.,][]{kolokythas22,jennings25}.

Brightest Group Galaxies (BGGs) serve as excellent laboratories for studying the evolution of galaxy groups and massive galaxies \citep[e.g.,][]{gozaliasl16, gozaliasl18}. These galaxies are typically highly luminous, old ellipticals located near the centers of the IGrM halo \citep[e.g.,][]{vonderlinden07,stott10}. Their stellar kinematics span a wide range, from field elliptical-like properties to those resembling BCGs \citep{loubser18, Gozaliasl24}, and their nuclei host supermassive black holes \citep[e.g.,][]{rafferty06}. Many BGGs exhibit AGN activity, with some launching powerful radio jets that inject energy back into the IGrM over distances of hundreds of kiloparsecs \citep[e.g.,][]{mcnamara07, hardcastle20}. The fraction of massive galaxies hosting detectable radio emission has increased with survey depth. The LoTSS survey at 150\,MHz shows that galaxies with mass $> 10^{11}\,M_\odot$, essentially $100\%$ display radio AGN activity above a luminosity threshold of $L_{150\,\mathrm{MHz}} \geq 10^{21}\,\mathrm{W\,Hz^{-1}}$ \citep{sabater19}, superseding the much smaller sample by \citet{dunn10}. 
 Earlier, shallower surveys \citep[e.g.,][]{best05,shabala08} reported much lower detection fractions ($\sim 30\%$), consistent with their higher flux limits and different sample selections. The Complete Local-Volume Groups Sample (CLoGS) study found a radio detection rates of 92\% for the high-richness X-ray bright groups and 87\% for the X-ray faint systems \citep{Osullivan17, kolokythas18, kolokythas19}, further highlighting the strong link between radio AGN activity and group environment. In addition to AGN fueling through cooling flows from the IGrM, the link between radio AGN activity and dense galaxy environments \citep[e.g.,][]{Lilly09,bardelli10,malavasi15} may be influenced by large-scale mergers (e.g., group infall into clusters) or galaxy-galaxy interactions within groups \citep[e.g.,][]{miles04,taylor05}. These interactions and mergers can funnel the gas toward the central AGN, triggering radio emission and launching large-scale jets, but are less studied so far due to the dearth of deep enough X-ray observations.

The \textit{XMM-Newton} Group AGN Project (X-GAP\footnote{\url{https://www.astro.unige.ch/xgap/blog/people}}, \citealt{eckert24}) seeks to quantify the effects of AGN feedback in a complete sample of 49 galaxy groups, with estimated masses ranging from 10$^{13} \lesssim \rm M_{200} \lesssim 10^{14}$ M$_{\odot}$, having a narrow redshift range of $0.025 < z < 0.06$. These groups have been observed with \textit{XMM-Newton} for a total exposure\footnote{Exposure time varies per system.} of approximately 852 ks. They were selected from the ROSAT All-Sky Survey (RASS), and cross-matched with spectroscopic Friends-of-Friends groups from SDSS \citep{tempel17, damsted24}. The selection criteria and detailed methodologies are described in \citet{eckert24}. The X-GAP sample has been observed by LOw Frequency ARray (LOFAR) at 144 MHz \citep[LoTSS;][]{shimwell22}. In data release 2 (DR2) and the ongoing analysis for DR3 of LoTSS, images are available for most of the X-GAP groups, providing radio information on both extended and compact emission associated with the BGG (Brienza et al., in prep.).

This paper presents the radio properties of the BGGs of three galaxy groups (Table~\ref{source-list}), using new data from the upgraded Giant Metrewave Radio Telescope (uGMRT) observations at 400 MHz, and images from LOFAR 144 MHz observations. We present the properties of the central radio sources, examine their environment, and provide a qualitative comparison between the radio data and X-ray data for each group. The paper is organised as follows: In Section~\ref{obs-data-analysis}, we describe the uGMRT and X-ray observations and the approach followed for the data reduction. In Section~\ref{radio_results}, we present the radio images and the spectral characteristics of the individual sources, and in Section~\ref{xray_results}, their X-ray properties (2D thermodynamic maps). Section~\ref{discussion} contains a discussion of our results, focusing on the environmental properties of the radio sources and the spectral index behaviour in those extreme systems. The summary and conclusions are given in Section~\ref{summary}. Throughout the paper, we adopt the $\Lambda$CDM cosmology with H$_{0}$ = 70 km~s$^{-1}$~Mpc$^{-1}$, $\Omega_{\rm m}$ = 0.27, and $\Omega_{\Lambda}$ = 0.73. All the radio images are corrected for the primary beam attenuation.

\begin{table*}
  \centering
  \caption{The list of galaxy groups.}
  \begin{tabular}{@{}cccccccccc@{}}
    \hline\hline

& Group & BGG name & RA & DEC & Redshift & Velocity dispersion &  M$_{200}$ & X-ray Luminosity & Scale \\
 & &  & (deg) & (deg) &  & (km~s$^{-1}$) & (10$^{13}$M$_{\odot}$) & (10$^{42}$erg~s$^{-1}$) & (kpc/$''$) \\

\hline

 & SDSSTG8102 & MCG+07-33-011 & 238.76 & 41.58 & 0.033 & 492 & 15.7 $\pm$ 1.4 & 4.5 $\pm$ 1.6 & 0.662 \\

&SDSSTG16393& MCG+09-17-036 & 152.71  & 54.21 & 0.047 & 295 & 4.8 $\pm$ 0.7 & 19.6 $\pm$ 3.1 &  0.928 \\

&SDSSTG28674 & MCG+06-30-029 & 203.24  & 32.61 & 0.037 & 282 & 11.9 $\pm$ 1.2 & 4.2 $\pm$ 0.9 & 0.739 \\

 \hline
  \end{tabular}
  \label{source-list}
\end{table*}

\begin{table*}
  \centering
  \caption{Details of the radio observations and image properties.}
  \begin{tabular}{@{}cccccccc@{}}
    \hline\hline

& Source & Frequency & bandwidth & Integration time & Beam & Rms noise  & Source size  \\
&  & (MHz) & (MHz) & (Hr.) & ($''$) & ($\mu$Jy~beam$^{-1}$) & (kpc $\times$ kpc)  \\
    \hline

& SDSSTG8102 & 144  & 48 & 8 & 8.5 $\times$ 8.5 & 110.0 & 93.7 $\times$ 51.9 \\

&  & 400  & 200 & 5 & 8.5 $\times$ 8.5 & 30.5 & 72.4 $\times$ 51.9 \\

& SDSSTG16393 & 144  & 48 & 8 & 8.0 $\times$ 8.0 & 92.0 & 119.2 $\times$ 82.8 \\

&  & 400  & 200 & 5 & 8.0 $\times$ 8.0 & 32.0 & 151.0 $\times$ 96.1 \\

& SDSSTG28674 & 144  & 48 & 8 & 7.0 $\times$ 7.0 & 85.1 & 81.5 $\times$ 38.6 \\

&  & 400  & 200 & 5 & 7.0 $\times$ 7.0 & 28.7 & 55.9 $\times$ 32.9 \\

 \hline
  \end{tabular}
  \label{obs-tab}
\end{table*}

\section{Observations and Data analysis} \label{obs-data-analysis}

Three groups were observed with the upgraded GMRT (observation code: 43\_072) (see Table~\ref{obs-tab}). Real-time radio frequency interference (RFI) online filtering, developed to mitigate broadband RFI \citep{buch22,buch23}, was applied during the uGMRT observations. The LOFAR observation details are also provided in Table~\ref{obs-tab}.

\subsection{LOFAR}

The X-GAP targets were observed using LOFAR as part of the LOFAR Two-metre Sky Survey \citep[LoTSS DR2;][]{shimwell17, shimwell19, shimwell22}. The observations were conducted in High Band Antenna (HBA) dual inner mode, with each target being observed in the field for an 8-hour single pointing. Data reduction and calibration were performed using the standard LoTSS DR2 pipeline \citep{Tasse21}, which incorporates both the \texttt{Pre-Factor} pipeline \citep{vweeren16, william16} and the \texttt{DDF pipeline} \citep{shimwell19, Tasse21}. The \texttt{Pre-Factor} pipeline corrects for direction-independent effects, such as ionospheric Faraday rotation, phase offsets between XX and YY, and clock discrepancies. The DDF pipeline performs direction-dependent calibration to address ionospheric distortions at low frequencies. For improved calibration and re-imaging, the data were processed using the \texttt{extraction$+$self-calibration} method \citep{vweeren21}, in which all sources outside a square region centered on the target were removed from the visibilities, followed by multiple self-calibration loops. The final imaging is done using the \texttt{WSClean} \citep{offringa14}.

\subsection{uGMRT}

We processed the uGMRT observations of the three groups using the \texttt{CAPTURE}\footnote{\url{https://github.com/ruta-k/CAPTURE-CASA6}} pipeline \citep{kale21}, a \texttt{CASA}-based continuum pipeline, specifically developed for the reduction of the GMRT continuum data. After this initial flagging, the flux density of the primary calibrator was set according to the flux scale of \citet{perley&butler17}. After following the standard calibration routines and flagging using the automated flagger (\texttt{tfcrop} and \texttt{rflag}), the calibration solutions were applied to the target field, and calibrated target source data were split and further flagged. Special care was taken to remove the narrow-band RFI using the \texttt{aoflagger} \citep{offringa13}, and some manual flagging has also been done at the small baselines. To reduce the volume of the data but still prevent bandwidth smearing, the data were averaged to $\sim$1 MHz frequency resolution. The target visibilities were imaged using the \texttt{CASA} task \texttt{tclean} with wide-field and wide-band imaging algorithms, with \texttt{nterms=2}, and \texttt{robust = 0}. We carried out 5 rounds of phase only and 3 rounds of phase and amplitude calibration to obtain a good gain solution for the target. For further details on the data analysis procedures for uGMRT, please follow \citet{kale21,kale22}.

\subsection{Comparison of the flux density scale}

The overall flux scale for all observations was verified by comparing the spectra of compact sources within the field of view with those available in the NVSS and TGSS catalogues. The LOFAR 144 MHz data points were already multiplied by a correction factor, mentioned in \citet{shimwell22}, to match the flux density scale of \cite{scaife12}. The uncertainty in the flux density measurements ($\Delta S$) was estimated as

\begin{equation}
    \Delta S = \sqrt{(f\cdot S)^{2} + N_{\rm beam} \cdot(\sigma_{\rm rms})^{2}},
    \label{eq-flux-err}
\end{equation}

where $S$ is the flux density, $f$ is the absolute flux density calibration error,  $N_{\rm beam}$ is the number of beams, and $\sigma_{\rm rms}$ is the rms noise. We assumed absolute flux density uncertainties of 10\% for LOFAR HBA data \cite[]{shimwell22} and uGMRT \cite[]{chandra17}. The images are corrected for the primary beam using the task \texttt{ugmrtpb}\footnote{\url{https://github.com/ruta-k/uGMRTprimarybeam-CASA6}} for the GMRT.

\subsection{XMM-Newton}

The \textit{XMM-Newton} observations of these groups were processed using the \texttt{XMM-SAS} software package, version 21, and the X-COP data analysis pipeline \citep{ghirardini19}. After applying standard event screening procedures, we extracted light curves from the field of view and the unexposed corners of the three detectors of the European Photon Imaging Camera (EPIC) to remove time periods affected by flaring background. The total good observing time for each group, after filtering out the flaring periods (energy range of $10-12$\,keV, count rate threshold of 0.4 cts/s) is provided in Table~\ref{fig:xmm_exposures}. For the two MOS detectors, we excluded chips operating in anomalous mode (CCD \# 4 for \texttt{MOS1} and CCD \# 5 for \texttt{MOS2}). From the clean event lists, we extracted images from all three cameras in the [0.7$-$1.2] keV band, which optimises the signal-to-background ratio \citep{ettori12}. Effective exposure maps for the three detectors, including the telescope’s vignetting, were generated using the task \texttt{eexpmap}. The contribution from residual soft protons was estimated using an empirical relation between the difference in high-energy count rates inside and outside the field of view and the soft proton component's normalisation \citep{salvetti17}. Finally, we combined the EPIC maps by summing the individual maps from the three detectors, the exposure maps, and the non X-ray background maps. Detailed data analysis procedures of the X-GAP groups have been discussed in \cite{eckert25}.

\begin{table}
\centering
\caption{Effective exposure time of each \textit{XMM-Newton} EPIC observation\label{fig:xmm_exposures}.}
\begin{tabular}{lccc}
\hline\hline
Objects & EMOS1 & EMOS2 & EPN \\

& (ks) & (ks) & (ks) \\
\hline
SDSSTG8102 & 13.8 & 15.5 & 8.8 \\ 
SDSSTG16393 & 13.5 & 13.5 & 13.9 \\ 
SDSSTG28674 & 10.9 & 14.7 & 18.7 \\ 
\hline

\end{tabular}
\end{table}

\section{Results: Radio observations} \label{radio_results}

\begin{figure*}
    \centering
    \begin{tabular}{ccc}
     
        \includegraphics[width=0.33\textwidth]{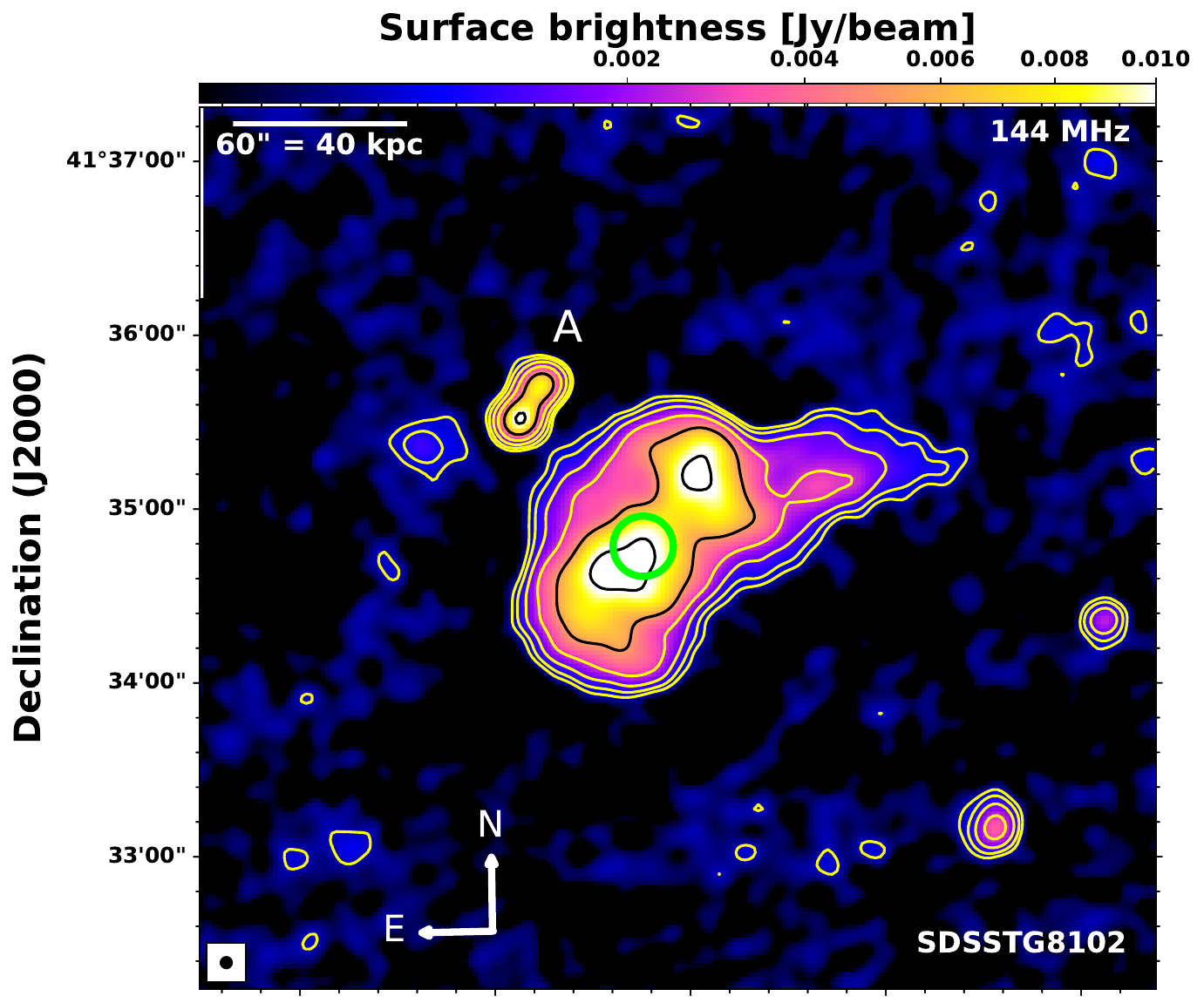} &
        \includegraphics[width=0.285\textwidth]{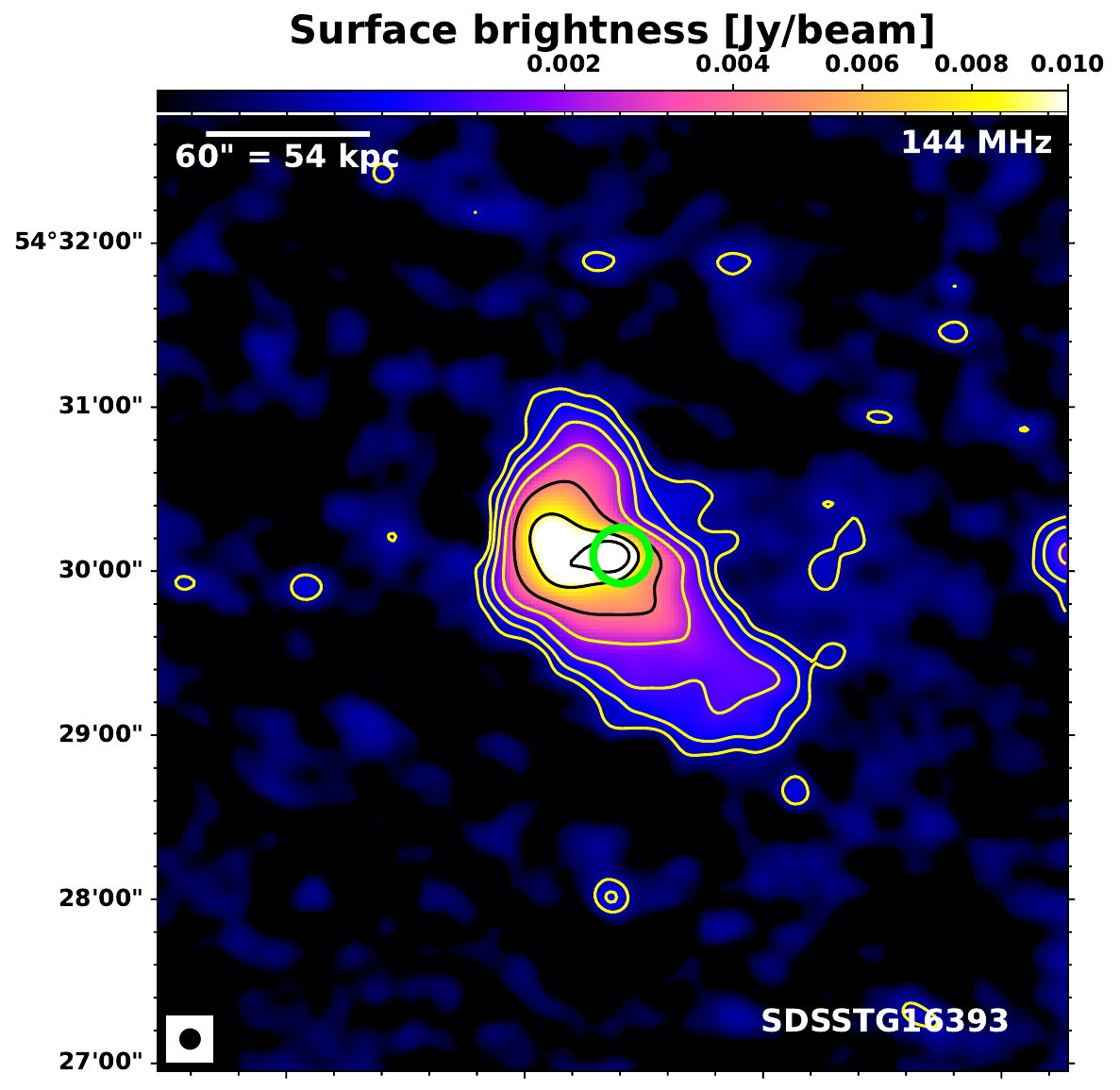} &
        
         \includegraphics[width=0.30\textwidth]{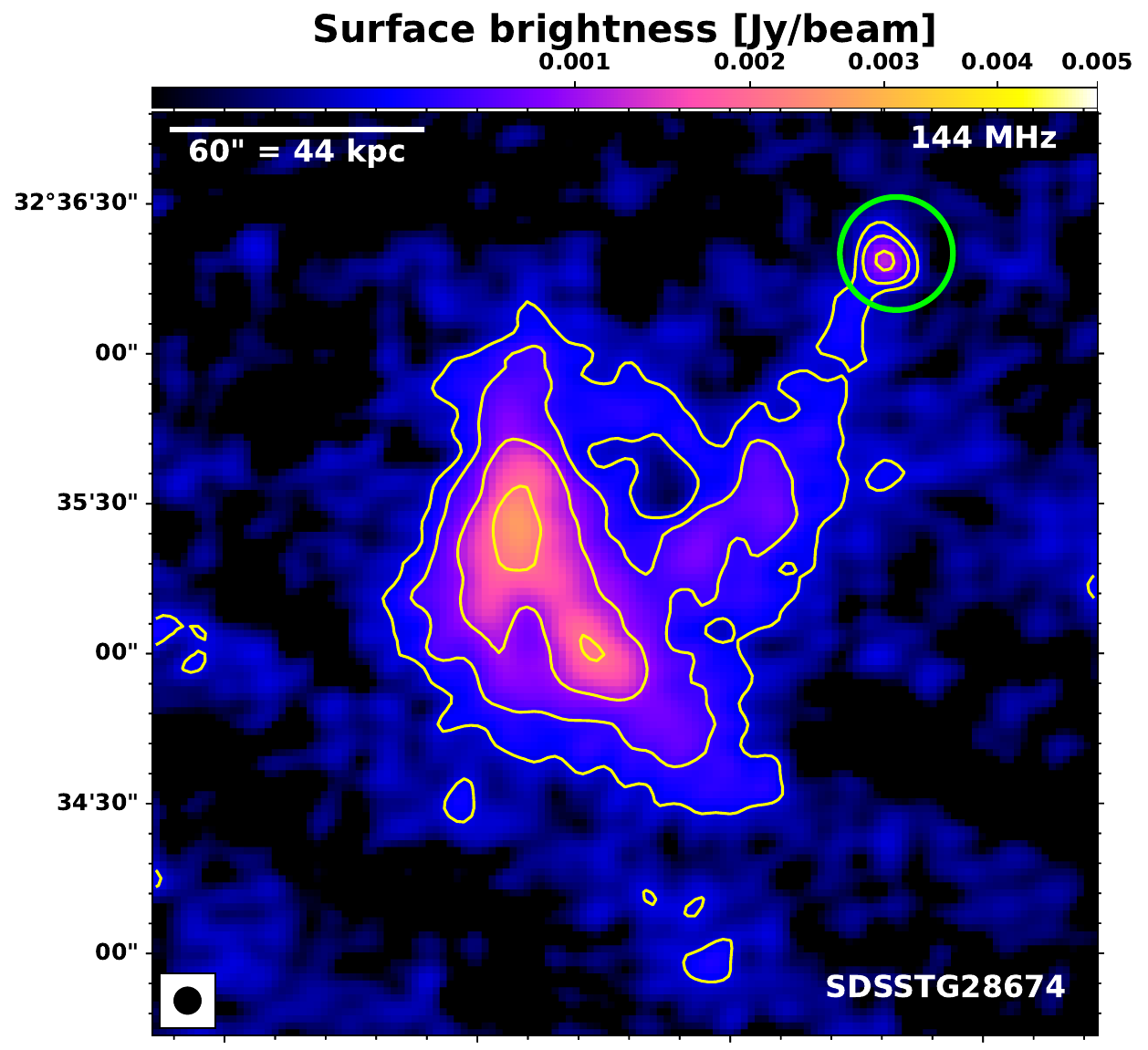}\\

        \includegraphics[width=0.33\textwidth]{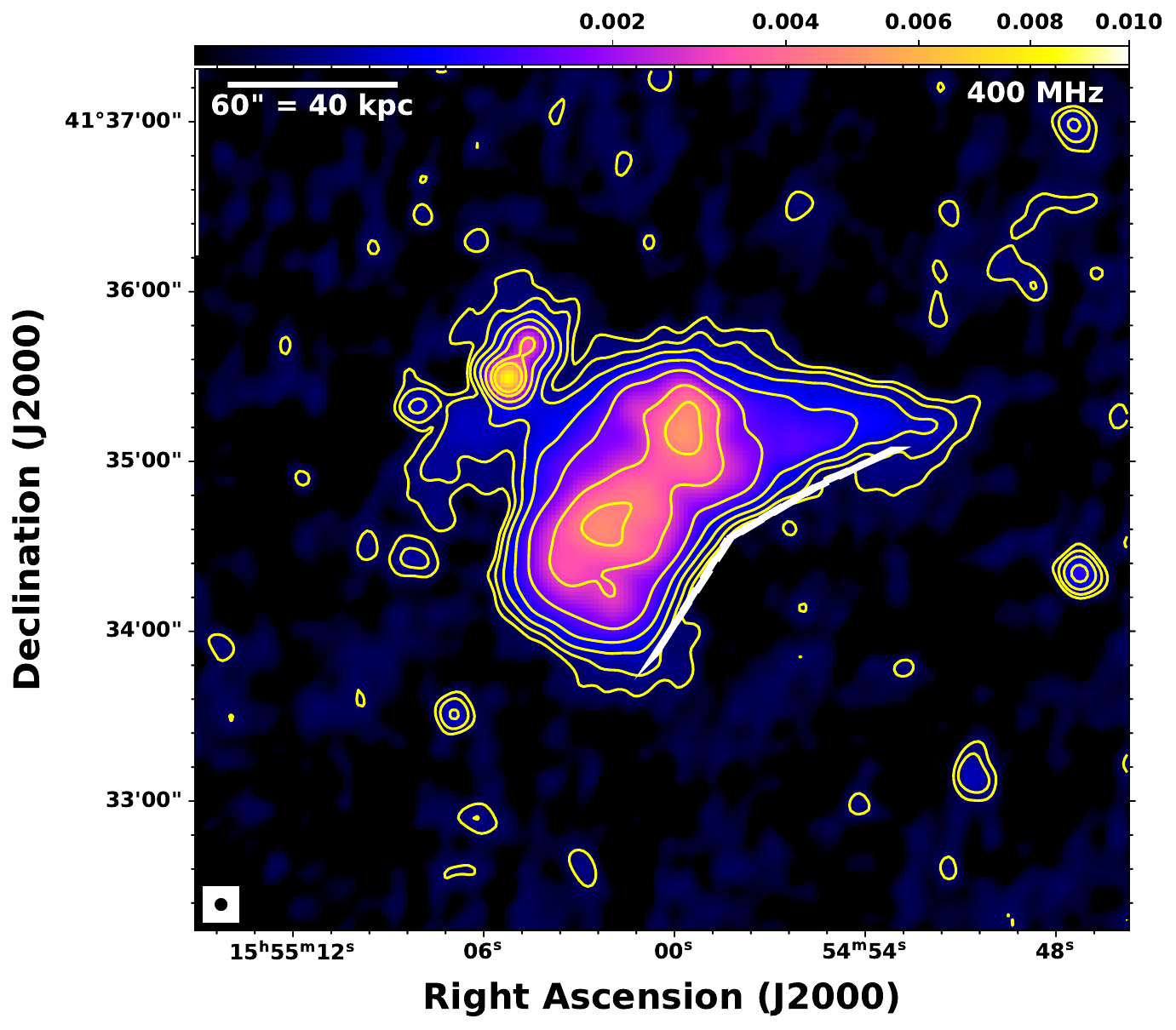} &
        \includegraphics[width=0.285\textwidth]{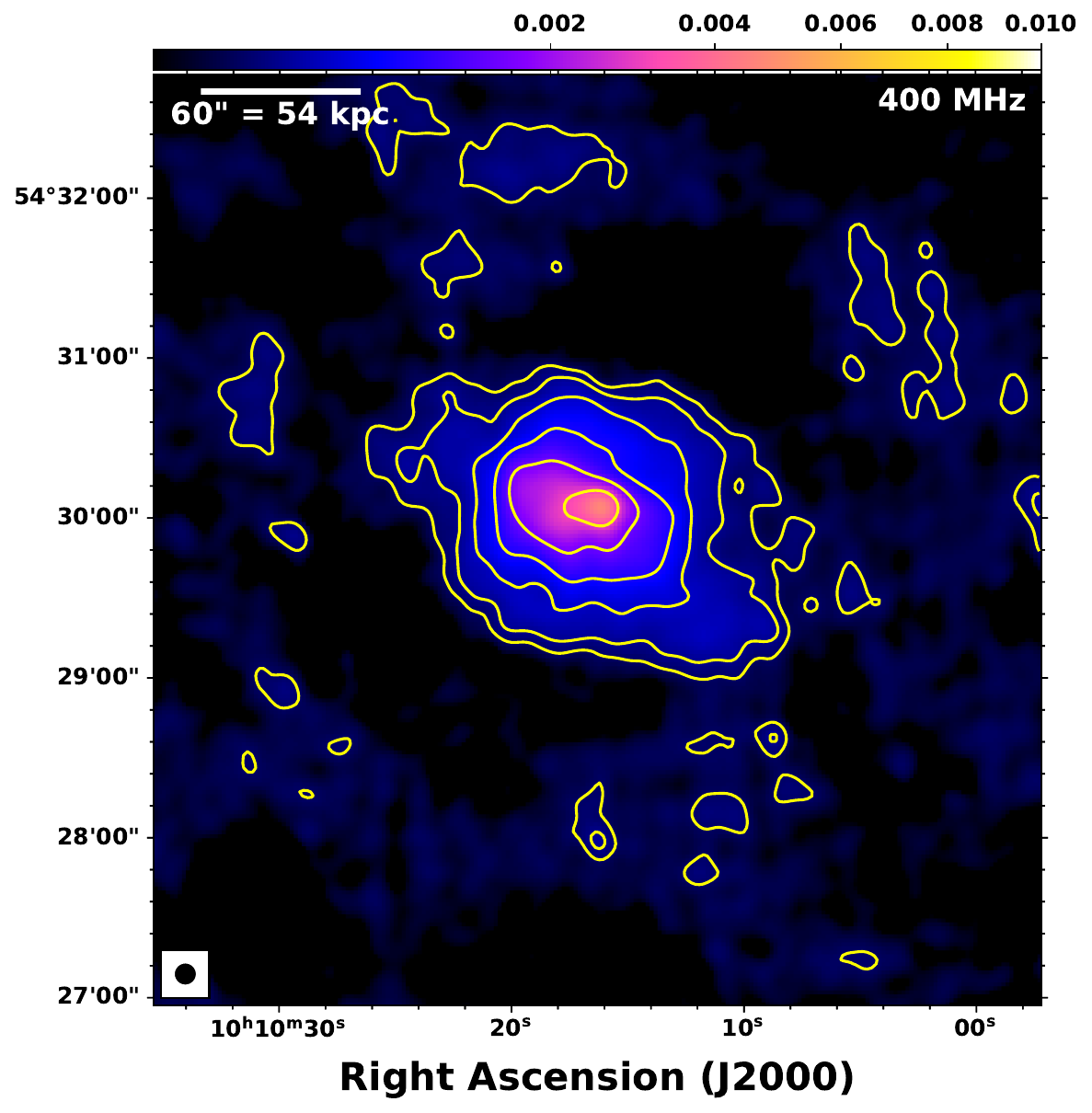} &
         \includegraphics[width=0.30\textwidth]{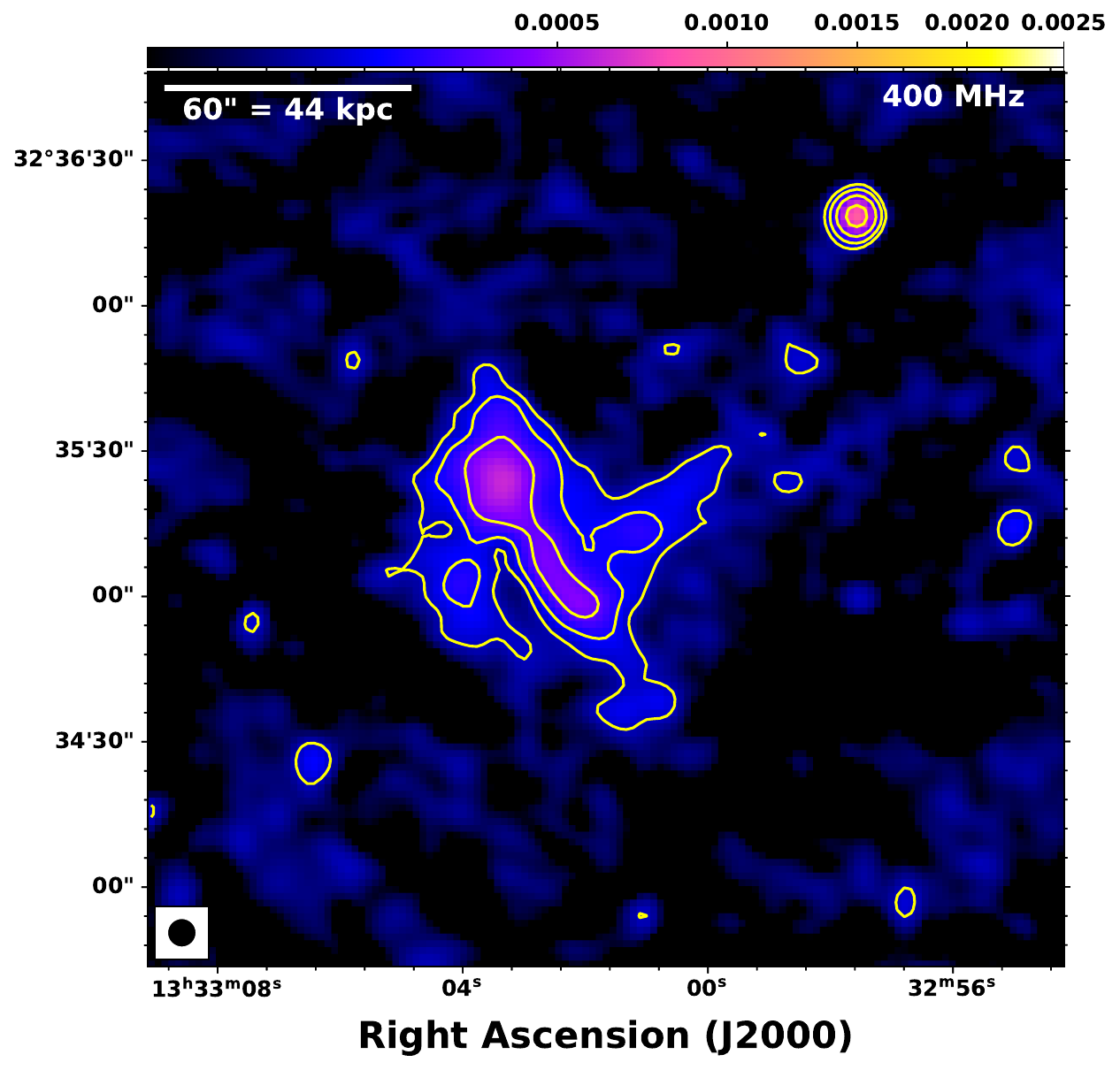}\\
    \end{tabular}
    \caption{\textit{Left panel:} The LOFAR (upper) and uGMRT (lower) surface brightness maps of SDSSTG8102 are shown in color, and the contours start with 3$\sigma_{\rm rms}$ $\times$ [1,2,4,...], with $\sigma_{\rm rms} = 110\mu$Jy~beam$^{-1}$ (at 144 MHz), and 30.5 $\mu$Jy~beam$^{-1}$ (at 400 MHz) respectively. The green circle marks the BGG of the group. The white line in the lower image shows the location of the change in surface brightness. \textit{Middle panel:} The radio surface brightness map for SDSSTG16393 (at 8$''$ resolution) is shown at both frequencies with a similar contour level as the upper panel, with rms values, $\sigma_{144 MHz} = 92 \mu$Jy~beam$^{-1}$ (upper), $\sigma_{400 MHz} = 32 \mu$Jy~beam$^{-1}$ (lower). \textit{Right panel:} The same is shown for SDSSTG28674 (at 7$''$ resolution), with similar contour levels as the left panel with $\sigma_{144 MHz} = 85.1 \mu$Jy~beam$^{-1}$ (upper), $\sigma_{400 MHz} = 28.7 \mu$Jy~beam$^{-1}$ (lower).}
    \label{xgap-radio-images}
\end{figure*}

\subsection{Continuum Images}

Our new multi-frequency images provide a comprehensive view of the radio emission from the central regions of these groups. We have achieved an rms noise of 28 - 32 $\mu$Jy~beam$^{-1}$ at 400 MHz with uGMRT, and 85 $-$ 110 $\mu$Jy~beam$^{-1}$ at 144 MHz with LOFAR; LOFAR becomes more sensitive than uGMRT for sources with spectral indices steeper than $\alpha \leq$ $-$1.2. We present the uGMRT observations of these groups for the first time. Here, we will discuss the most prominent structures in the various regions of those individual groups.

\subsubsection{SDSSTG8102}

SDSSTG8102 is the most massive among the three groups in our study (redshift $z$ = 0.033), with a mass of 1.5$\times$10$^{14}$M$_{\odot}$, consistent with a low-mass cluster, and a lower X-ray luminosity, compared to the others (Table~\ref{source-list}). The BGG and most of the group members identified from SDSS are early-type galaxies in this group. This group consists of 26 member galaxies with a velocity dispersion of $\sim$500 km~s$^{-1}$. We do not see any sign of tidal interactions between the BGG and its nearby companion galaxies.

The uGMRT 400 MHz and LOFAR 144 MHz images (Figure~\ref{xgap-radio-images}, left panel) reveal central radio emission extending several tens of kpc from the group center. At 144 MHz, the morphology suggests two lobe-like features extending northwest and southeast. Higher-resolution and high-frequency data from FIRST (5$''$) at 1400 MHz \citep{becker94} and VLASS ($2.1''$) at 3000 MHz \citep{lacy20} show only a patch of diffuse emission, with no identifiable jet or core (green contour from VLASS $2''$ resolution in Figure~\ref{op_maps}). The peak of the radio emission from uGMRT and LOFAR coincides with the BGG of the group. The total emission spans a projected length of $\sim$93 kpc at 144 MHz and $\sim$72 kpc at 400 MHz, with a projected width of $\sim$52 kpc at both frequencies, indicating that uGMRT and LOFAR recover similar spatial scales (using 3$\sigma$ contours as a reference). The emission is predominantly extended, with no compact radio sources embedded within it (Figure~\ref{op_maps}). Notably, the northern extension bends westward over $\sim$56 kpc ($\sim$85$''$), suggesting either intrinsic motion of the BGG at the central part of this group toward the east, `leaving behind' radio emission or an external factor that has influenced the radio-emitting plasma. While no filamentary substructures are detected at any frequency, the inner region shows a double-peaked brightness structure separated by $\sim$23 kpc. The southern peak lies slightly closer to the BGG, which overall appears roughly centered between the two bright spots; the southern feature is more extended, suggesting a projection effect (with the jet axis inclined either in SE–NW direction or toward us). Assuming the jet is oriented in the plane of the sky, the differences between the two peaks would point to varying plasma conditions, likely influenced by the environment. Beyond this double inner structure, the emission gradually fades, except for a sharp drop in surface brightness (by a factor of 5$-$6) across a $\sim$10 kpc ($\sim$15$''$) boundary in the southwest. Furthermore, in the uGMRT image, we observe radio emission extending to the northeast, close to source A, up to 50$''$ (33 kpc), with a surface brightness approximately three times fainter than (at 3$\sigma$) that of the central radio emission. The source A, adjacent to the extended emission, shows a double structure in the LOFAR and uGMRT images, and has no optical counterpart in the SDSS image; therefore, it is a background radio source.

\subsubsection{SDSSTG16393}

SDSSTG16393 has the lowest mass among the three, 4.8 $\times$ 10$^{13}$M$_{\odot}$, and the highest X-ray luminosity (Table~\ref{source-list}). This group consists of 46 members and is dominated by early-type galaxies, with relatively high velocity dispersion of approximately 295 km~s$^{-1}$. MCG+09-17-036, the (giant elliptical) BGG, is located at $z$=0.047, consistent with the group's redshift (Figure ~\ref{op_maps}). UGC~5479, a spiral galaxy at a similar redshift, lies $\sim$54~kpc west of the BGG in projection, and while no clear signs of interaction are observed, its proximity may suggest some environmental influence affecting the BGG. Two additional group members, located close to the north and south of the BGG, however, the evidence of interactions is limited with the current data.

The uGMRT 400 MHz and LOFAR 144 MHz images (middle panel, Figure~\ref{xgap-radio-images}) reveal an extended radio source centred on the galaxy group. The morphology includes a bright central region, marginally extended towards the east ($\sim$20 kpc), surrounded by more diffuse emission. However, a distinct core–jet–lobe structure is not observed. High-resolution data from VLASS confirm a compact, unresolved core at the BGG position (green contour Figure~\ref{op_maps}), but no jet-like features or extended lobes are detected at these resolutions and sensitivities. Aside from the core, no additional compact sources are embedded within the diffuse lobe-like region. The surface brightness distribution is smooth, with no identifiable filamentary substructures. It peaks at the centre, coincident with the BGG (Figure~\ref{op_maps}), and declines gradually toward the outskirts. The uGMRT image shows broadly round morphology with mild elongation along the northeast–southwest axis, whereas the LOFAR image appears more asymmetric, with emission primarily extending north and southwest. The diffuse emission extends up to $\sim 100$ kpc southwestward in both images, while along the northeast direction, uGMRT recovers slightly more emission ($\sim 92$ kpc) than LOFAR ($\sim 82$ kpc).

\subsubsection{SDSSTG28674}

\begin{figure*}[ht!]
    \centering
    \includegraphics[width=0.32\textwidth]{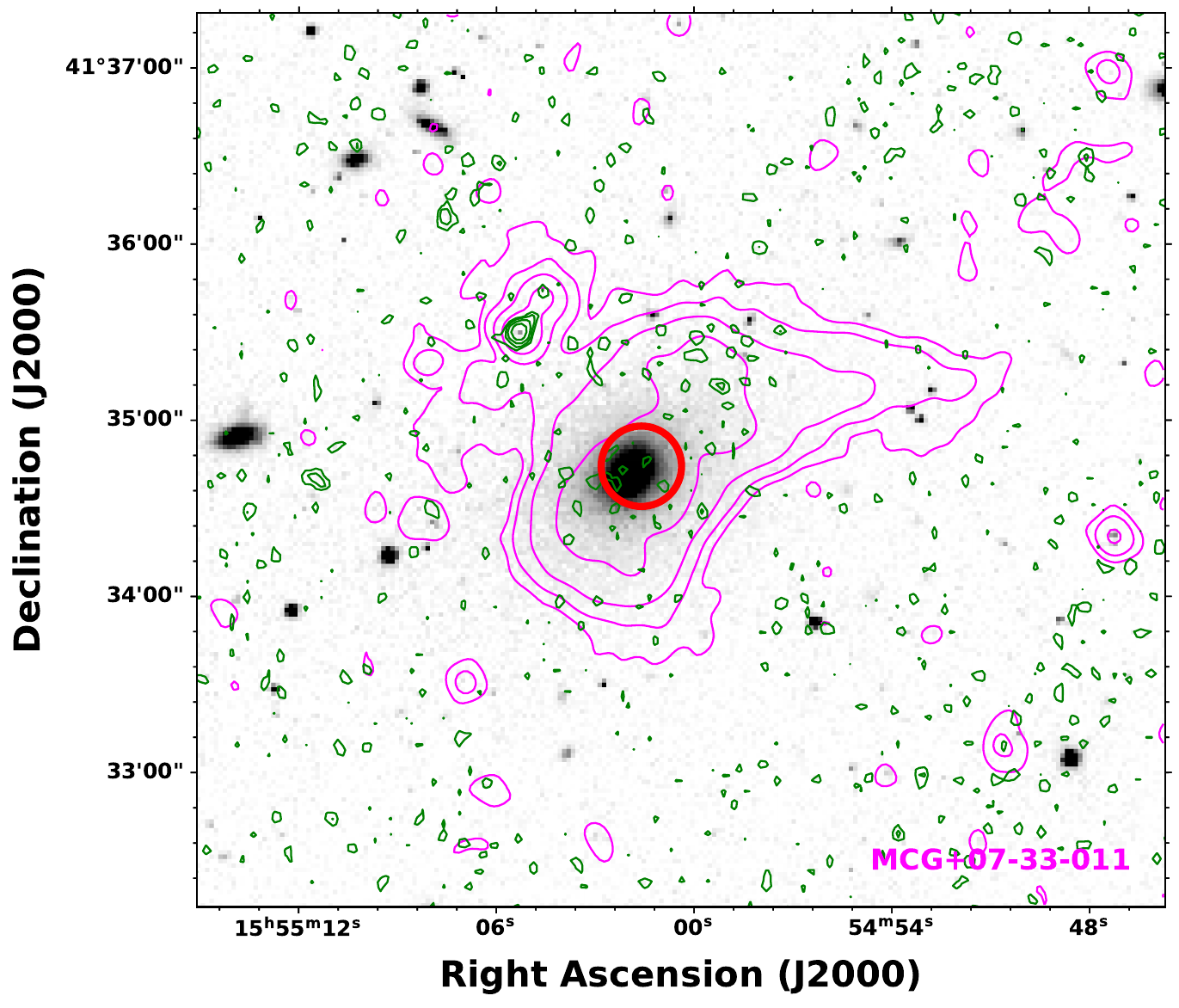} \includegraphics[width=0.27\textwidth]{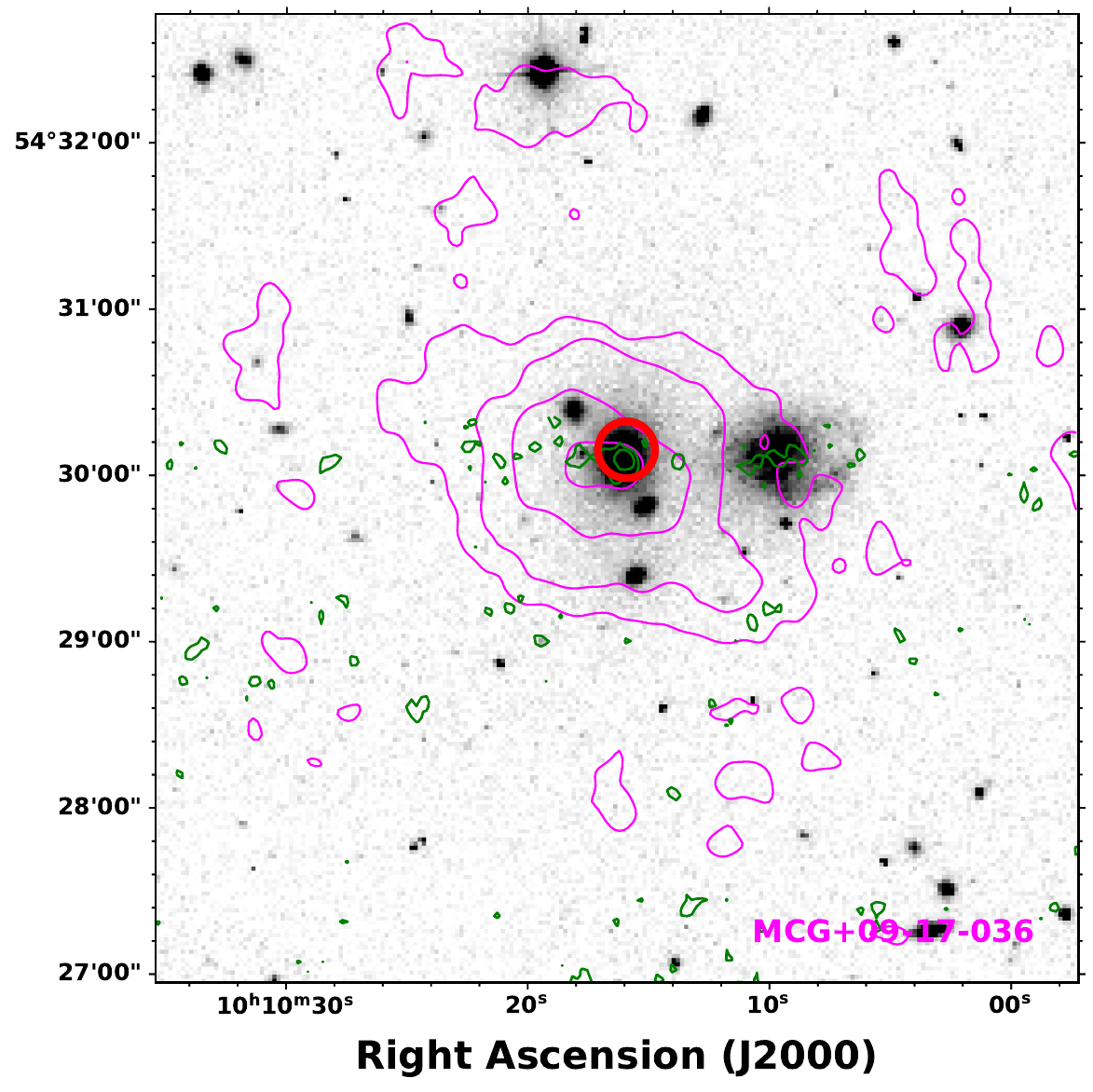} \includegraphics[width=0.28\textwidth]{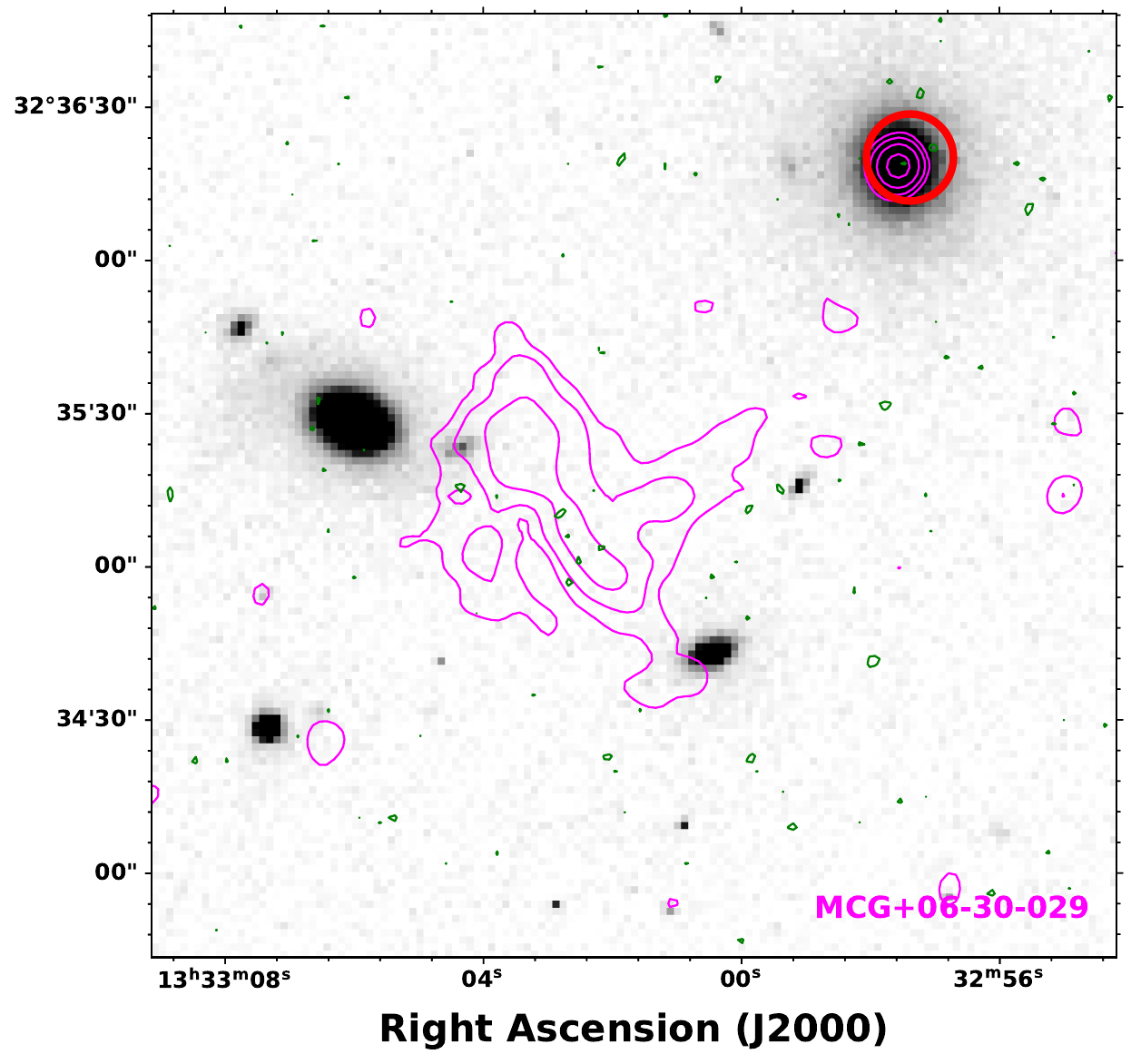} 
    
    \caption{\textit{Left:}The SDSS i-band image of the SDSSTG8102 is shown in colour (grey), and the overlaid magenta contours are from uGMRT 400 MHz, starting with 3$\sigma_{\rm rms}$ $\times$ [1,2,4,...], where $\sigma_{\rm rms} = 30.5 \mu$Jy~beam$^{-1}$. The green contours are from the VLASS 3.0 GHz image at 2$''$ resolution, where the contour levels are 3$\sigma_{\rm rms}$ $\times$ [1,2,4], with $\sigma_{\rm rms}=108\mu$Jy~beam$^{-1}$. The red circle indicates the BGG of the group. \textit{Middle:} The same is shown for the SDSSTG16393, with the same contour level from uGMRT with an rms of 32$\mu$Jy~beam$^{-1}$. The green contours are from the VLASS 3.0 GHz map. \textit{Right:} The same is shown for the SDSSTG28674, with the same contour level as the upper panels, with an rms of 28.7$\mu$Jy~beam$^{-1}$. The green contours are from the VLASS map, at a similar significance.}
    \label{op_maps}  
    
\end{figure*}

SDSSTG28674 has a mass of 1.2 $\times$ 10$^{14}$M$_{\odot}$, and lowest X-ray luminosity among the three groups (Table~\ref{source-list}). It is a system that appears to be dominated by several bright galaxies along with one giant elliptical, the BGG, at the center. This group comprises 20 galaxy members with a velocity dispersion of $\sim 300$ km~s$^{-1}$. MCG+06-30-029 is the BGG (lower right panel, Figure~\ref{op_maps}) situated at a redshift ($z$ = 0.0371) similar to that of the galaxy group. We do not see any signs of tidal interaction between the BGG and its nearby companion galaxies.

The uGMRT full-resolution image at 400 MHz and the LOFAR 144 MHz image (lower panel, Figure~\ref{xgap-radio-images}) reveal a compact source associated with the BGG and an extended low-surface-brightness source to the southeast of the BGG. No compact discrete sources are embedded within the diffuse emission, and the radio emission from the BGG is undetected in the VLASS image (green contour Figure~\ref{op_maps}). In the innermost region of the extended emission, two brightness peaks are observed, separated by $\sim$20 kpc and embedded within a broader, mushroom-shaped distribution. The radio emission is elongated along the northeast–southwest axis, with projected sizes of $\sim 81$ and 56 kpc in length, and $\sim 39$ and 33 kpc in width, at 144 and 400 MHz, respectively. At 144 MHz, a faint trail of emission is detected connecting the diffuse structure to the BGG, extending $\sim$55 kpc along the northwest–southeast direction. The BGG also shows compact radio emission detected at a significant level ($\sim$ 6$\sigma$), with the peak of optical and radio emissions well coinciding.

\subsection{Integrated spectrum} \label{sec:int-spec}

\begin{figure}
    \centering
        \includegraphics[width = \columnwidth]{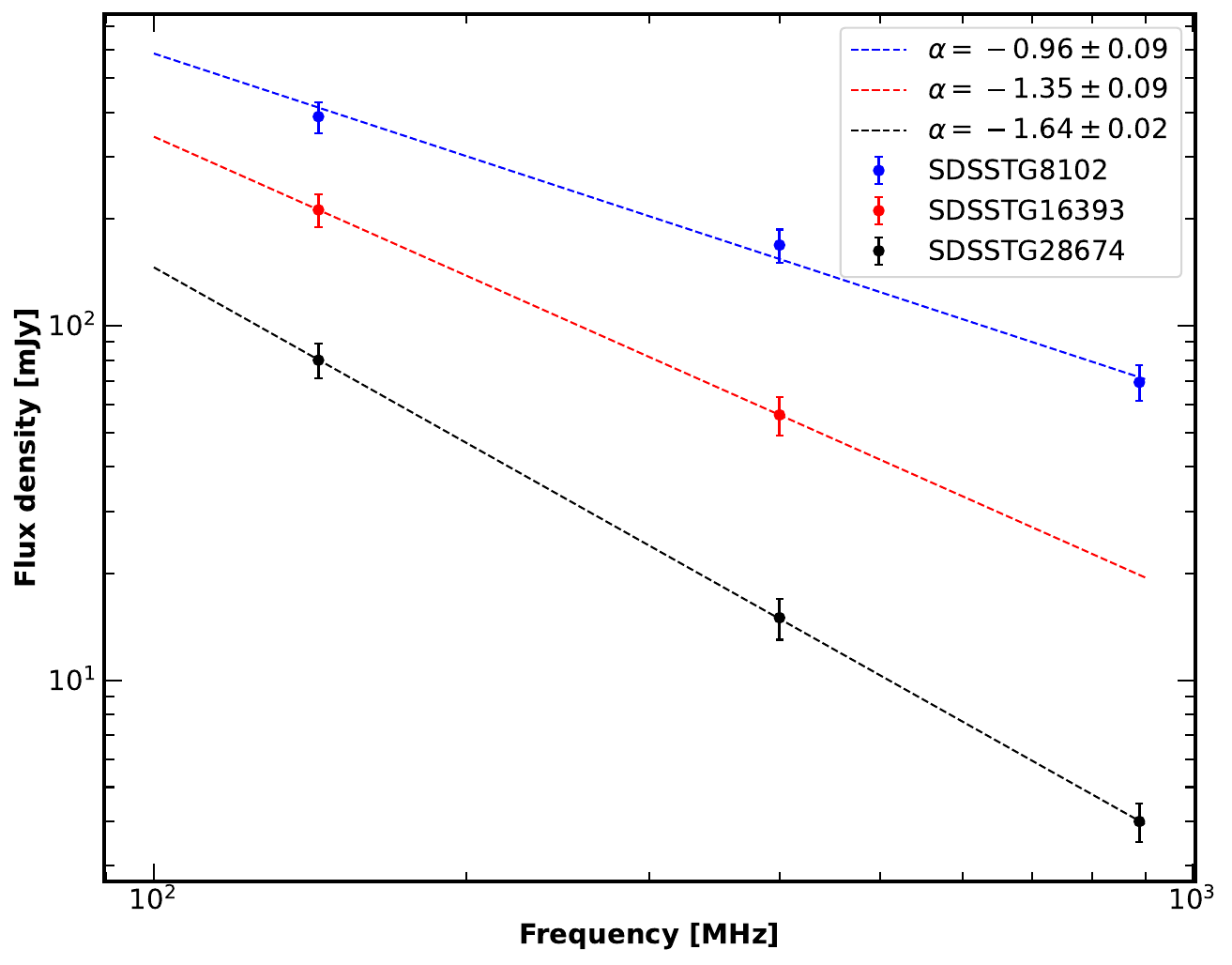}
        \caption{Integrated spectrum of the central radio emission for each of the groups (at different colours) is shown between 144 MHz and 887 MHz. The dashed lines represent the fitted single power law. We have estimated the average flux density from a common region at each frequency for individual groups.}
    \label{img:int-spec}
\end{figure}

We combined our new uGMRT and LOFAR observations with other archival data to investigate the integrated spectral profile of the central diffuse sources. All radio images (LOFAR, uGMRT, and RACS-Low) were corrected to a common flux density scale using \cite{perley&butler17}. The largest angular scale of the diffuse radio emission among the three groups is 2.79$'$, which is well below the maximum recoverable scales of our data (LOFAR HBA: 70$'$, uGMRT Band3: 20$'$ $-$ 30$'$). As diffuse radio emission in all three groups, apart from the core, is free of compact sources, we estimate the flux density for each group by convolving the radio images to the lowest available resolution across all frequencies, selecting a common area (following a 3$\sigma$ contour) for their flux density measurement. The integrated flux density values are provided in Table~\ref{int-spec-tab}. The integrated spectra of the radio emission for each group are shown in Figure~\ref{img:int-spec}. 

SDSSTG8102 displays no evidence of curvature or a spectral break. Given that only three flux density measurements are available (LOFAR, uGMRT, and RACS-low), we do not attempt spectral modelling, since the lack of higher-frequency points prevents meaningful constraints on radiative ageing models. Instead, we estimate the integrated spectral index to be $-0.96 \pm 0.11$ using these three frequencies. Examining the northwest and southeast components separately, we find similar values: $-0.92 \pm 0.13$ (northwest) and $-0.82 \pm 0.11$ (southeast). The radio power of the BGG at 400 MHz is 4.6 $\times$ 10$^{24}$ W~Hz$^{-1}$.

The integrated spectral index for SDSSTG16393, between 144 and 400 MHz, is estimated to be $-1.35 \pm 0.09$, based on a well-fitted single power law (Figure~\ref{img:int-spec}). Although the 3.0 GHz VLA observations (Figure~\ref{op_maps}) indicate some emission, the VLASS observations are insensitive to angular scales larger than $\sim 30 ''$, and a substantial portion of the extended emission is resolved out. Consequently, the flux density at this frequency is underestimated, and no information about possible high-frequency spectral behaviour can be obtained. The radio power of the BGG at 400 MHz is 5.1 $\times$ 10$^{23}$ W~Hz$^{-1}$.

The radio emission (compact radio source associated with the BGG and the extended source) for SDSSTG28674 is well-fitted with a single power law with an integrated spectral index of -1.64 $\pm$ 0.02, between 144 and 887.5 MHz. The spectral index of the compact source at the BGG between 144 and 400 MHz is $-0.62 \pm 0.06$. The radio power at 400 MHz of the BGG is 7.5 $\times$ 10$^{26}$ W~Hz$^{-1}$.

\begin{table}
  \centering
  \caption{Flux density estimates for radio sources of each group.}
\begin{tabular}{@{}cccc@{}}
    \hline
     Source & Freq. & Flux density (mJy) & Ref. \\
      \hline\hline

    SDSSTG8102 &144 & 389.07 $\pm$ 39.00 &  This work  \\
      & 400 & 184.94 $\pm$ 18.00 & This work \\

    &887.5 & 64.29 $\pm$ 8.00 &  RACS-low \\

    \hline

    SDSSTG16393 &144 & 212.33 $\pm$ 23.00 &  This work  \\
      & 400 & 62.07 $\pm$ 7.00 & This work \\

    \hline
    
    SDSSTG28674 &144 & 80.03 $\pm$ 9.00 &  This work  \\
      & 400 & 15.02 $\pm$ 2.00 & This work\\
    
    &887.5 & 4.00 $\pm$ 0.50 &  RACS-low \\

    \hline
\end{tabular}
     
     \tablefoot{References for the RACS-low is \citet{mcconnell20,hale21}.}

  \label{int-spec-tab}
\end{table}

\begin{figure*}[ht!]
    \centering
    \includegraphics[width=0.33\textwidth]{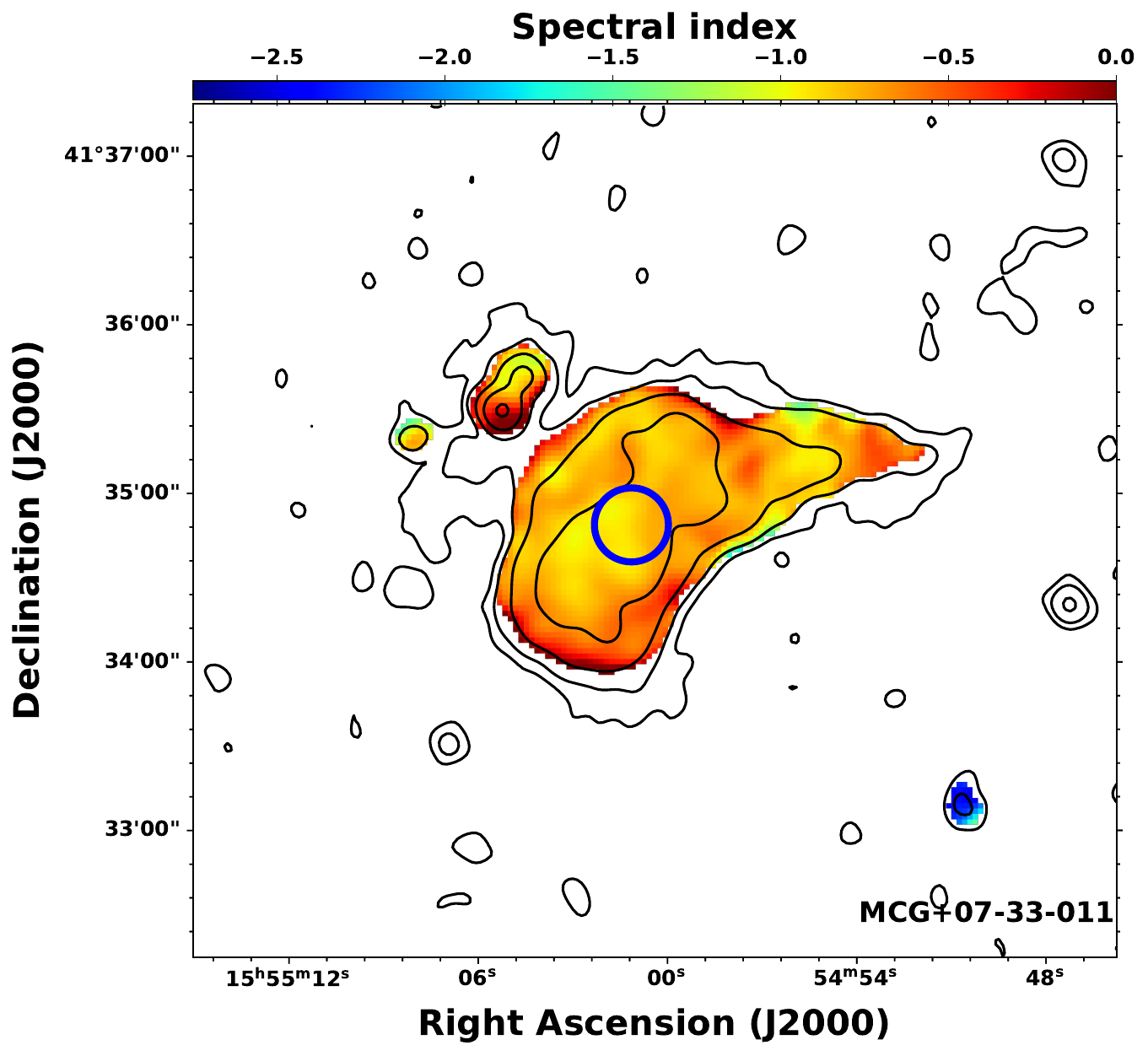} \includegraphics[width=0.29\textwidth]{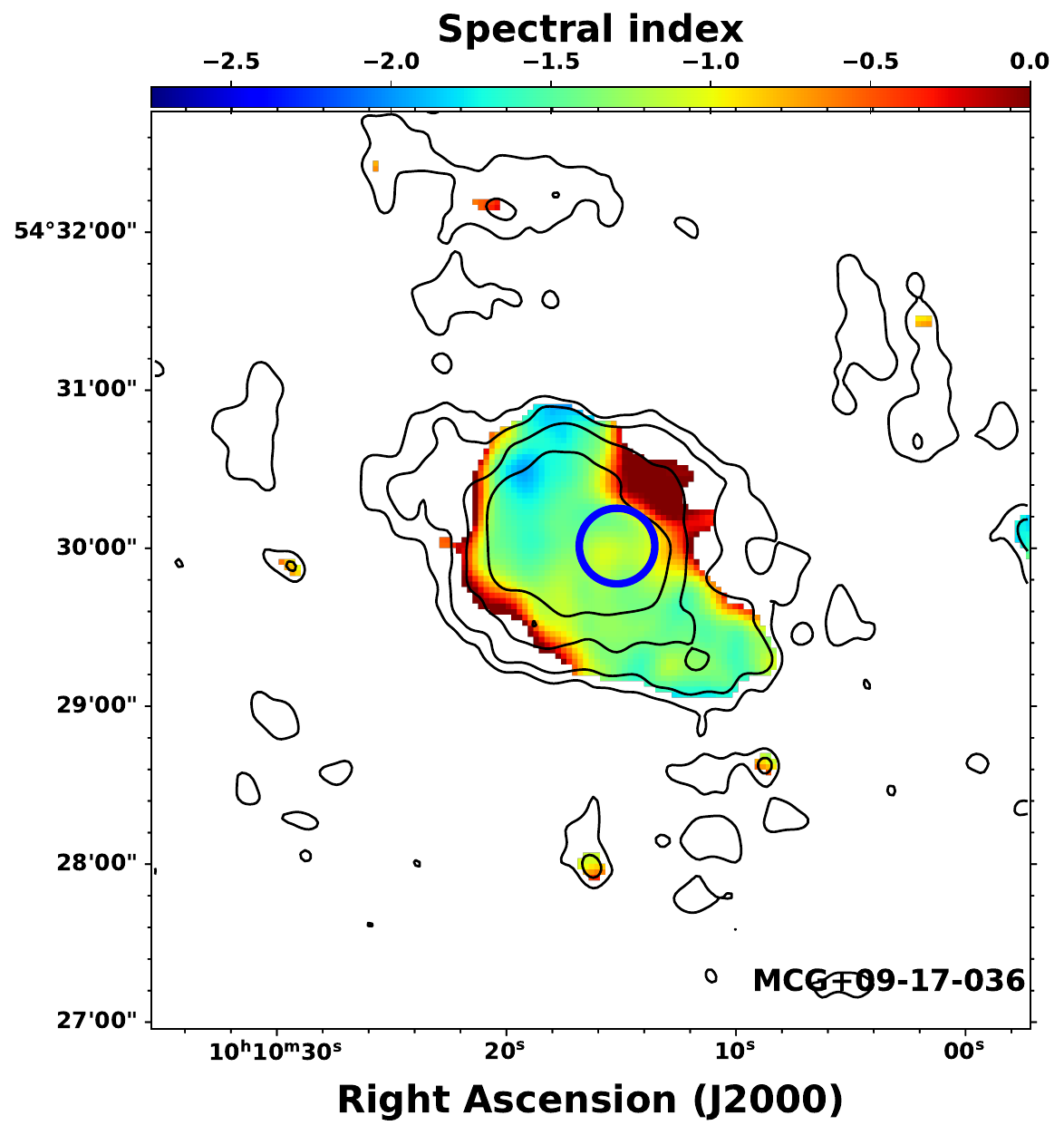} \includegraphics[width=0.305\textwidth]{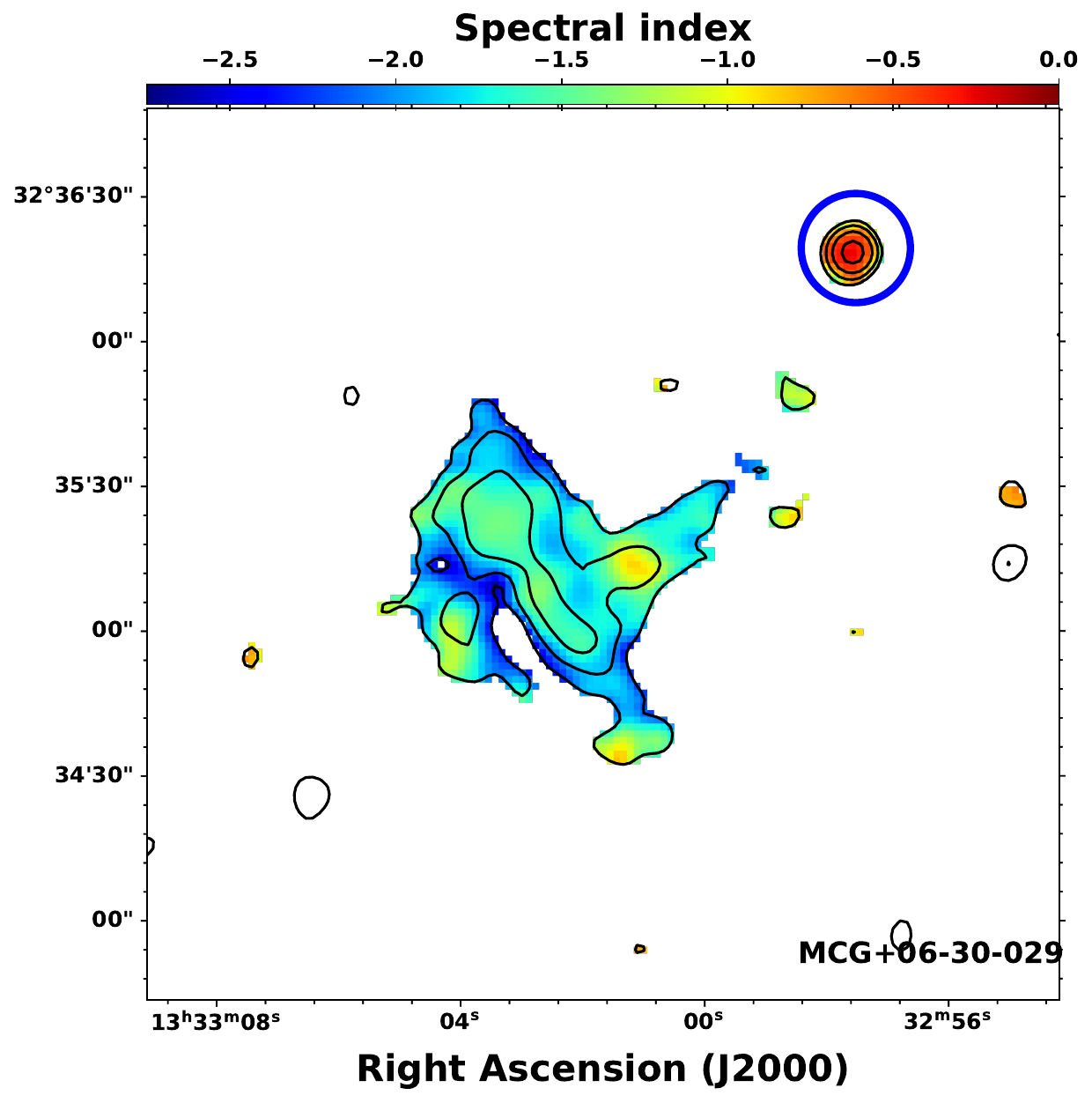}

    \caption{\textit{Left:} The spectral index map between 144 and 400 MHz is shown for SDSSTG8102. The overlaid black contours are from uGMRT 400 MHz, starting with 3$\sigma_{\rm rms}$ $\times$ [1,2,4,...], where $\sigma_{\rm rms} = 30.5 \mu$Jy~beam$^{-1}$, and the blue circle indicates the BGG of the group. \textit{middle:} The spectral index map is shown for SDSSTG16393. The contour level is similar to the upper panel, and the $\sigma_{\rm rms} = 32 \mu$Jy~beam$^{-1}$. \textit{Right:} The same is shown for SDSSTG28674, having similar contour levels, with an $\sigma_{\rm rms} = 28.7 \mu$Jy~beam$^{-1}$.}
    \label{spec-map}
\end{figure*}

\subsection{Resolved spectral index map}

Spatially resolved spectral indices provide valuable insights into the particle acceleration mechanisms behind group-scale extended radio AGN \citep[e.g.,][]{kolokythas20,rajpurohit2024,pasini25,riseley25}. Using the 400- and 144-MHz images, we produced a resolved spectral index map for the extended emissions in each group. These spectral index maps were generated by convolving each image to a common resolution, which enhances the signal-to-noise ratio (S/N) of the extended emission and preserves any small-scale features that would otherwise average out. The procedure for creating the spectral index maps follows the method described in \cite{degasperin17}, where the spectral index values were estimated as the logarithmic ratio of the two images, and to estimate the uncertainties, the flux density values of each pixel are obtained 1000 times from a Gaussian distribution that has the same mean as the estimated flux density value and the same standard deviation as the background rms of the images. The individual properties and a detailed discussion of the spectral index maps are presented below for individual groups:

\textbf{SDSSTG8102:} The spectral index map appears to be relatively uniform throughout the extent of the emission (Left panel Figure~\ref{spec-map}). In the central region, the average spectral index ranges from $\sim$ $-$0.7 to $-$0.9. This general spectral index distribution aligns with the integrated spectral index measurement for total diffuse emission (Section~\ref{sec:int-spec}). Different spectral indices are observed for the inner bimodal surface brightness distribution, with the southern part showing an average spectral index of $\sim$ $-$0.94 $\pm$ 0.09 and the northern part showing $\sim$ $-$0.75 $\pm$ 0.11. The flatter spectral indices suggest that these regions are predominantly filled with younger radio plasma. At the BGG location, the spectral index is mainly uniform with $\sim-$ 0.7. The northwest extension also exhibits a uniform spectral index of $\sim-$0.75. The spectral index error ranges from $\sim 0.01$ near the core to $\sim$0.1 in the outer regions (shown in Figure.~\ref{spec-map-err}, see Appendix~\ref{app:a} for the spectral index error maps).

\textbf{SDSSTG16393:} The spectral index map for the extended emission does not show any clear trends or small-scale features (middle panel, Figure~\ref{spec-map}). However, the overall spectral index distribution is steep (below $-1.0$) over the extent of the emission from the southwest to the northeast. In the core region, surrounding the BGG, the average spectral index is $\sim-$1.0, while in other parts of the extended emission, the spectral index steepens to $\sim-$1.6. We observe a steepening of the spectral index moving from the central region to outwards. Such a steep spectrum suggests that these regions may contain considerably older radio plasma, possibly originating from some remnant radio lobes. The error in the spectral index ranges from $\sim 0.01$ near the core to $\sim 0.2$ in the outer regions. However, the errors in the southwest part are higher compared to northern part. 

\textbf{SDSSTG28674:} The spectral index distribution shows spatial variations 
(right panel, Figure~\ref{spec-map}). In particular, the spectral index steepens toward the eastern part of the emission, while remaining relatively uniform ($\sim -$1.5) along the northeast–southwest axis. A notably steeper region, with a spectral index of $\sim -$1.9, is observed surrounding the main streak. The BGG itself exhibits a flatter spectral index, averaging greater than $-$0.5, contrasting with the diffuse emission, which generally has a steeper index of $\sim -$1.6, consistent with the integrated spectral index (Section~\ref{sec:int-spec}). Additionally, a patch with a spectral index of $\sim -$1.0 is located at the junction of the ‘L’-shaped structure, although no associated galaxy or point source is found at that position. Overall, the spectral index errors range from $\sim 0.05$ in the brighter central regions to $\sim 0.25$ in the outskirts.

\begin{figure*}[ht!]
    \centering
    \includegraphics[width=0.27\textwidth]{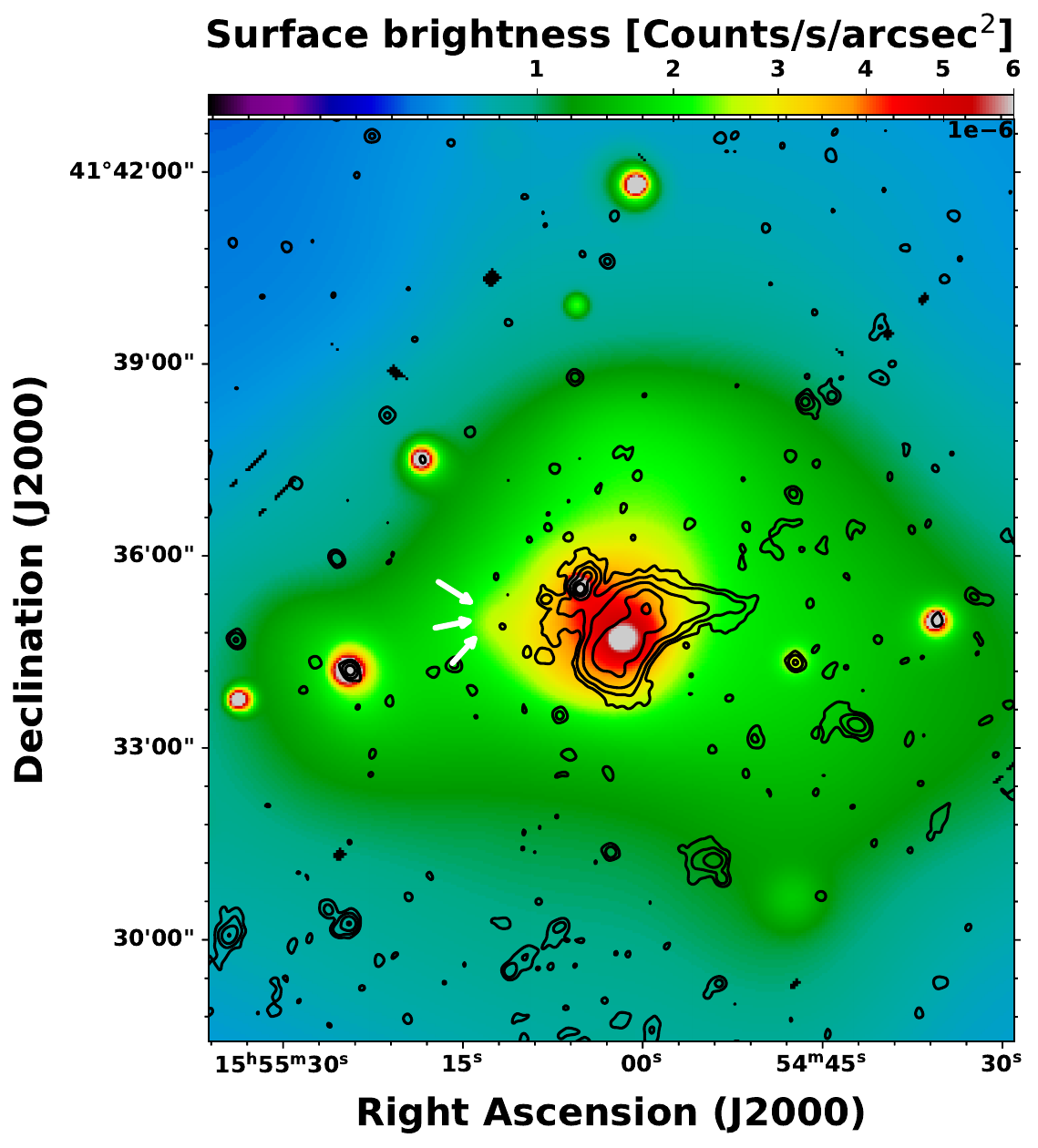}
    \includegraphics[width=0.27\textwidth]{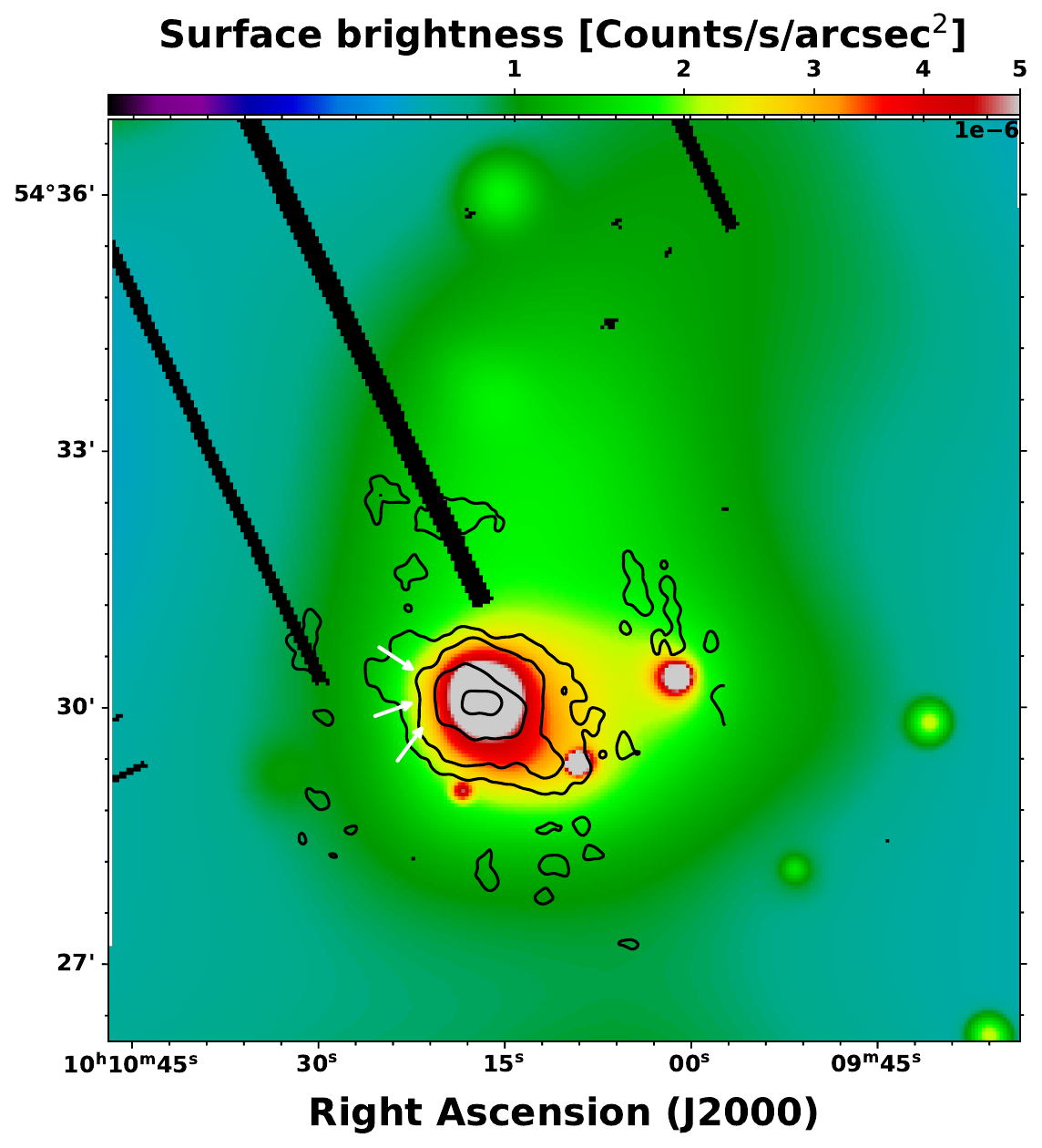}
    \includegraphics[width=0.26\textwidth]{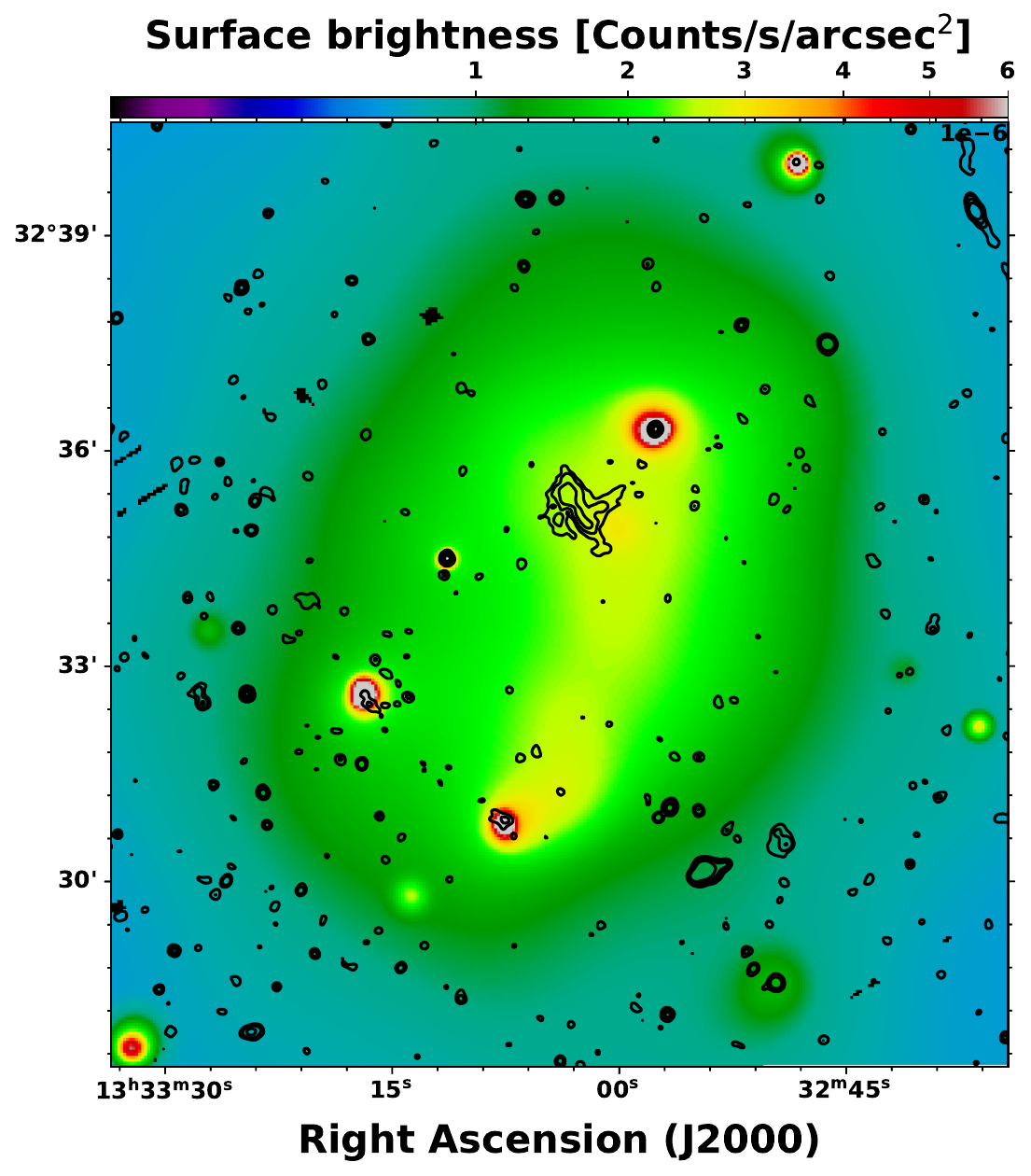}
    \caption{\textit{Left:} Adaptively smoothed 0.7 $-$ 1.2 keV image from the \textit{XMM-Newton} is shown in color for SDSSTG8102, and the overlaid contours are from uGMRT 400 MHz, starting with 3$\sigma_{\rm rms}$ $\times$ [1,2,4,...], with $\sigma_{\rm rms} = 30.5 \mu$Jy~beam$^{-1}$. \textit{Middle:} The same is shown for the SDSSTG16393 at 0.7 $-$ 1.2 keV range, with similar contour spacings, starting with an rms of 32$\mu$Jy~beam$^{-1}$. The black strips are the chip gaps for the \textit{XMM-Newton} CCDs. \textit{Right:} The X-ray map is shown for the SDSSTG28674, with the same contour level as the previous panels, with an rms of 28.7$\mu$Jy~beam$^{-1}$.}
    \label{x-ray-images}
\end{figure*}

\section{Results: X-ray observations} \label{xray_results}

This section presents the X-ray surface brightness maps from the \textit{XMM-Newton} for the three galaxy groups. Their individual properties, along with the detailed description of the thermal emission, are discussed as follows: 

\subsection{Surface brightness map}

\textbf{SDSSTG8102:} The \textit{XMM-Newton} observation of this galaxy group reveals a disturbed morphology of the diffuse emission from the IGrM (left panel, Figure~\ref{x-ray-images}). The extended emission is elongated along the east-west direction, with a projected size of approximately 300 kpc, and the overall shape of the emission is asymmetric. The brightest X-ray emission is closely aligned with the position of the BGG. However, another bright compact source is located about 40 kpc away from the BGG in the northeast direction, and the inner surface brightness distribution shows slight elongation in that direction, suggesting some interaction between the two. An extended feature is visible on the eastern side (white arrows); however, the resolution of the current \textit{XMM-Newton} image limits the ability to analyse for any potential surface brightness edge. The peak of the radio emission is coincident with the X-ray peak, and the radio emission at both frequencies is confined to the group core. Morphologically, the central X-ray emission is elongated eastward, while the radio emission is oriented in the northwest direction.

\textbf{SDSSTG16393:} X-ray observations of this group reveals disturbed IGrM emission (Figure~\ref{x-ray-images}, middle). The diffuse morphology, elongated northward ($\sim 250$ kpc), forms a broad triangular shape with the brightest peak centred on the BGG. A second bright group galaxy, $\sim 150$ kpc west of the BGG, may indicate a group-scale interaction or gravitational influence, driving the asymmetry. On the eastern side of the BGG, a sharp surface brightness drop by a factor of 2$-$3 is observed across $\sim 18$ kpc (white arrows). Further south, at 398 kpc away from the BGG, a $\sim 700$ kpc patch of extended diffuse X-ray emission is detected (Appendix ~\ref{diff-regions}), hosting three bright unresolved sources but no clear link to a concentrated galaxy population. Although some galaxies in this region share the group redshift, the nature of the emission remains uncertain, potentially linked to a separate infalling group.

The radio morphology closely follows the thermal emission, with both 144 and 400 MHz emission largely confined to the core. At 144 MHz, the emission extends northward, similar to the X-ray distribution, but is bounded on the eastern side by a sharp edge where the inner jet structure is deflected along northward; beyond this boundary, no 144 MHz emission is seen. However, at 400 MHz, some emission is detected past this edge.

\textbf{SDSSTG28674:} The X-ray emission from this group reveals two peaks and a disturbed IGrM. The diffuse emission is elongated along the northwest–southeast axis, spans $\sim$350 kpc in projection, and exhibits an overall elliptical morphology. The IGrM bridges the two galaxy cores and aligns with the inferred merger axis. Of these, the northwest galaxy is the BGG, while the southeast galaxy is a spectroscopically confirmed member ($z\sim0.0359$), supporting the scenario of an internal group merger rather than a projection with a background system. Although multiple galaxies are visible in the SDSS i-band image, most of them are not detected in X-rays.

Compact radio sources are detected coincident with the BGG and the dominant galaxy of the southern core. The extended radio structure is not coincident with any group member, and does not seem to be associated with any definite IGrM structure, although it lies close to the merger axis. The X-ray emission is significantly extended beyond the radio emission, and much of the thermal gas lacks corresponding radio emission.

\begin{figure*}
    \centering
    \begin{tabular}{ccc}
        \includegraphics[width=0.33\textwidth]{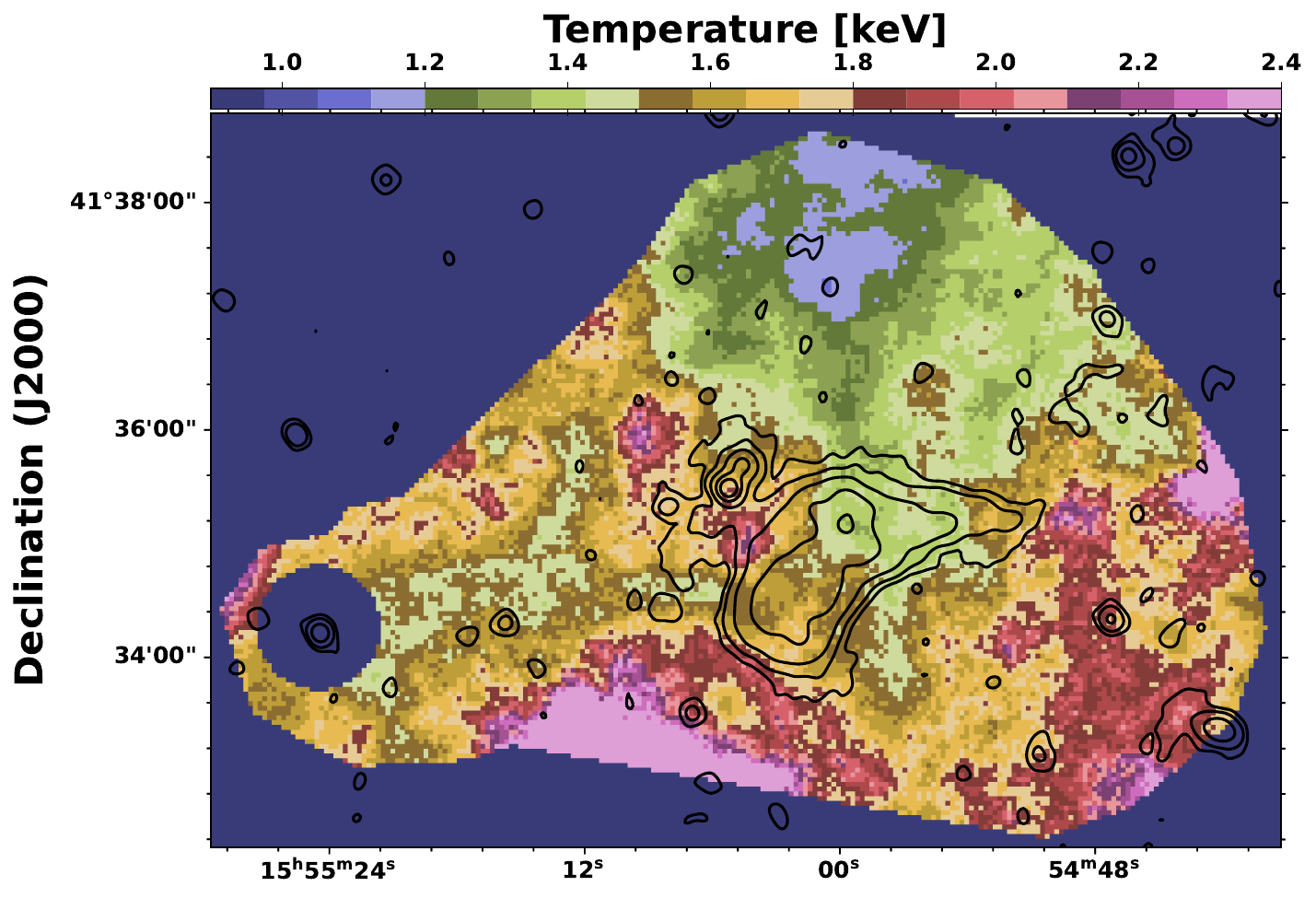} &
        \includegraphics[width=0.27\textwidth]{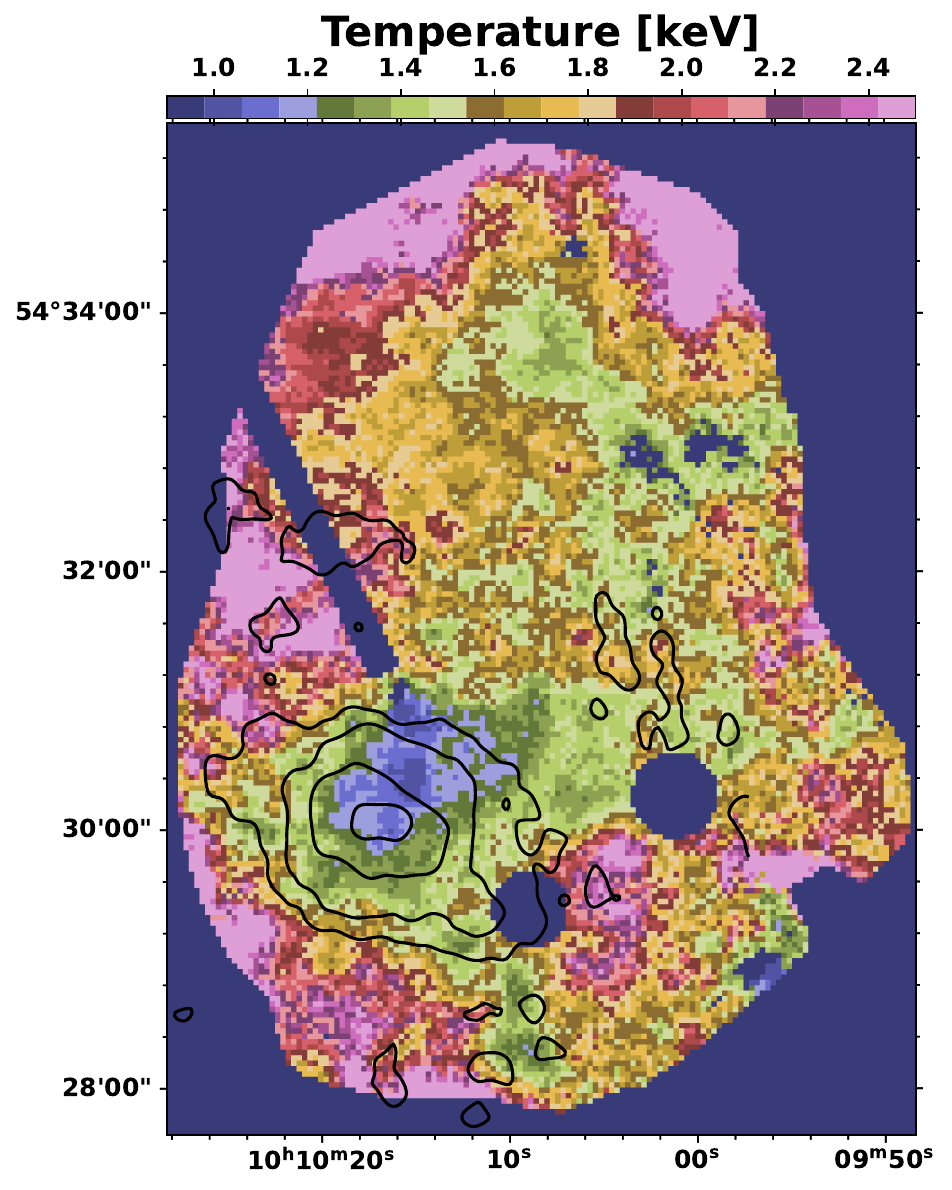} &
        
         \includegraphics[width=0.27\textwidth]{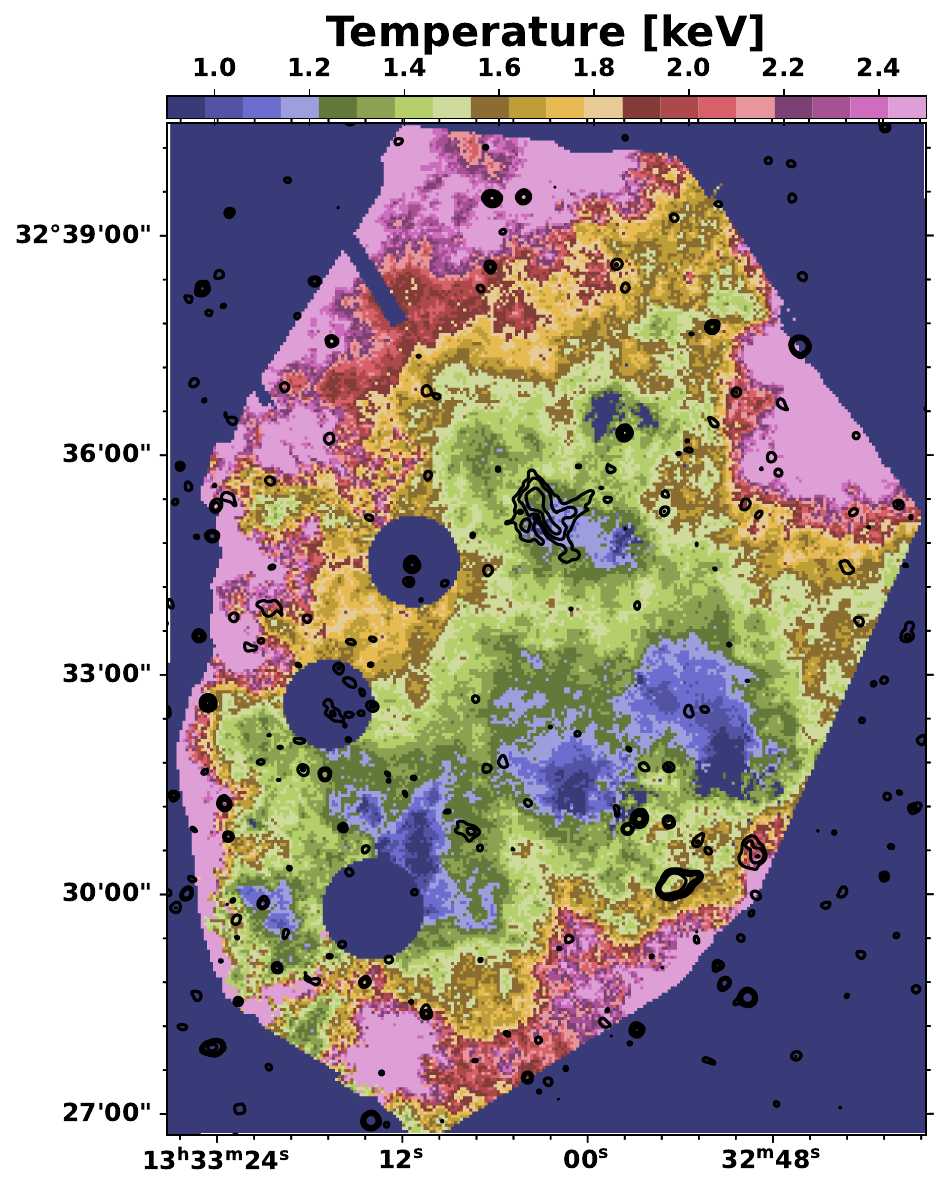}\\

        \includegraphics[width=0.33\textwidth]{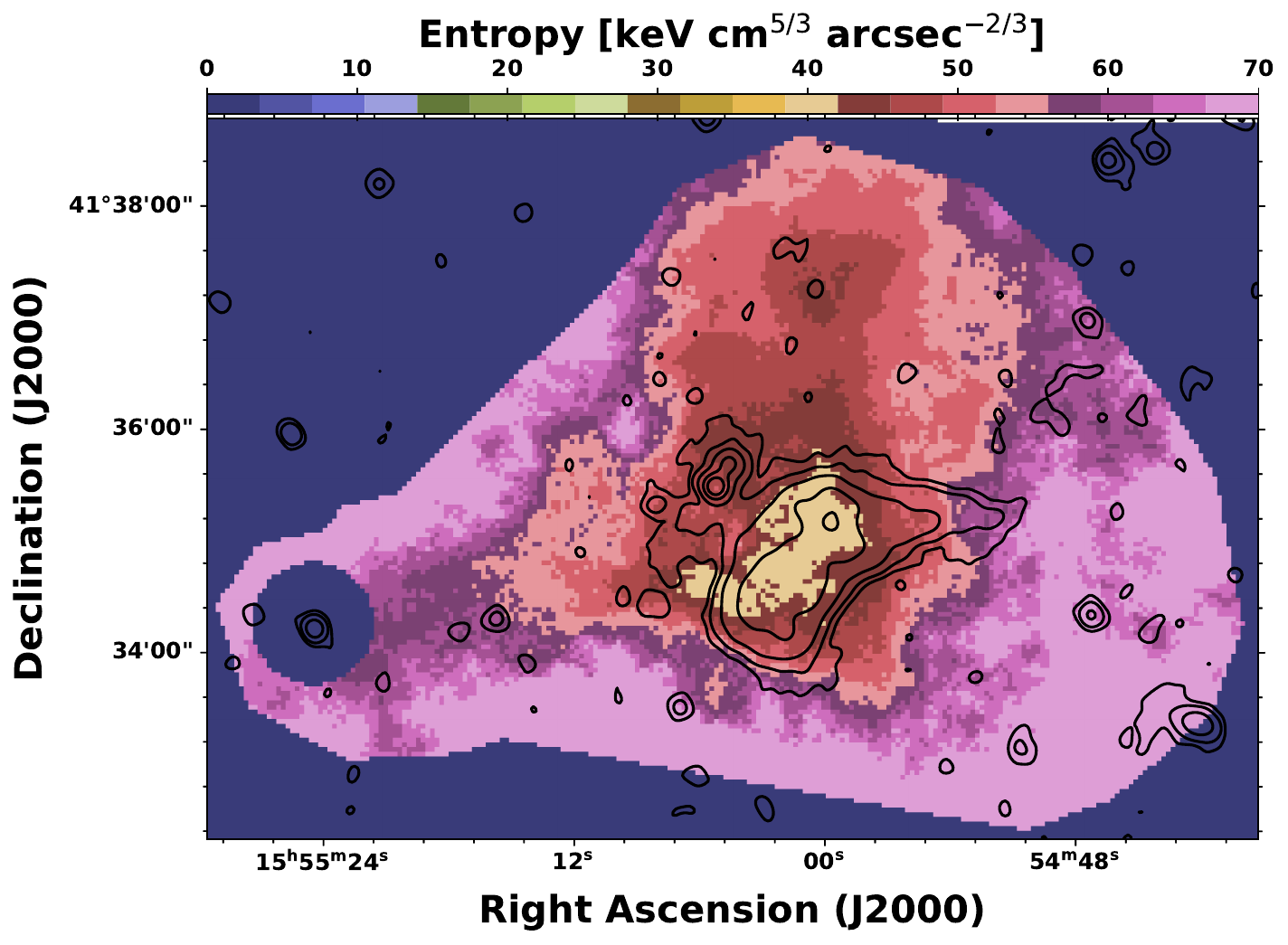} &
        \includegraphics[width=0.27\textwidth]{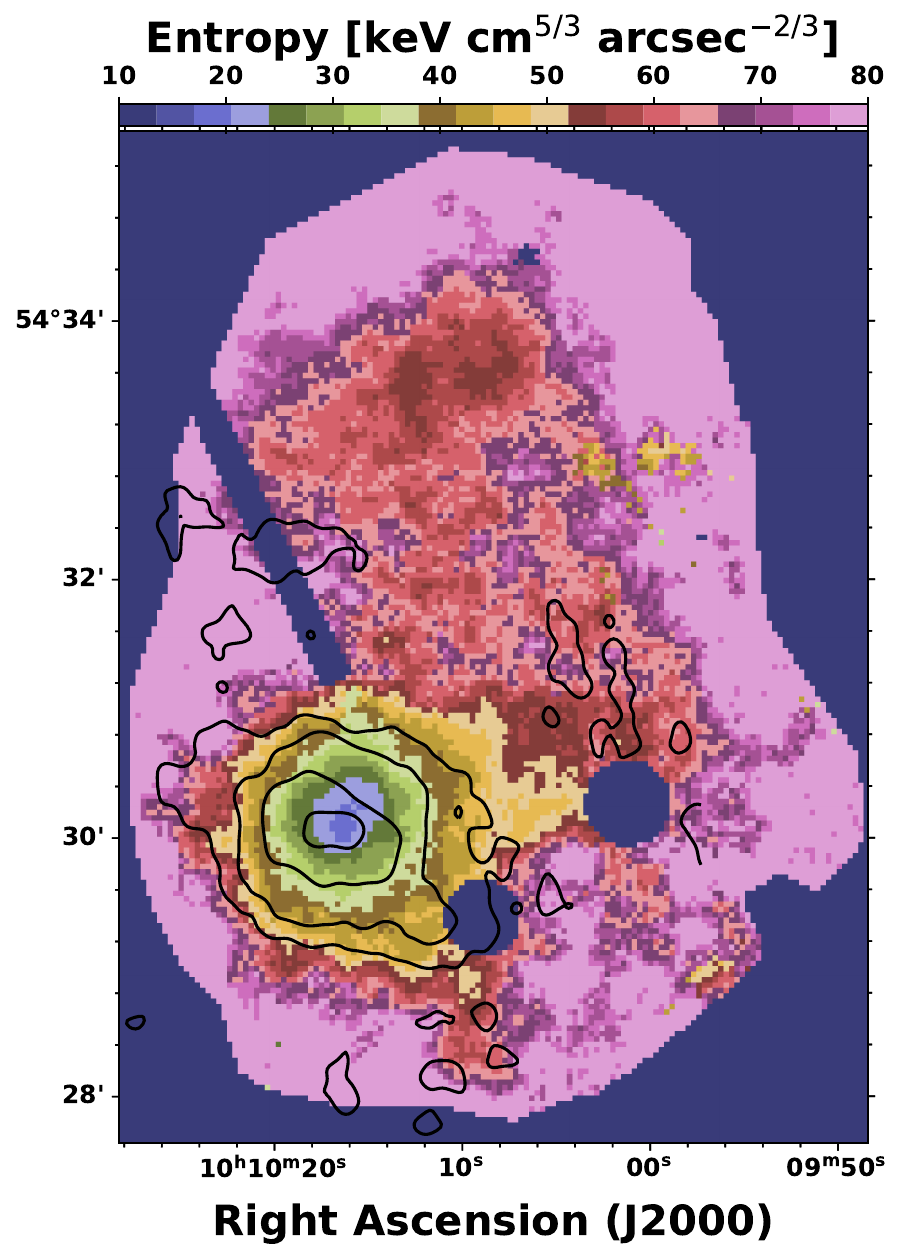} &
         \includegraphics[width=0.28\textwidth]{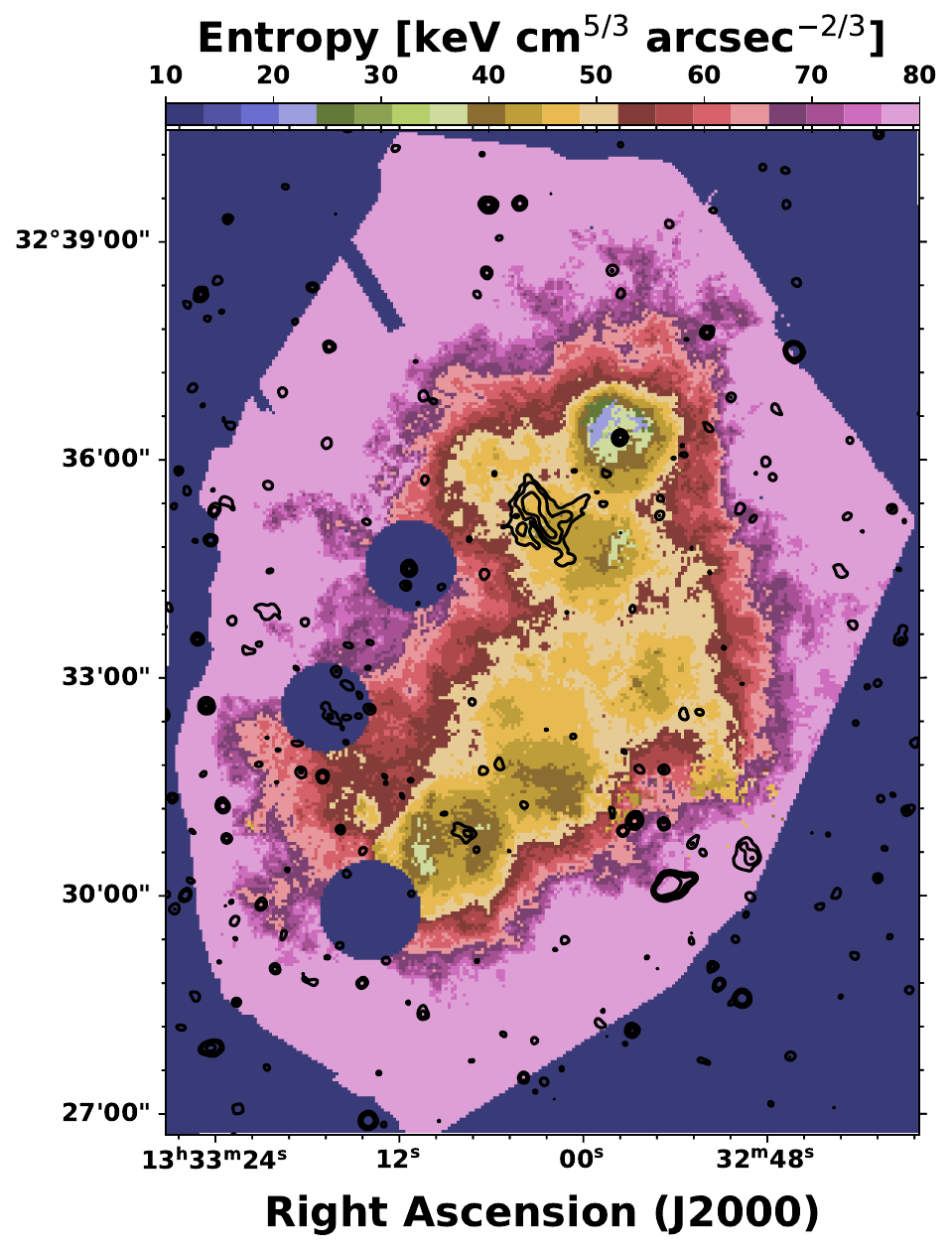}\\
    \end{tabular}
    \caption{\textit{Left:} The 2D temperature (upper) and corresponding entropy map (below) is shown for SDSSTG8102. The overlaid yellow contours are from uGMRT 400 MHz radio emission, starting with 3$\sigma_{\rm rms}$ $\times$ [1,2,4,...], where $\sigma_{\rm rms} = 30.5 \mu$Jy~beam$^{-1}$. \textit{Middle:} The temperature and entropy maps are shown for the group SDSSTG16393. The contour level is similar to the upper panel with an rms noise of $\sigma_{\rm rms} = 32 \mu$Jy~beam$^{-1}$. \textit{Right:} The same is shown for SDSSTG28674, having similar for the contour levels, with an $\sigma_{\rm rms} = 28.7 \mu$Jy~beam$^{-1}$.}
    \label{therm-maps}
\end{figure*}

\subsection{2D thermodynamical maps}

We applied the Adaptive Circular Binning (ACB) algorithm to map the 2D distribution of projected IGrM thermodynamical properties. Spectra were fitted with an absorbed APEC model, with hydrogen column density fixed to the Galactic value and abundance to 0.3 Z$_{\odot}$. Background spectra were taken from source-free CCD regions and supplemented with ROSAT All-Sky Survey data, with a model (local hot bubble, Galactic halo, cosmic X-ray background) jointly fitted to EPIC and RASS spectra. The procedure for producing \textit{XMM-Newton} temperature maps is detailed in \citet{botteon24}. The properties of the individual groups are discussed below.

\textbf{SDSSTG8102:} The 2D temperature and pseudoentropy distributions for this group are shown in Figure~\ref{therm-maps} (left). The temperature distribution is non-uniform with localised high-temperature patches. The central regions have an average temperature of $\sim 1.40$ keV, decreasing to $\sim 1.15$ keV along northward, while toward the east it remains uniform at $\sim 1.50$ keV. Some central regions show a temperature $\sim 2$ keV. Temperature errors are $\sim$ 8-12\% in the core, rising to $\sim 17$\% along eastward. The pseudoentropy map also shows largely uniform distributions, with a minimum of 35 keV~cm$^{2}$ at the core (uncertainty $\sim 10$\%). Overall entropy rises to $\sim 50$ keV~cm$^{2}$ in the northward and $\sim 60$ keV~cm$^{2}$ in the eastward, with uncertainties of $\sim 13$\%.

The radio emission, overlaid on the temperature map, lies in a region of largely uniform temperature, though smoothing from the adaptive binning algorithm may affect fine-scale variations. The radio peak coincides with a low entropy region. The southern part of the source has a slightly higher average temperature ($\sim 1.53$ keV) than the northern part ($\sim 1.30$ keV), but the difference is within 2$\sigma$. Most of the radio emission is co-spatial with regions of entropy $\sim 35$ keV~cm$^{2}$ along the western side, while the lobes also extend into higher-entropy regions ($50$–$55$ keV~cm$^{2}$).

\textbf{SDSSTG16393:} The X-ray surface brightness distribution of this group shows elongated emission toward the north, consistent with the thermodynamic maps (Figure~\ref{therm-maps}; middle panel). The temperature distribution is non-uniform, marked by several high-temperature patches. In the central regions, the average temperature is slightly lower ($\sim$1.15 keV) than in the surroundings, while toward the north it rises to $\sim$1.7 keV. The elongation of the X-ray emission in this direction is accompanied by high-temperature gas. Near a discrete source close to the centre on the southwest side, a patch of hotter gas reaches to $\sim$2.1 keV. Temperature uncertainties remain relatively uniform ($\sim$8–12\%) near the core, increasing northward up to $\sim$20\%. The pseudoentropy map shows a central minimum of 25 keV~cm$^{2}$ (uncertainty $\sim$9\%), rising along northward to $\sim$80 keV~cm$^{2}$ with uncertainties of 10–13\%.

When overlaid on the temperature map, the radio emission coincides with a cool patch extending northwest from the BGG, and with hotter emission along the northeast and southwest. However, the core is relatively cool, and the temperature rises to $\sim$1.25 keV at $\sim$60 kpc, matching the radio extent. The brighter radio emission is co-located with the low-entropy central regions, and along the southwest direction, the temperature increase is accompanied by an entropy rise to $\sim$50 keV~cm$^{2}$.

\textbf{SDSSTG28674:} Figure~\ref{therm-maps} (right) shows the temperature map of this group from \textit{XMM-Newton} multi-band images. The temperature distribution is not azimuthally symmetric: the coolest regions (0.8–1.0 keV) lie near the northwest BGG, while no such low-temperature regions are seen around the southeast galaxy. The average IGrM temperature is $\sim$1.4 keV, with patches of slightly lower values ($\sim$1.1~keV). Pockets of higher temperature suggest interactions or heating of the gas, consistent with the X-ray surface brightness map (Figure~\ref{x-ray-images}). The pseudoentropy map shows the lowest values at the BGG ($\sim$30 keV~cm$^{2}$), rising to $\sim$50–60 keV~cm$^{2}$ in the surroundings. No clear trend is seen along the merger axis, though some low-temperature regions exhibit higher entropy than their surroundings. Typical uncertainties are 5–10\% in the core, increasing beyond 20\% at larger radii due to low S/N. The diffuse radio structure coincides with a small cool region, southwest of the BGG.

\section{Discussion} \label{discussion}

In this paper, we focus on three galaxy groups from the X-GAP sample, presenting the detailed radio and X-ray analyses of their diffuse emission. The three groups are dynamically disturbed, and the radio emission in and around the BGG shows the signatures of an aged relativistic plasma. These three groups form the first part of our detailed investigation of the entire X-GAP sample. We present an interpretation of the individual systems and discuss their properties in the context of the other well-studied groups and samples.

\subsection{BGGs in the three groups}

\textbf{SDSSTG8102:} The BGG hosts radio emission extending to $\sim 90$ kpc, with clear signs of northwest lobes being displaced westward by the motions of IGrM. Such features are not unusual, but they indicate that the group core is not fully relaxed. The BGG often shows peculiar velocities comparable to, or even slightly exceeding, the group velocity dispersion \citep[e.g.,][]{gozaliasl20}. The uGMRT images reveal additional extensions toward the southwest and an overall westward bend of the lobes, both consistent with IGrM flows drawing out the radio plasma. This interpretation is supported by the \textit{XMM-Newton} images, which show the gas density elongated along the east–west axis. The uGMRT maps, however, reveal that the lobes extend to the northeast, which does not neatly fit this picture. In dynamically active systems, bulk IGrM motions can distort the radio lobes and promote energy redistribution. Simulations \citep{bourne19, bourne21} show that such motions enhance mixing between jet plasma and the ambient gas, leading to more isotropic and efficient heating, effects that may also be operating here. The emission has a steep spectrum ($\alpha \sim -$0.9), implying the presence of an aged relativistic electron population.

\textbf{SDSSTG16393:} The BGG hosts central radio emission beyond 100 kpc. The uGMRT image reveals a nearly symmetrical source with a slight SW–NE extension, while the LOFAR map highlights a distortion on the northern side, suggesting displacement by motions of the IGrM. This interpretation is also supported by the X-ray data, where a drop in surface brightness (white arrow in Figure~\ref{x-ray-images}) is observed at the same location, indicating that the gas motions may be affecting the radio plasma. The temperature map reveals hotter regions toward the SW, coinciding with the radio extent and suggesting local IGrM heating by the radio source, similar to what has been reported from numerical simulations by \citet{bourne23}. The emission has a steep spectrum ($\alpha \sim -1.6$), consistent with an aged relativistic electron population, most plausibly the remnant lobes of past AGN activity.

\textbf{SDSSTG28674:} The BGG, located at the northern peak among the two X-ray peaks in the group, hosts a compact source linked by a faint bridge to an extended source. One interpretation is that it represents a single lobe connected to the BGG by a fading jet, with the absence of a counter-lobe explained by gas motions from the ongoing merger compressing and re-energising only one side \citep[e.g.,][]{randriamanakoto20, shulevski24}. Alternatively, the emission could trace an old pair of lobes deposited in a previous cycle of activity, now connected back to the BGG by a faint trail as the galaxy moved (similar to NGC 5419; \citealt{subrahmanyan03}). The steep spectral index also raises the possibility of a radio phoenix, where fossil plasma is revived by compression or shocks in the IGrM, similar to the ultra-steep phoenix in Abell 2877 \citep{hodgson21}. Another scenario is that the BGG has drifted away from the centre of a once-symmetric source \citep[e.g.,][]{shulevski24}, as observed in offset remnants such as 3C 338 \citep{timmerman22, antas24}. In all cases, the link with the merger axis suggests that the group’s dynamical state has strongly influenced the emission, making this an intriguing example of IGrM-driven radio morphology. Further data, especially polarisation, is required to distinguish between a re-energised lobe and a remnant structure.

\subsection{Comparison with other groups}

\subsubsection{Presence of extended lobes}
Extended radio emission is increasingly being observed in galaxy groups, providing important insights into the interaction between AGN outflows and the IGrM \citep[e.g.,][]{schellenberger17, kolokythas20, rajpurohit2024, pasini25, brienza25}. BGGs are detected in the radio more frequently than the general galaxy population \citep[e.g.,][]{grossova22, wojtowicz23}, but comparisons across group samples must be made with caution, as selection criteria differ in redshift, luminosity, and other parameters. 

Detecting extended diffuse emission in groups requires high sensitivity and angular resolution, since such features are faint and relatively uncommon ($\sim$9\% or 5/53 in the local Universe; \citealt{Osullivan17, kolokythas19}). \citet{kolokythas19} argues that their presence depends primarily on the X-ray properties of the group core (e.g., cool cores fueling AGN) and the nature of the BGG, with less influence from the stellar population mix. Within the CLoGS sample \citep{kolokythas18, kolokythas19}, diffuse radio sources are found in both spiral-rich and spiral-poor groups, ranging from remnant jets to amorphous lobes. In the three systems studied here, two sources (SDSSTG8102 and SDSSTG16393) show morphological alignments between their extended radio and X-ray emission, and both are elongated along the same axis (Figure~\ref{x-ray-images}). No cavities or shocks are seen, possibly due to the sensitivity limits of current X-ray data, but both sources host central diffuse emission coincident with the BGG. High-resolution maps reveal compact cores in SDSSTG16393, but no clear jets. Both broadly resemble a central radio galaxy, with distortions suggestive of IGrM bulk motions. In contrast, SDSSTG28674 hosts a more unusual source, offset from the BGG by $\sim 63$ kpc and lacking a compact AGN core or symmetric lobes. A faint 144 MHz bridge connects the diffuse emission to the BGG, but the strongly one-sided morphology remains puzzling. Comparisons to systems such as NGC~677 and NGC~5903 in CLoGS \citep{kolokythas18, Osullivan18, kolokythas19} highlight both similarities and differences: the X-GAP sources extend to larger scales (80–120 kpc) and often lack classical jet–lobes morphologies, resembling remnant jet systems like NGC~1407 but showing greater structural diversity. Overall, these results suggest that diffuse central radio sources in groups can arise through multiple pathways: direct AGN outbursts, buoyantly rising lobes, or phoenix-like revivals of fossil plasma. The detection of such extended diffuse structures in both the CLoGS and X-GAP samples suggests that their appearance requires specific environmental conditions. The conditions may depend not only on whether the group is relaxed or merging, but also on the interplay between the AGN activity, IGrM, and the large-scale group dynamics.

\subsubsection{Spectral index distribution}

 Spectral index provides a powerful insight into the age and energetics of the radio plasma, as well as the environmental processes that shape its evolution \citep[e.g.,][]{lisenfeld00, konar06, jamrozy08}. While compact AGN cores often display flat spectra due to synchrotron self-absorption and ongoing particle injection, extended lobes and diffuse components generally steepen over time as radiative and adiabatic losses dominate \citep[e.g.,][]{mckean16}. In dense group environments, surrounding IGrM can confine the plasma and prolong the visibility of steep-spectrum lobes ($\alpha \sim -1$ to -2; \citealt{giacintucci11}), whereas in lower-density environments the plasma expands more freely, leading to faster fading of relic structures \citep[e.g.,][]{2025PASA...42..152S}.

Across the three X-GAP systems, we find a clear diversity in the spectral behaviour, reflecting different stages of AGN life cycles. SDSSTG8102 exhibits a spectral index of approximately $-0.9$, which is steep, and some radio remnants have been observed with similar spectra below 1 GHz \citep[e.g.,][]{brienza16, brienza17}. The BGG in SDSSTG8102 remains a candidate remnant radio galaxy. In contrast, the BGG in for SDSSTG16393, the spectrum is much steeper ($\alpha \lesssim -1.0$), characteristics of a strongly aged synchrotron plasma and indicative of fading activity. The BGG in SDSSTG28674 shows a flat spectrum core emission, with an extended lobe having a very steep spectrum ($-$1.74), indicating a restarted system, where an active core coexists with extended, aged emission from a previous activity cycle. The BGG in SDSSTG16393 shows classical ultra–steep-spectrum characteristics consistent with an evolved remnant; the other two represent earlier or more complex remnant phases. For SDSSTG8102, although the low-frequency spectral index (144–400 MHz) is not ultra-steep, it is already relatively steep given the frequency range probed, and the absence of compact features, jets, or a radio core, together with the amorphous morphology, suggests that large-scale jet activity has ceased. This source is therefore consistent with a young remnant, in line with both observational studies and simulations predicting a population of remnants whose spectral break lies at higher frequencies. SDSSTG28674 likely represents a restarted system, where an active core coexists with extended, aged emission from a previous activity cycle. This spread implies the coexistence of different remnant phases of central radio sources in groups, with the spectral properties strongly shaped by duty cycle and the availability of gas to fuel the activity. In CLoGS sample it was found that, while typical integrated spectral indices in X-ray bright groups lie around $-0.6$ to $-0.9$, systems with spectral index of $\alpha_{235}^{610}=-1.38\pm0.07$ and $\alpha_{235}^{1400}=-1.11\pm0.07$ were also found \citep{kolokythas19}. The ultra steep indices ($\lesssim -$1.0) in two of our X-GAP groups (SDSSTG16393, SDSSTG28674) are consistent with this latter class.

Another key trend observed in the CLoGS sample is that large-scale jet ($\gtrsim 50$ kpc) systems show distinct spectral properties based on their evolutionary stage and interaction with the surrounding medium. Environmental factors, such as jet bending or confinement by IGrM, can also contribute to spectral steepening \citep{kolokythas18,kolokythas19}. This supports the idea that a combination of source age, AGN history, and environmental conditions plays a central role in shaping the spectral evolution of radio sources in galaxy groups, as is seen in the three X-GAP groups.

\subsubsection{X-ray environmental effects}

The X-ray environment plays a key role in shaping the morphology and energetics of the group-central radio galaxies \citep[e.g.,][]{shin16, liu19}. In all three X-GAP systems, the IGrM is bright and extended to $\sim$300 kpc, and each hosts a central radio source with emission exceeding 50 kpc in length. These emissions remain confined within the X-ray halos, supporting the idea that dense hot gas both stabilises the radio outflows and provides a reservoir for AGN fueling. Energy transfer from the jets to the IGrM—through shocks, turbulence, or cosmic ray diffusion—may help to offset the radiative cooling and maintain thermal balance \citep{fabian12, eckert21, hlavacek22}. This agrees with the CLoGS survey, where jet-mode AGN are almost exclusively found in X-ray bright groups \citep{kolokythas18}, while X-ray faint groups usually host only compact or weak radio sources. Such an environmental dependence suggests that a substantial hot gas halo is essential for sustaining large-scale radio-mode AGN activity, most likely via cold chaotic accretion \citep[CCA;][]{gaspari13, gaspari19}, but alternative fueling processes remain possible.

Mergers disturb the IGrM, enhancing the turbulence and mixing, and potentially triggering multiphase gas condensation and SMBH fueling \citep[e.g.,][]{brienza22, shulevski17}. Although major group–group mergers are rare, typically occurring on $\sim$5–6 Gyr timescales \citep{rodriguez15, pasini21, sharma24}, their signatures are evident in these X-GAP groups. Each of the three X-GAP groups shows elongated X-ray halos or asymmetric distributions, accompanied by correspondingly distorted radio structures. This morphological coupling suggests that merger-driven gas motions have played a role in shaping both the thermal and non-thermal components. In particular, one system shows radio jets aligned with an elongated X-ray halo, another exhibits asymmetric lobes within a disturbed north–south gas distribution, and a third reveals diffuse, fading emission coincident with a binary group merger. Together, these lead to an evolutionary sequence where mergers initially drive gas inflows that enhance AGN activity, followed by fading radio structures as fueling diminishes. Another physical aspect is that large-scale, bright radio lobes cannot be sustained without confinement by a substantial hot halo. The X-ray emitting IGrM provides the dominant external thermal pressure; if the internal lobe pressure greatly exceeds this value, the lobes will expand supersonically and rapidly fade until equilibrium is reached. Since radio surface brightness traces the line-of-sight non-thermal pressure, while X-ray brightness traces the integrated thermal pressure, richer X-ray environments naturally host brighter and more extended radio lobes, even when the fueling mechanisms are similar.

Compared with CLoGS systems that experienced gas-poor (dry) mergers and lack extended radio emission \citep[e.g.,][]{kolokythas19}, the X-GAP groups stand out as gas-rich, dynamically active environments capable of sustaining large-scale diffuse structures. The differences could be due to different sample selection strategies, as CLoGS targets nearby, optically-selected groups with evolved BGGs, while X-GAP preferentially includes X-ray bright systems, where residual gas reservoirs are still present. Therefore, merger type plays a crucial role in shaping the radio properties of galaxy groups, and some mergers produce diffuse emission, and others do not. This suggests that the presence of large-scale jets depends on the environment, the group's evolutionary stage, the nature of the merger itself, and the efficiency of the gas accretion on the central SMBH. The three merging X-GAP groups offer a valuable opportunity to refine our understanding of these processes, especially by revealing merging systems with diffuse emission rarely observed.

\section{Summary} \label{summary}

We present the results from the uGMRT 400 MHz observations for the three X-GAP galaxy groups. These sensitive uGMRT observations at 400 MHz enable us to obtain detailed spectral characteristics of the central extended sources by combining them with LOFAR observations at 144 MHz. Our radio results, corroborated with available X-ray observations (\textit{XMM-Newton}), provide profound physical insights into the thermal and non-thermal connections in the IGrM. The overall findings are summarised below:

\begin{enumerate}

    \item All three groups show large-scale lobes spanning from $\sim$ 50 - 150 kpc at both 400 and 144 MHz. Double-lobed radio emission is associated with the BGG for two groups, except SDSSTG28674, where the BGG hosting a compact radio source is connected with the extended emission with a faint bridge of emission.

    \item The radio emission in SDSSTG8102 follows a single power law with an average spectral index of $\sim-0.9$, consistent across its northern and southern components, and shows a uniform distribution ($\sim-0.8$) in the spectral index map. SDSSTG16393 has an integrated spectral index of $\sim-1.2$ (144--400 MHz) and a spatially varying spectral index, with steeper values ($\sim-1.6$) in the outskirts and a slightly flatter index ($\sim-0.95$) in the central region. SDSSTG28674 has the steepest integrated spectrum of $-1.75$, indicative of a rapidly fading remnant lobe that is undetected at higher frequencies.

    \item The X-ray maps reveal elongated morphologies in all three groups, with SDSSTG28674 showing clear signs of a binary merger. No cavities or discontinuities are detected, likely due to the limited depth of the \textit{XMM-Newton} data. In SDSSTG8102 and SDSSTG16393, the central radio emission aligns well with the X-ray emission and BGG position.

    \item We present 2D thermodynamic maps of the groups, highlighting the thermal structure of the IGrM. SDSSTG8102 shows an average central temperature of $\sim$1.4 keV, increasing to $\sim$1.5 keV toward the west. SDSSTG16393 has a cooler core ($\sim$0.9 keV) with a hotter region ($\sim$1.7 keV) to the north, where the gas temperature increases up to 60 kpc from the center toward the radio emission. SDSSTG28674 exhibits signatures of merging activity, with the BGG region at 0.7 - 1.0 keV and hotter gas ($\sim$1.5 keV) along the merger axis.

    \item We discuss similarities of each of these systems with other groups that have been studied earlier regarding their linear extents, spectral index distribution, and X-ray properties. 

    \item This study showcases the behaviour of radio emission associated with the BGGs in merging galaxy groups. In the future, the study will be extended to the complete sample of 49 groups in X-GAP.
    
\end{enumerate}

In conclusion, our multi-wavelength study of three merging galaxy groups, using uGMRT, LOFAR, and \textit{XMM-Newton}, reveals significant insights into the interplay between AGN activity and the IGrM. The radio emission observed in these groups appears to be influenced by interactions with the surrounding IGrM. The X-ray and radio emissions show distinct correlations, providing evidence for the complex processes at play in these merging systems. The findings highlight the importance of environmental conditions in shaping the radio properties and evolution of galaxy groups. These results motivate deeper multi-wavelength follow-up of X-GAP groups to confirm the presence of cold gas and test whether wet merger signatures consistently accompany large-scale AGN feedback.


\begin{acknowledgements}
  We thank the anonymous referee for their constructive comments that have improved the clarity of the paper. R.S. and R.K. acknowledge the support of the Department of Atomic Energy, Government of India, under project no. 12-R\&D-TFR-5.02-0700. MB acknowledges financial support from Next Generation EU funds within the National Recovery and Resilience Plan (PNRR), Mission 4 – Education and Research, Component 2 – From Research to Business (M4C2), Investment Line 3.1 – Strengthening and creation of Research Infrastructures, Project IR0000034 – ``STILES – Strengthening the Italian Leadership in ELT and SKA'', from INAF under the Large GO 2024 funding scheme (project ``MeerKAT and Euclid Team up: Exploring the galaxy-halo connection at cosmic noon''), the Large Grant 2022 funding scheme (project ``MeerKAT and LOFAR Team up: a Unique Radio Window on Galaxy/AGN co-Evolution'') and the Mini Grant 2023 funding scheme (project `Low radio frequencies as a probe of AGN jet feedback at low and high redshift'). M.A.B. is supported by a UKRI Stephen Hawking Fellowship (EP/X04257X/1). LOFAR is the Low Frequency Array designed and constructed by ASTRON. It has observing, data processing, and data storage facilities in several countries, which are owned by various parties (each with their own funding sources), and which are collectively operated by the LOFAR ERIC under a joint scientific policy. The LOFAR resources have benefited from the following recent major funding sources: CNRS-INSU, Observatoire de Paris and Université d'Orléans, France; BMBF, MIWF-NRW, MPG, Germany; Science Foundation Ireland (SFI), Department of Business, Enterprise and Innovation (DBEI), Ireland; NWO, The Netherlands; The Science and Technology Facilities Council, UK; Ministry of Science and Higher Education, Poland; The Istituto Nazionale di Astrofisica (INAF), Italy. This research made use of the Dutch national e-infrastructure with support of the SURF Cooperative (e-infra 180169) and the LOFAR e-infra group. The Jülich LOFAR Long Term Archive and the German LOFAR network are both coordinated and operated by the Jülich Supercomputing Centre (JSC), and computing resources on the supercomputer JUWELS at JSC were provided by the Gauss Centre for Supercomputing e.V. (grant CHTB00) through the John von Neumann Institute for Computing (NIC). This research made use of the University of Hertfordshire high-performance computing facility and the LOFAR-UK computing facility located at the University of Hertfordshire and supported by STFC [ST/P000096/1], and of the Italian LOFAR IT computing infrastructure supported and operated by INAF, and by the Physics Department of Turin university (under an agreement with Consorzio Interuniversitario per la Fisica Spaziale) at the C3S Supercomputing Centre, Italy. This research is part of the project LOFAR Data Valorization (LDV) [project numbers 2020.031, 2022.033, and 2024.047] of the research programme Computing Time on National Computer Facilities using SPIDER that is (co-)funded by the Dutch Research Council (NWO), hosted by SURF through the call for proposals of Computing Time on National Computer Facilities. We thank the staff of the GMRT who made these observations possible. The GMRT is run by the National Centre for Radio Astrophysics (NCRA) of the Tata Institute of Fundamental Research (TIFR). This research made use of the NASA/IPAC Extragalactic Database (NED), which is operated by the Jet Propulsion Laboratory, California Institute of Technology, under contract with the National Aeronautics and Space Administration. Based on observations obtained with \textit{XMM-Newton}, an ESA science mission with instruments and contributions directly funded by ESA Member States and NASA. 
\end{acknowledgements}

%
%
\bibliography{sample631}{}

\begin{thebibliography}{120}
\expandafter\ifx\csname natexlab\endcsname\relax\def\natexlab#1{#1}\fi

\bibitem[{{Alonso} {et~al.}(2012){Alonso}, {Mesa}, {Padilla}, \& {Lambas}}]{alonso12}
{Alonso}, S., {Mesa}, V., {Padilla}, N., \& {Lambas}, D.~G. 2012, \aap, 539, A46

\bibitem[{{Antas} {et~al.}(2024){Antas}, {Caproni}, {Machado}, {Lagan{\'a}}, \& {Souza}}]{antas24}
{Antas}, A.~S.~R., {Caproni}, A., {Machado}, R.~E.~G., {Lagan{\'a}}, T.~F., \& {Souza}, G.~S. 2024, \mnras, 533, 1341

\bibitem[{{Bardelli} {et~al.}(2010){Bardelli}, {Schinnerer}, {Smol{\v{c}}ic}, {Zamorani}, {Zucca}, {Mignoli}, {Halliday}, {Kova{\v{c}}}, {Ciliegi}, {Caputi}, {Koekemoer}, {Bongiorno}, {Bondi}, {Bolzonella}, {Vergani}, {Pozzetti}, {Carollo}, {Contini}, {Kneib}, {Le F{\`e}vre}, {Lilly}, {Mainieri}, {Renzini}, {Scodeggio}, {Coppa}, {Cucciati}, {de la Torre}, {de Ravel}, {Franzetti}, {Garilli}, {Iovino}, {Kampczyk}, {Knobel}, {Lamareille}, {Le Borgne}, {Le Brun}, {Maier}, {Pell{\`o}}, {Peng}, {Perez-Montero}, {Ricciardelli}, {Silverman}, {Tanaka}, {Tasca}, {Tresse}, {Abbas}, {Bottini}, {Cappi}, {Cassata}, {Cimatti}, {Guzzo}, {Leauthaud}, {Maccagni}, {Marinoni}, {McCracken}, {Memeo}, {Meneux}, {Oesch}, {Porciani}, {Scaramella}, {Capak}, {Sanders}, {Scoville}, {Taniguchi}, \& {Jahnke}}]{bardelli10}
{Bardelli}, S., {Schinnerer}, E., {Smol{\v{c}}ic}, V., {et~al.} 2010, \aap, 511, A1

\bibitem[{{Beck} {et~al.}(1999){Beck}, {Ehle}, {Shoutenkov}, {Shukurov}, \& {Sokoloff}}]{beck1999}
{Beck}, R., {Ehle}, M., {Shoutenkov}, V., {Shukurov}, A., \& {Sokoloff}, D. 1999, \nat, 397, 324

\bibitem[{{Becker} {et~al.}(1994){Becker}, {White}, \& {Helfand}}]{becker94}
{Becker}, R.~H., {White}, R.~L., \& {Helfand}, D.~J. 1994, in Astronomical Society of the Pacific Conference Series, Vol.~61, Astronomical Data Analysis Software and Systems III, ed. D.~R. {Crabtree}, R.~J. {Hanisch}, \& J.~{Barnes}, 165

\bibitem[{Best {et~al.}(2005)Best, Kauffmann, Heckman, Brinchmann, Charlot, Ivezić, \& White}]{best05}
Best, P.~N., Kauffmann, G., Heckman, T.~M., {et~al.} 2005, Monthly Notices of the Royal Astronomical Society, 362, 25

\bibitem[{{B{\^\i}rzan} {et~al.}(2012){B{\^\i}rzan}, {Rafferty}, {Nulsen}, {McNamara}, {R{\"o}ttgering}, {Wise}, \& {Mittal}}]{birzan12}
{B{\^\i}rzan}, L., {Rafferty}, D.~A., {Nulsen}, P.~E.~J., {et~al.} 2012, \mnras, 427, 3468

\bibitem[{{Botteon} {et~al.}(2024){Botteon}, {van Weeren}, {Eckert}, {Gastaldello}, {Markevitch}, {Giacintucci}, {Brunetti}, {Kale}, \& {Venturi}}]{botteon24}
{Botteon}, A., {van Weeren}, R.~J., {Eckert}, D., {et~al.} 2024, \aap, 690, A222

\bibitem[{{Bourne} \& {Sijacki}(2021)}]{bourne21}
{Bourne}, M.~A. \& {Sijacki}, D. 2021, \mnras, 506, 488

\bibitem[{{Bourne} {et~al.}(2019){Bourne}, {Sijacki}, \& {Puchwein}}]{bourne19}
{Bourne}, M.~A., {Sijacki}, D., \& {Puchwein}, E. 2019, \mnras, 490, 343

\bibitem[{{Bourne} \& {Yang}(2023)}]{bourne23}
{Bourne}, M.~A. \& {Yang}, H.-Y.~K. 2023, Galaxies, 11, 73

\bibitem[{{Brienza} {et~al.}(2016){Brienza}, {Godfrey}, \& {Morganti}}]{brienza16}
{Brienza}, M., {Godfrey}, L., \& {Morganti}, R. 2016, in Active Galactic Nuclei: What's in a Name?, 102

\bibitem[{{Brienza} {et~al.}(2017){Brienza}, {Godfrey}, {Morganti}, {Prandoni}, {Harwood}, {Mahony}, {Hardcastle}, {Murgia}, {R{\"o}ttgering}, {Shimwell}, \& {Shulevski}}]{brienza17}
{Brienza}, M., {Godfrey}, L., {Morganti}, R., {et~al.} 2017, \aap, 606, A98

\bibitem[{{Brienza} {et~al.}(2022){Brienza}, {Lovisari}, {Rajpurohit}, {Bonafede}, {Gastaldello}, {Murgia}, {Vazza}, {Bonnassieux}, {Botteon}, {Brunetti}, {Drabent}, {Hardcastle}, {Pasini}, {Riseley}, {R{\"o}ttgering}, {Shimwell}, {Simionescu}, \& {van Weeren}}]{brienza22}
{Brienza}, M., {Lovisari}, L., {Rajpurohit}, K., {et~al.} 2022, \aap, 661, A92

\bibitem[{{Brienza} {et~al.}(2025){Brienza}, {Rajpurohit}, {Churazov}, {Heywood}, {Br{\"u}ggen}, {Hoeft}, {Vazza}, {Bonafede}, {Botteon}, {Brunetti}, {Gastaldello}, {Khabibullin}, {Lyskova}, {Majumder}, {R{\"o}ttgering}, {Shimwell}, {Simionescu}, \& {van Weeren}}]{brienza25}
{Brienza}, M., {Rajpurohit}, K., {Churazov}, E., {et~al.} 2025, arXiv e-prints, arXiv:2502.18244

\bibitem[{{Brienza} {et~al.}(2021){Brienza}, {Shimwell}, {de Gasperin}, {Bikmaev}, {Bonafede}, {Botteon}, {Br{\"u}ggen}, {Brunetti}, {Burenin}, {Capetti}, {Churazov}, {Hardcastle}, {Khabibullin}, {Lyskova}, {R{\"o}ttgering}, {Sunyaev}, {van Weeren}, {Gastaldello}, {Mandal}, {Purser}, {Simionescu}, \& {Tasse}}]{brienza21}
{Brienza}, M., {Shimwell}, T.~W., {de Gasperin}, F., {et~al.} 2021, Nature Astronomy, 5, 1261

\bibitem[{{Buch} {et~al.}(2023){Buch}, {Kale}, {Muley}, {Kudale}, \& {Ajithkumar}}]{buch23}
{Buch}, K.~D., {Kale}, R., {Muley}, M., {Kudale}, S., \& {Ajithkumar}, B. 2023, Journal of Astrophysics and Astronomy, 44, 37

\bibitem[{{Buch} {et~al.}(2022){Buch}, {Kale}, {Naik}, {Aragade}, {Muley}, {Kudale}, \& {Ajith Kumar}}]{buch22}
{Buch}, K.~D., {Kale}, R., {Naik}, K.~D., {et~al.} 2022, Journal of Astronomical Instrumentation, 11, 2250008

\bibitem[{{Chandra} \& {Kanekar}(2017)}]{chandra17}
{Chandra}, P. \& {Kanekar}, N. 2017, \apj, 846, 111

\bibitem[{{Damsted} {et~al.}(2024){Damsted}, {Finoguenov}, {Lietzen}, {Mamon}, {Comparat}, {Tempel}, {Dmitrieva}, {Clerc}, {Collins}, {Gozaliasl}, \& {Eckert}}]{damsted24}
{Damsted}, S., {Finoguenov}, A., {Lietzen}, H., {et~al.} 2024, arXiv e-prints, arXiv:2403.17055

\bibitem[{{de Gasperin} {et~al.}(2017){de Gasperin}, {Intema}, {Shimwell}, {Brunetti}, {Br{\"u}ggen}, {En{\ss}lin}, {van Weeren}, {Bonafede}, \& {R{\"o}ttgering}}]{degasperin17}
{de Gasperin}, F., {Intema}, H.~T., {Shimwell}, T.~W., {et~al.} 2017, Science Advances, 3, e1701634

\bibitem[{Dunn {et~al.}(2010)Dunn, Allen, Taylor, Shurkin, Gentile, Fabian, \& Reynolds}]{dunn10}
Dunn, R. J.~H., Allen, S.~W., Taylor, G.~B., {et~al.} 2010, Monthly Notices of the Royal Astronomical Society, 404, 180

\bibitem[{{Eckert} {et~al.}(2021){Eckert}, {Gaspari}, {Gastaldello}, {Le Brun}, \& {O'Sullivan}}]{eckert21}
{Eckert}, D., {Gaspari}, M., {Gastaldello}, F., {Le Brun}, A. M.~C., \& {O'Sullivan}, E. 2021, Universe, 7, 142

\bibitem[{{Eckert} {et~al.}(2025){Eckert}, {Gastaldello}, {Lovisari}, {McGee}, {Pasini}, {Brienza}, {Kolokythas}, {O'Sullivan}, {Simionescu}, {Sun}, {Ayromlou}, {Bourne}, {Chen}, {Cui}, {Ettori}, {Finoguenov}, {Gozaliasl}, {Kale}, {Mernier}, {Oppenheimer}, {Schellenberger}, {Seppi}, \& {Tempel}}]{eckert25}
{Eckert}, D., {Gastaldello}, F., {Lovisari}, L., {et~al.} 2025, arXiv e-prints, arXiv:2506.13907

\bibitem[{{Eckert} {et~al.}(2024){Eckert}, {Gastaldello}, {O'Sullivan}, {Finoguenov}, {Brienza}, \& {X-GAP Collaboration}}]{eckert24}
{Eckert}, D., {Gastaldello}, F., {O'Sullivan}, E., {et~al.} 2024, Galaxies, 12, 24

\bibitem[{{Eckert} {et~al.}(2012){Eckert}, {Vazza}, {Ettori}, {Molendi}, {Nagai}, {Lau}, {Roncarelli}, {Rossetti}, {Snowden}, \& {Gastaldello}}]{eckert12}
{Eckert}, D., {Vazza}, F., {Ettori}, S., {et~al.} 2012, \aap, 541, A57

\bibitem[{{Eke} {et~al.}(2004){Eke}, {Baugh}, {Cole}, {Frenk}, {Norberg}, {Peacock}, {Baldry}, {Bland-Hawthorn}, {Bridges}, {Cannon}, {Colless}, {Collins}, {Couch}, {Dalton}, {de Propris}, {Driver}, {Efstathiou}, {Ellis}, {Glazebrook}, {Jackson}, {Lahav}, {Lewis}, {Lumsden}, {Maddox}, {Madgwick}, {Peterson}, {Sutherland}, \& {Taylor}}]{eke04}
{Eke}, V.~R., {Baugh}, C.~M., {Cole}, S., {et~al.} 2004, \mnras, 348, 866

\bibitem[{{Ettori} {et~al.}(2012){Ettori}, {Rasia}, {Fabjan}, {Borgani}, \& {Dolag}}]{ettori12}
{Ettori}, S., {Rasia}, E., {Fabjan}, D., {Borgani}, S., \& {Dolag}, K. 2012, \mnras, 420, 2058

\bibitem[{{Fabian}(2012)}]{fabian12}
{Fabian}, A.~C. 2012, \araa, 50, 455

\bibitem[{{Finoguenov} {et~al.}(2003){Finoguenov}, {Borgani}, {Tornatore}, \& {B{\"o}hringer}}]{finoguenov03}
{Finoguenov}, A., {Borgani}, S., {Tornatore}, L., \& {B{\"o}hringer}, H. 2003, \aap, 398, L35

\bibitem[{{Forbes} {et~al.}(2006){Forbes}, {Ponman}, {Pearce}, {Osmond}, {Kilborn}, {Brough}, {Raychaudhury}, {Mundell}, {Miles}, \& {Kern}}]{forbes06}
{Forbes}, D.~A., {Ponman}, T., {Pearce}, F., {et~al.} 2006, \pasa, 23, 38

\bibitem[{{Gaspari} {et~al.}(2019){Gaspari}, {Eckert}, {Ettori}, {Tozzi}, {Bassini}, {Rasia}, {Brighenti}, {Sun}, {Borgani}, {Johnson}, {Tremblay}, {Stone}, {Temi}, {Yang}, {Tombesi}, \& {Cappi}}]{gaspari19}
{Gaspari}, M., {Eckert}, D., {Ettori}, S., {et~al.} 2019, \apj, 884, 169

\bibitem[{{Gaspari} {et~al.}(2013){Gaspari}, {Ruszkowski}, \& {Oh}}]{gaspari13}
{Gaspari}, M., {Ruszkowski}, M., \& {Oh}, S.~P. 2013, \mnras, 432, 3401

\bibitem[{{Geller} \& {Huchra}(1983)}]{geller83}
{Geller}, M.~J. \& {Huchra}, J.~P. 1983, \apjs, 52, 61

\bibitem[{{Ghirardini} {et~al.}(2019){Ghirardini}, {Eckert}, {Ettori}, {Pointecouteau}, {Molendi}, {Gaspari}, {Rossetti}, {De Grandi}, {Roncarelli}, {Bourdin}, {Mazzotta}, {Rasia}, \& {Vazza}}]{ghirardini19}
{Ghirardini}, V., {Eckert}, D., {Ettori}, S., {et~al.} 2019, \aap, 621, A41

\bibitem[{{Giacintucci} {et~al.}(2011){Giacintucci}, {O'Sullivan}, {Vrtilek}, {David}, {Raychaudhury}, {Venturi}, {Athreya}, {Clarke}, {Murgia}, {Mazzotta}, {Gitti}, {Ponman}, {Ishwara-Chandra}, {Jones}, \& {Forman}}]{giacintucci11}
{Giacintucci}, S., {O'Sullivan}, E., {Vrtilek}, J., {et~al.} 2011, \apj, 732, 95

\bibitem[{{Giodini} {et~al.}(2009){Giodini}, {Pierini}, {Finoguenov}, {Pratt}, {Boehringer}, {Leauthaud}, {Guzzo}, {Aussel}, {Bolzonella}, {Capak}, {Elvis}, {Hasinger}, {Ilbert}, {Kartaltepe}, {Koekemoer}, {Lilly}, {Massey}, {McCracken}, {Rhodes}, {Salvato}, {Sanders}, {Scoville}, {Sasaki}, {Smolcic}, {Taniguchi}, {Thompson}, \& {COSMOS Collaboration}}]{giodini09}
{Giodini}, S., {Pierini}, D., {Finoguenov}, A., {et~al.} 2009, \apj, 703, 982

\bibitem[{{Gozaliasl}(2016)}]{gozaliasl16}
{Gozaliasl}, G. 2016, PhD thesis, University of Helsinki, Finland

\bibitem[{{Gozaliasl} {et~al.}(2024){Gozaliasl}, {Finoguenov}, {Babul}, {Ilbert}, {Sargent}, {Vardoulaki}, {Faisst}, {Liu}, {Shuntov}, {Cooper}, {Dolag}, {Toft}, {Magdis}, {Toni}, {Mobasher}, {Barr{\'e}}, {Cui}, \& {Rennehan}}]{Gozaliasl24}
{Gozaliasl}, G., {Finoguenov}, A., {Babul}, A., {et~al.} 2024, \aap, 690, A315

\bibitem[{{Gozaliasl} {et~al.}(2018){Gozaliasl}, {Finoguenov}, {Khosroshahi}, {Henriques}, {Tanaka}, {Ilbert}, {Wuyts}, {McCracken}, \& {Montanari}}]{gozaliasl18}
{Gozaliasl}, G., {Finoguenov}, A., {Khosroshahi}, H.~G., {et~al.} 2018, \mnras, 475, 2787

\bibitem[{{Gozaliasl} {et~al.}(2020){Gozaliasl}, {Finoguenov}, {Khosroshahi}, {Laigle}, {Kirkpatrick}, {Kiiveri}, {Devriendt}, {Dubois}, \& {Ahoranta}}]{gozaliasl20}
{Gozaliasl}, G., {Finoguenov}, A., {Khosroshahi}, H.~G., {et~al.} 2020, \aap, 635, A36

\bibitem[{{Grogin} \& {Geller}(2000)}]{grogin00}
{Grogin}, N.~A. \& {Geller}, M.~J. 2000, \aj, 119, 32

\bibitem[{{Grossov{\'a}} {et~al.}(2022){Grossov{\'a}}, {Werner}, {Massaro}, {Lakhchaura}, {Pl{\v{s}}ek}, {Gab{\'a}nyi}, {Rajpurohit}, {Canning}, {Nulsen}, {O'Sullivan}, {Allen}, \& {Fabian}}]{grossova22}
{Grossov{\'a}}, R., {Werner}, N., {Massaro}, F., {et~al.} 2022, \apjs, 258, 30

\bibitem[{{Haines} {et~al.}(2018){Haines}, {Finoguenov}, {Smith}, {Babul}, {Egami}, {Mazzotta}, {Okabe}, {Pereira}, {Bianconi}, {McGee}, {Ziparo}, {Campusano}, \& {Loyola}}]{haines18}
{Haines}, C.~P., {Finoguenov}, A., {Smith}, G.~P., {et~al.} 2018, \mnras, 477, 4931

\bibitem[{{Hale} {et~al.}(2021){Hale}, {McConnell}, {Thomson}, {Lenc}, {Heald}, {Hotan}, {Leung}, {Moss}, {Murphy}, {Pritchard}, {Sadler}, {Stewart}, \& {Whiting}}]{hale21}
{Hale}, C.~L., {McConnell}, D., {Thomson}, A.~J.~M., {et~al.} 2021, \pasa, 38, e058

\bibitem[{{Hardcastle} \& {Croston}(2020)}]{hardcastle20}
{Hardcastle}, M.~J. \& {Croston}, J.~H. 2020, \nar, 88, 101539

\bibitem[{{Hlavacek-Larrondo} {et~al.}(2022){Hlavacek-Larrondo}, {Li}, \& {Churazov}}]{hlavacek22}
{Hlavacek-Larrondo}, J., {Li}, Y., \& {Churazov}, E. 2022, in Handbook of X-ray and Gamma-ray Astrophysics, ed. C.~{Bambi} \& A.~{Sangangelo}, 5

\bibitem[{{Hodgson} {et~al.}(2021){Hodgson}, {Bartalucci}, {Johnston-Hollitt}, {McKinley}, {Vazza}, \& {Wittor}}]{hodgson21}
{Hodgson}, T., {Bartalucci}, I., {Johnston-Hollitt}, M., {et~al.} 2021, \apj, 909, 198

\bibitem[{{Jamrozy} {et~al.}(2008){Jamrozy}, {Konar}, {Machalski}, \& {Saikia}}]{jamrozy08}
{Jamrozy}, M., {Konar}, C., {Machalski}, J., \& {Saikia}, D.~J. 2008, \mnras, 385, 1286

\bibitem[{{Jennings} {et~al.}(2025){Jennings}, {Babul}, {Dav{\'e}}, {Cui}, \& {Rennehan}}]{jennings25}
{Jennings}, F.~J., {Babul}, A., {Dav{\'e}}, R., {Cui}, W., \& {Rennehan}, D. 2025, \mnras, 536, 145

\bibitem[{{Kale} \& {Ishwara-Chandra}(2021)}]{kale21}
{Kale}, R. \& {Ishwara-Chandra}, C.~H. 2021, Experimental Astronomy, 51, 95

\bibitem[{{Kale} {et~al.}(2022){Kale}, {Parekh}, {Rahaman}, {Joshi}, {Venturi}, {Kolokythas}, {Chibueze}, {Sikhosana}, {Pillay}, \& {Knowles}}]{kale22}
{Kale}, R., {Parekh}, V., {Rahaman}, M., {et~al.} 2022, \mnras, 514, 5969

\bibitem[{{Kolokythas} {et~al.}(2015){Kolokythas}, {O'Sullivan}, {Giacintucci}, {Raychaudhury}, {Ishwara-Chandra}, {Worrall}, \& {Birkinshaw}}]{kolokythas15}
{Kolokythas}, K., {O'Sullivan}, E., {Giacintucci}, S., {et~al.} 2015, \mnras, 450, 1732

\bibitem[{{Kolokythas} {et~al.}(2020){Kolokythas}, {O'Sullivan}, {Giacintucci}, {Worrall}, {Birkinshaw}, {Raychaudhury}, {Horellou}, {Intema}, \& {Loubser}}]{kolokythas20}
{Kolokythas}, K., {O'Sullivan}, E., {Giacintucci}, S., {et~al.} 2020, \mnras, 496, 1471

\bibitem[{{Kolokythas} {et~al.}(2019){Kolokythas}, {O'Sullivan}, {Intema}, {Raychaudhury}, {Babul}, {Giacintucci}, \& {Gitti}}]{kolokythas19}
{Kolokythas}, K., {O'Sullivan}, E., {Intema}, H., {et~al.} 2019, \mnras, 489, 2488

\bibitem[{{Kolokythas} {et~al.}(2018){Kolokythas}, {O'Sullivan}, {Raychaudhury}, {Giacintucci}, {Gitti}, \& {Babul}}]{kolokythas18}
{Kolokythas}, K., {O'Sullivan}, E., {Raychaudhury}, S., {et~al.} 2018, \mnras, 481, 1550

\bibitem[{{Kolokythas} {et~al.}(2022){Kolokythas}, {Vaddi}, {O'Sullivan}, {Loubser}, {Babul}, {Raychaudhury}, {Lagos}, \& {Jarrett}}]{kolokythas22}
{Kolokythas}, K., {Vaddi}, S., {O'Sullivan}, E., {et~al.} 2022, \mnras, 510, 4191

\bibitem[{{Konar} {et~al.}(2006){Konar}, {Saikia}, {Jamrozy}, \& {Machalski}}]{konar06}
{Konar}, C., {Saikia}, D.~J., {Jamrozy}, M., \& {Machalski}, J. 2006, \mnras, 372, 693

\bibitem[{{Lacy} {et~al.}(2020){Lacy}, {Baum}, {Chandler}, {Chatterjee}, {Clarke}, {Deustua}, {English}, {Farnes}, {Gaensler}, {Gugliucci}, {Hallinan}, {Kent}, {Kimball}, {Law}, {Lazio}, {Marvil}, {Mao}, {Medlin}, {Mooley}, {Murphy}, {Myers}, {Osten}, {Richards}, {Rosolowsky}, {Rudnick}, {Schinzel}, {Sivakoff}, {Sjouwerman}, {Taylor}, {White}, {Wrobel}, {Andernach}, {Beasley}, {Berger}, {Bhatnager}, {Birkinshaw}, {Bower}, {Brandt}, {Brown}, {Burke-Spolaor}, {Butler}, {Comerford}, {Demorest}, {Fu}, {Giacintucci}, {Golap}, {G{\"u}th}, {Hales}, {Hiriart}, {Hodge}, {Horesh}, {Ivezi{\'c}}, {Jarvis}, {Kamble}, {Kassim}, {Liu}, {Loinard}, {Lyons}, {Masters}, {Mezcua}, {Moellenbrock}, {Mroczkowski}, {Nyland}, {O'Dea}, {O'Sullivan}, {Peters}, {Radford}, {Rao}, {Robnett}, {Salcido}, {Shen}, {Sobotka}, {Witz}, {Vaccari}, {van Weeren}, {Vargas}, {Williams}, \& {Yoon}}]{lacy20}
{Lacy}, M., {Baum}, S.~A., {Chandler}, C.~J., {et~al.} 2020, \pasp, 132, 035001

\bibitem[{Lilly {et~al.}(2009)Lilly, Le~Brun, Maier, Mainieri, Mignoli, Scodeggio, Zamorani, Carollo, Contini, Kneib, Le~Fèvre, Renzini, Bardelli, Bolzonella, Bongiorno, Caputi, Coppa, Cucciati, de~la Torre, de~Ravel, Franzetti, Garilli, Iovino, Kampczyk, Kovac, Knobel, Lamareille, Le~Borgne, Pello, Peng, Pérez-Montero, Ricciardelli, Silverman, Tanaka, Tasca, Tresse, Vergani, Zucca, Ilbert, Salvato, Oesch, Abbas, Bottini, Capak, Cappi, Cassata, Cimatti, Elvis, Fumana, Guzzo, Hasinger, Koekemoer, Leauthaud, Maccagni, Marinoni, McCracken, Memeo, Meneux, Porciani, Pozzetti, Sanders, Scaramella, Scarlata, Scoville, Shopbell, \& Taniguchi}]{Lilly09}
Lilly, S.~J., Le~Brun, V., Maier, C., {et~al.} 2009, The Astrophysical Journal Supplement Series, 184, 218

\bibitem[{{Lisenfeld} \& {V{\"o}lk}(2000)}]{lisenfeld00}
{Lisenfeld}, U. \& {V{\"o}lk}, H.~J. 2000, \aap, 354, 423

\bibitem[{{Liu} {et~al.}(2019){Liu}, {Sun}, {Nulsen}, {Clarke}, {Sarazin}, {Forman}, {Gaspari}, {Giacintucci}, {Lal}, \& {Edge}}]{liu19}
{Liu}, W., {Sun}, M., {Nulsen}, P., {et~al.} 2019, \mnras, 484, 3376

\bibitem[{Loubser {et~al.}(2018)Loubser, Hoekstra, Babul, \& O'Sullivan}]{loubser18}
Loubser, S.~I., Hoekstra, H., Babul, A., \& O'Sullivan, E. 2018, Monthly Notices of the Royal Astronomical Society, 477, 335

\bibitem[{{Lovisari} {et~al.}(2021){Lovisari}, {Ettori}, {Gaspari}, \& {Giles}}]{lovisari21}
{Lovisari}, L., {Ettori}, S., {Gaspari}, M., \& {Giles}, P.~A. 2021, Universe, 7, 139

\bibitem[{{Lovisari} {et~al.}(2015){Lovisari}, {Reiprich}, \& {Schellenberger}}]{lovisari15}
{Lovisari}, L., {Reiprich}, T.~H., \& {Schellenberger}, G. 2015, \aap, 573, A118

\bibitem[{{Malavasi} {et~al.}(2015){Malavasi}, {Bardelli}, {Ciliegi}, {Ilbert}, {Pozzetti}, \& {Zucca}}]{malavasi15}
{Malavasi}, N., {Bardelli}, S., {Ciliegi}, P., {et~al.} 2015, \aap, 576, A101

\bibitem[{{McConnell} {et~al.}(2020){McConnell}, {Hale}, {Lenc}, {Banfield}, {Heald}, {Hotan}, {Leung}, {Moss}, {Murphy}, {O'Brien}, {Pritchard}, {Raja}, {Sadler}, {Stewart}, {Thomson}, {Whiting}, {Allison}, {Amy}, {Anderson}, {Ball}, {Bannister}, {Bell}, {Bock}, {Bolton}, {Bunton}, {Chippendale}, {Collier}, {Cooray}, {Cornwell}, {Diamond}, {Edwards}, {Gupta}, {Hayman}, {Heywood}, {Jackson}, {Koribalski}, {Lee-Waddell}, {McClure-Griffiths}, {Ng}, {Norris}, {Phillips}, {Reynolds}, {Roxby}, {Schinckel}, {Shields}, {Tremblay}, {Tzioumis}, {Voronkov}, \& {Westmeier}}]{mcconnell20}
{McConnell}, D., {Hale}, C.~L., {Lenc}, E., {et~al.} 2020, \pasa, 37, e048

\bibitem[{{McIntosh} {et~al.}(2008){McIntosh}, {Guo}, {Hertzberg}, {Katz}, {Mo}, {van den Bosch}, \& {Yang}}]{msintosh08}
{McIntosh}, D.~H., {Guo}, Y., {Hertzberg}, J., {et~al.} 2008, \mnras, 388, 1537

\bibitem[{{McKean} {et~al.}(2016){McKean}, {Godfrey}, {Vegetti}, {Wise}, {Morganti}, {Hardcastle}, {Rafferty}, {Anderson}, {Avruch}, {Beck}, {Bell}, {van Bemmel}, {Bentum}, {Bernardi}, {Best}, {Blaauw}, {Bonafede}, {Breitling}, {Broderick}, {Br{\"u}ggen}, {Cerrigone}, {Ciardi}, {de Gasperin}, {Deller}, {Duscha}, {Engels}, {Falcke}, {Fallows}, {Frieswijk}, {Garrett}, {Grie{\ss}meier}, {van Haarlem}, {Heald}, {Hoeft}, {Horst}, {Iacobelli}, {Intema}, {Juette}, {Karastergiou}, {Kondratiev}, {Koopmans}, {Kuniyoshi}, {Kuper}, {van Leeuwen}, {Maat}, {Mann}, {Markoff}, {McFadden}, {McKay-Bukowski}, {Mulcahy}, {Munk}, {Nelles}, {Orru}, {Paas}, {Pandey-Pommier}, {Pietka}, {Pizzo}, {Polatidis}, {Reich}, {R{\"o}ttgering}, {Rowlinson}, {Scaife}, {Serylak}, {Shulevski}, {Sluman}, {Smirnov}, {Steinmetz}, {Stewart}, {Swinbank}, {Tagger}, {Thoudam}, {Toribio}, {Vermeulen}, {Vocks}, {van Weeren}, {Wucknitz}, {Yatawatta}, \& {Zarka}}]{mckean16}
{McKean}, J.~P., {Godfrey}, L.~E.~H., {Vegetti}, S., {et~al.} 2016, \mnras, 463, 3143

\bibitem[{{McNamara} \& {Nulsen}(2007)}]{mcnamara07}
{McNamara}, B.~R. \& {Nulsen}, P.~E.~J. 2007, \araa, 45, 117

\bibitem[{{Miles} {et~al.}(2004){Miles}, {Raychaudhury}, {Forbes}, {Goudfrooij}, {Ponman}, \& {Kozhurina-Platais}}]{miles04}
{Miles}, T.~A., {Raychaudhury}, S., {Forbes}, D.~A., {et~al.} 2004, \mnras, 355, 785

\bibitem[{{Morganti} {et~al.}(2013){Morganti}, {Frieswijk}, {Oonk}, {Oosterloo}, \& {Tadhunter}}]{morganti13}
{Morganti}, R., {Frieswijk}, W., {Oonk}, R.~J.~B., {Oosterloo}, T., \& {Tadhunter}, C. 2013, \aap, 552, L4

\bibitem[{{Nelson} {et~al.}(2024){Nelson}, {Pillepich}, {Ayromlou}, {Lee}, {Lehle}, {Rohr}, \& {Truong}}]{nelson24}
{Nelson}, D., {Pillepich}, A., {Ayromlou}, M., {et~al.} 2024, \aap, 686, A157

\bibitem[{{Offringa} {et~al.}(2013){Offringa}, {de Bruyn}, {Zaroubi}, {van Diepen}, {Martinez-Ruby}, {Labropoulos}, {Brentjens}, {Ciardi}, {Daiboo}, {Harker}, {Jeli{\'c}}, {Kazemi}, {Koopmans}, {Mellema}, {Pandey}, {Pizzo}, {Schaye}, {Vedantham}, {Veligatla}, {Wijnholds}, {Yatawatta}, {Zarka}, {Alexov}, {Anderson}, {Asgekar}, {Avruch}, {Beck}, {Bell}, {Bell}, {Bentum}, {Bernardi}, {Best}, {Birzan}, {Bonafede}, {Breitling}, {Broderick}, {Br{\"u}ggen}, {Butcher}, {Conway}, {de Vos}, {Dettmar}, {Eisloeffel}, {Falcke}, {Fender}, {Frieswijk}, {Gerbers}, {Griessmeier}, {Gunst}, {Hassall}, {Heald}, {Hessels}, {Hoeft}, {Horneffer}, {Karastergiou}, {Kondratiev}, {Koopman}, {Kuniyoshi}, {Kuper}, {Maat}, {Mann}, {McKean}, {Meulman}, {Mevius}, {Mol}, {Nijboer}, {Noordam}, {Norden}, {Paas}, {Pandey}, {Pizzo}, {Polatidis}, {Rafferty}, {Rawlings}, {Reich}, {R{\"o}ttgering}, {Schoenmakers}, {Sluman}, {Smirnov}, {Sobey}, {Stappers}, {Steinmetz}, {Swinbank}, {Tagger}, {Tang}, {Tasse}, {van Ardenne}, {van Cappellen}, {van
  Duin}, {van Haarlem}, {van Leeuwen}, {van Weeren}, {Vermeulen}, {Vocks}, {Wijers}, {Wise}, \& {Wucknitz}}]{offringa13}
{Offringa}, A.~R., {de Bruyn}, A.~G., {Zaroubi}, S., {et~al.} 2013, \aap, 549, A11

\bibitem[{{Offringa} {et~al.}(2014){Offringa}, {McKinley}, {Hurley-Walker}, {Briggs}, {Wayth}, {Kaplan}, {Bell}, {Feng}, {Neben}, {Hughes}, {Rhee}, {Murphy}, {Bhat}, {Bernardi}, {Bowman}, {Cappallo}, {Corey}, {Deshpande}, {Emrich}, {Ewall-Wice}, {Gaensler}, {Goeke}, {Greenhill}, {Hazelton}, {Hindson}, {Johnston-Hollitt}, {Jacobs}, {Kasper}, {Kratzenberg}, {Lenc}, {Lonsdale}, {Lynch}, {McWhirter}, {Mitchell}, {Morales}, {Morgan}, {Kudryavtseva}, {Oberoi}, {Ord}, {Pindor}, {Procopio}, {Prabu}, {Riding}, {Roshi}, {Shankar}, {Srivani}, {Subrahmanyan}, {Tingay}, {Waterson}, {Webster}, {Whitney}, {Williams}, \& {Williams}}]{offringa14}
{Offringa}, A.~R., {McKinley}, B., {Hurley-Walker}, N., {et~al.} 2014, \mnras, 444, 606

\bibitem[{{O'Sullivan} {et~al.}(2018){O'Sullivan}, {Kolokythas}, {Kantharia}, {Raychaudhury}, {David}, \& {Vrtilek}}]{Osullivan18}
{O'Sullivan}, E., {Kolokythas}, K., {Kantharia}, N.~G., {et~al.} 2018, \mnras, 473, 5248

\bibitem[{{O'Sullivan} {et~al.}(2017){O'Sullivan}, {Ponman}, {Kolokythas}, {Raychaudhury}, {Babul}, {Vrtilek}, {David}, {Giacintucci}, {Gitti}, \& {Haines}}]{Osullivan17}
{O'Sullivan}, E., {Ponman}, T.~J., {Kolokythas}, K., {et~al.} 2017, \mnras, 472, 1482

\bibitem[{{O'Sullivan} {et~al.}(2024){O'Sullivan}, {Rajpurohit}, {Schellenberger}, {Vrtilek}, {David}, {Babul}, {Olivares}, {Ubertosi}, {Kolokythas}, {Babyk}, \& {Loubser}}]{osullivan24}
{O'Sullivan}, E., {Rajpurohit}, K., {Schellenberger}, G., {et~al.} 2024, \apj, 970, 65

\bibitem[{{O'Sullivan} {et~al.}(2011){O'Sullivan}, {Worrall}, {Birkinshaw}, {Trinchieri}, {Wolter}, {Zezas}, \& {Giacintucci}}]{Osullivan11}
{O'Sullivan}, E., {Worrall}, D.~M., {Birkinshaw}, M., {et~al.} 2011, \mnras, 416, 2916

\bibitem[{Panagoulia {et~al.}(2013)Panagoulia, Fabian, \& Sanders}]{panagoulia14}
Panagoulia, E.~K., Fabian, A.~C., \& Sanders, J.~S. 2013, Monthly Notices of the Royal Astronomical Society, 438, 2341

\bibitem[{{Pasini} {et~al.}(2021){Pasini}, {Finoguenov}, {Br{\"u}ggen}, {Gaspari}, {de Gasperin}, \& {Gozaliasl}}]{pasini21}
{Pasini}, T., {Finoguenov}, A., {Br{\"u}ggen}, M., {et~al.} 2021, \mnras, 505, 2628

\bibitem[{{Pasini} {et~al.}(2025){Pasini}, {Mahatma}, {Brienza}, {Kolokythas}, {Eckert}, {de Gasperin}, {van Weeren}, {Gastaldello}, {Hoang}, \& {Santra}}]{pasini25}
{Pasini}, T., {Mahatma}, V.~H., {Brienza}, M., {et~al.} 2025, \aap, 693, A94

\bibitem[{{Pearson} {et~al.}(2024){Pearson}, {Santos}, {Goto}, {Huang}, {Kim}, {Matsuhara}, {Pollo}, {Ho}, {Hwang}, {Ma{\l}ek}, {Nakagawa}, {Romano}, {Serjeant}, {Suelves}, {Shim}, \& {White}}]{pearson24}
{Pearson}, W.~J., {Santos}, D.~J.~D., {Goto}, T., {et~al.} 2024, \aap, 686, A94

\bibitem[{{Perley} \& {Butler}(2017)}]{perley&butler17}
{Perley}, R.~A. \& {Butler}, B.~J. 2017, \apjs, 230, 7

\bibitem[{{Ponman} {et~al.}(1999){Ponman}, {Cannon}, \& {Navarro}}]{ponman99}
{Ponman}, T.~J., {Cannon}, D.~B., \& {Navarro}, J.~F. 1999, \nat, 397, 135

\bibitem[{{Ponman} {et~al.}(2003){Ponman}, {Sanderson}, \& {Finoguenov}}]{ponman03}
{Ponman}, T.~J., {Sanderson}, A.~J.~R., \& {Finoguenov}, A. 2003, \mnras, 343, 331

\bibitem[{{Rafferty} {et~al.}(2006){Rafferty}, {McNamara}, {Nulsen}, \& {Wise}}]{rafferty06}
{Rafferty}, D.~A., {McNamara}, B.~R., {Nulsen}, P.~E.~J., \& {Wise}, M.~W. 2006, \apj, 652, 216

\bibitem[{{Rajpurohit} {et~al.}(2024){Rajpurohit}, {O'Sullivan}, {Schellenberger}, {Brienza}, {Vrtilek}, {Forman}, {David}, {Clarke}, {Botteon}, {Vazza}, {Giacintucci}, {Jones}, {Br{\"u}ggen}, {Shimwell}, {Drabent}, {Loi}, {Loubser}, {Kolokythas}, {Babyk}, \& {R{\"o}ttgering}}]{rajpurohit2024}
{Rajpurohit}, K., {O'Sullivan}, E., {Schellenberger}, G., {et~al.} 2024, arXiv e-prints, arXiv:2408.15197

\bibitem[{{Randriamanakoto} {et~al.}(2020){Randriamanakoto}, {Ishwara-Chandra}, \& {Taylor}}]{randriamanakoto20}
{Randriamanakoto}, Z., {Ishwara-Chandra}, C.~H., \& {Taylor}, A.~R. 2020, \mnras, 496, 3381

\bibitem[{{Riseley} {et~al.}(2025){Riseley}, {Vernstrom}, {Lovisari}, {O'Sullivan}, {Gastaldello}, {Brienza}, {Nayak}, {Bonafede}, {Carretti}, {Duchesne}, {Giacintucci}, {Hopkins}, {Koribalski}, {Loi}, {Pfrommer}, {Raja}, {Ross}, {Rubinur}, {Ruszkowski}, {Shimwell}, {de Villiers}, {West}, {Zovaro}, {Akahori}, {Anderson}, {Bomans}, {Drabent}, {Rudnick}, \& {Santra}}]{riseley25}
{Riseley}, C.~J., {Vernstrom}, T., {Lovisari}, L., {et~al.} 2025, arXiv e-prints, arXiv:2503.08840

\bibitem[{{Robotham} {et~al.}(2011){Robotham}, {Norberg}, {Driver}, {Baldry}, {Bamford}, {Hopkins}, {Liske}, {Loveday}, {Merson}, {Peacock}, {Brough}, {Cameron}, {Conselice}, {Croom}, {Frenk}, {Gunawardhana}, {Hill}, {Jones}, {Kelvin}, {Kuijken}, {Nichol}, {Parkinson}, {Pimbblet}, {Phillipps}, {Popescu}, {Prescott}, {Sharp}, {Sutherland}, {Taylor}, {Thomas}, {Tuffs}, {van Kampen}, \& {Wijesinghe}}]{robotham11}
{Robotham}, A.~S.~G., {Norberg}, P., {Driver}, S.~P., {et~al.} 2011, \mnras, 416, 2640

\bibitem[{{Rodriguez-Gomez} {et~al.}(2015){Rodriguez-Gomez}, {Genel}, {Vogelsberger}, {Sijacki}, {Pillepich}, {Sales}, {Torrey}, {Snyder}, {Nelson}, {Springel}, {Ma}, \& {Hernquist}}]{rodriguez15}
{Rodriguez-Gomez}, V., {Genel}, S., {Vogelsberger}, M., {et~al.} 2015, \mnras, 449, 49

\bibitem[{{Sabater} {et~al.}(2019){Sabater}, {Best}, {Hardcastle}, {Shimwell}, {Tasse}, {Williams}, {Br{\"u}ggen}, {Cochrane}, {Croston}, {de Gasperin}, {Duncan}, {G{\"u}rkan}, {Mechev}, {Morabito}, {Prandoni}, {R{\"o}ttgering}, {Smith}, {Harwood}, {Mingo}, {Mooney}, \& {Saxena}}]{sabater19}
{Sabater}, J., {Best}, P.~N., {Hardcastle}, M.~J., {et~al.} 2019, \aap, 622, A17

\bibitem[{{Salvetti} {et~al.}(2017){Salvetti}, {Marelli}, {Gastaldello}, {Ghizzardi}, {Molendi}, {De Luca}, {Moretti}, {Rossetti}, \& {Tiengo}}]{salvetti17}
{Salvetti}, D., {Marelli}, M., {Gastaldello}, F., {et~al.} 2017, Experimental Astronomy, 44, 309

\bibitem[{{Scaife} \& {Heald}(2012)}]{scaife12}
{Scaife}, A. M.~M. \& {Heald}, G.~H. 2012, \mnras, 423, L30

\bibitem[{{Schellenberger} {et~al.}(2017){Schellenberger}, {Vrtilek}, {David}, {O'Sullivan}, {Giacintucci}, {Johnston-Hollitt}, {Duchesne}, \& {Raychaudhury}}]{schellenberger17}
{Schellenberger}, G., {Vrtilek}, J.~M., {David}, L., {et~al.} 2017, \apj, 845, 84

\bibitem[{Shabala {et~al.}(2008)Shabala, Ash, Alexander, \& Riley}]{shabala08}
Shabala, S.~S., Ash, S., Alexander, P., \& Riley, J.~M. 2008, Monthly Notices of the Royal Astronomical Society, 388, 625

\bibitem[{{Sharma} {et~al.}(2024){Sharma}, {Choi}, {Somerville}, {Snyder}, {Jhee}, {Kocevski}, {Hirschmann}, {Moster}, {Naab}, {Narayanan}, {Ostriker}, \& {Rosario}}]{sharma24}
{Sharma}, R.~S., {Choi}, E., {Somerville}, R.~S., {et~al.} 2024, \mnras, 527, 9461

\bibitem[{{Shimwell} {et~al.}(2022){Shimwell}, {Hardcastle}, {Tasse}, {Best}, {R{\"o}ttgering}, {Williams}, {Botteon}, {Drabent}, {Mechev}, {Shulevski}, {van Weeren}, {Bester}, {Br{\"u}ggen}, {Brunetti}, {Callingham}, {Chy{\.z}y}, {Conway}, {Dijkema}, {Duncan}, {de Gasperin}, {Hale}, {Haverkorn}, {Hugo}, {Jackson}, {Mevius}, {Miley}, {Morabito}, {Morganti}, {Offringa}, {Oonk}, {Rafferty}, {Sabater}, {Smith}, {Schwarz}, {Smirnov}, {O'Sullivan}, {Vedantham}, {White}, {Albert}, {Alegre}, {Asabere}, {Bacon}, {Bonafede}, {Bonnassieux}, {Brienza}, {Bilicki}, {Bonato}, {Calistro Rivera}, {Cassano}, {Cochrane}, {Croston}, {Cuciti}, {Dallacasa}, {Danezi}, {Dettmar}, {Di Gennaro}, {Edler}, {En{\ss}lin}, {Emig}, {Franzen}, {Garc{\'\i}a-Vergara}, {Grange}, {G{\"u}rkan}, {Hajduk}, {Heald}, {Heesen}, {Hoang}, {Hoeft}, {Horellou}, {Iacobelli}, {Jamrozy}, {Jeli{\'c}}, {Kondapally}, {Kukreti}, {Kunert-Bajraszewska}, {Magliocchetti}, {Mahatma}, {Ma{\l}ek}, {Mandal}, {Massaro}, {Meyer-Zhao}, {Mingo}, {Mostert}, {Nair},
  {Nakoneczny}, {Nikiel-Wroczy{\'n}ski}, {Orr{\'u}}, {Pajdosz-{\'S}mierciak}, {Pasini}, {Prandoni}, {van Piggelen}, {Rajpurohit}, {Retana-Montenegro}, {Riseley}, {Rowlinson}, {Saxena}, {Schrijvers}, {Sweijen}, {Siewert}, {Timmerman}, {Vaccari}, {Vink}, {West}, {Wo{\l}owska}, {Zhang}, \& {Zheng}}]{shimwell22}
{Shimwell}, T.~W., {Hardcastle}, M.~J., {Tasse}, C., {et~al.} 2022, \aap, 659, A1

\bibitem[{{Shimwell} {et~al.}(2017){Shimwell}, {R{\"o}ttgering}, {Best}, {Williams}, {Dijkema}, {de Gasperin}, {Hardcastle}, {Heald}, {Hoang}, {Horneffer}, {Intema}, {Mahony}, {Mandal}, {Mechev}, {Morabito}, {Oonk}, {Rafferty}, {Retana-Montenegro}, {Sabater}, {Tasse}, {van Weeren}, {Br{\"u}ggen}, {Brunetti}, {Chy{\.z}y}, {Conway}, {Haverkorn}, {Jackson}, {Jarvis}, {McKean}, {Miley}, {Morganti}, {White}, {Wise}, {van Bemmel}, {Beck}, {Brienza}, {Bonafede}, {Calistro Rivera}, {Cassano}, {Clarke}, {Cseh}, {Deller}, {Drabent}, {van Driel}, {Engels}, {Falcke}, {Ferrari}, {Fr{\"o}hlich}, {Garrett}, {Harwood}, {Heesen}, {Hoeft}, {Horellou}, {Israel}, {Kapi{\'n}ska}, {Kunert-Bajraszewska}, {McKay}, {Mohan}, {Orr{\'u}}, {Pizzo}, {Prandoni}, {Schwarz}, {Shulevski}, {Sipior}, {Smith}, {Sridhar}, {Steinmetz}, {Stroe}, {Varenius}, {van der Werf}, {Zensus}, \& {Zwart}}]{shimwell17}
{Shimwell}, T.~W., {R{\"o}ttgering}, H.~J.~A., {Best}, P.~N., {et~al.} 2017, \aap, 598, A104

\bibitem[{{Shimwell} {et~al.}(2019){Shimwell}, {Tasse}, {Hardcastle}, {Mechev}, {Williams}, {Best}, {R{\"o}ttgering}, {Callingham}, {Dijkema}, {de Gasperin}, {Hoang}, {Hugo}, {Mirmont}, {Oonk}, {Prandoni}, {Rafferty}, {Sabater}, {Smirnov}, {van Weeren}, {White}, {Atemkeng}, {Bester}, {Bonnassieux}, {Br{\"u}ggen}, {Brunetti}, {Chy{\.z}y}, {Cochrane}, {Conway}, {Croston}, {Danezi}, {Duncan}, {Haverkorn}, {Heald}, {Iacobelli}, {Intema}, {Jackson}, {Jamrozy}, {Jarvis}, {Lakhoo}, {Mevius}, {Miley}, {Morabito}, {Morganti}, {Nisbet}, {Orr{\'u}}, {Perkins}, {Pizzo}, {Schrijvers}, {Smith}, {Vermeulen}, {Wise}, {Alegre}, {Bacon}, {van Bemmel}, {Beswick}, {Bonafede}, {Botteon}, {Bourke}, {Brienza}, {Calistro Rivera}, {Cassano}, {Clarke}, {Conselice}, {Dettmar}, {Drabent}, {Dumba}, {Emig}, {En{\ss}lin}, {Ferrari}, {Garrett}, {G{\'e}nova-Santos}, {Goyal}, {G{\"u}rkan}, {Hale}, {Harwood}, {Heesen}, {Hoeft}, {Horellou}, {Jackson}, {Kokotanekov}, {Kondapally}, {Kunert-Bajraszewska}, {Mahatma}, {Mahony}, {Mandal}, {McKean},
  {Merloni}, {Mingo}, {Miskolczi}, {Mooney}, {Nikiel-Wroczy{\'n}ski}, {O'Sullivan}, {Quinn}, {Reich}, {Roskowi{\'n}ski}, {Rowlinson}, {Savini}, {Saxena}, {Schwarz}, {Shulevski}, {Sridhar}, {Stacey}, {Urquhart}, {van der Wiel}, {Varenius}, {Webster}, \& {Wilber}}]{shimwell19}
{Shimwell}, T.~W., {Tasse}, C., {Hardcastle}, M.~J., {et~al.} 2019, \aap, 622, A1

\bibitem[{Shin {et~al.}(2016)Shin, Woo, \& Mulchaey}]{shin16}
Shin, J., Woo, J.-H., \& Mulchaey, J.~S. 2016, The Astrophysical Journal Supplement Series, 227, 31

\bibitem[{{Shulevski} {et~al.}(2024){Shulevski}, {Brienza}, {Massaro}, {Morganti}, {Intema}, {Oosterloo}, {De Gasperin}, {Rajpurohit}, {Pasini}, {Kutkin}, {Vohl}, {Adams}, {Adebahr}, {Br{\"u}ggen}, {Hess}, {Loose}, {Oostrum}, \& {Ziemke}}]{shulevski24}
{Shulevski}, A., {Brienza}, M., {Massaro}, F., {et~al.} 2024, \aap, 682, A171

\bibitem[{{Shulevski} {et~al.}(2017){Shulevski}, {Morganti}, {Harwood}, {Barthel}, {Jamrozy}, {Brienza}, {Brunetti}, {R{\"o}ttgering}, {Murgia}, {White}, {Croston}, \& {Br{\"u}ggen}}]{shulevski17}
{Shulevski}, A., {Morganti}, R., {Harwood}, J.~J., {et~al.} 2017, \aap, 600, A65

\bibitem[{{Solanes} {et~al.}(2018){Solanes}, {Perea}, \& {Valent{\'\i}-Rojas}}]{solanes18}
{Solanes}, J.~M., {Perea}, J.~D., \& {Valent{\'\i}-Rojas}, G. 2018, \aap, 614, A66

\bibitem[{{Stewart} {et~al.}(2025){Stewart}, {Shabala}, {Turner}, {Yates-Jones}, {Krause}, {Wong}, {Power}, \& {Hardcastle}}]{2025PASA...42..152S}
{Stewart}, G. S.~C., {Shabala}, S.~S., {Turner}, R.~J., {et~al.} 2025, \pasa, 42, e152

\bibitem[{{Stott} {et~al.}(2010){Stott}, {Collins}, {Sahl{\'e}n}, {Hilton}, {Lloyd-Davies}, {Capozzi}, {Hosmer}, {Liddle}, {Mehrtens}, {Miller}, {Romer}, {Stanford}, {Viana}, {Davidson}, {Hoyle}, {Kay}, \& {Nichol}}]{stott10}
{Stott}, J.~P., {Collins}, C.~A., {Sahl{\'e}n}, M., {et~al.} 2010, \apj, 718, 23

\bibitem[{{Subrahmanyan} {et~al.}(2003){Subrahmanyan}, {Beasley}, {Goss}, {Golap}, \& {Hunstead}}]{subrahmanyan03}
{Subrahmanyan}, R., {Beasley}, A.~J., {Goss}, W.~M., {Golap}, K., \& {Hunstead}, R.~W. 2003, \aj, 125, 1095

\bibitem[{{Sun}(2012)}]{sun12}
{Sun}, M. 2012, New Journal of Physics, 14, 045004

\bibitem[{{Sun} {et~al.}(2009){Sun}, {Voit}, {Donahue}, {Jones}, {Forman}, \& {Vikhlinin}}]{sun09}
{Sun}, M., {Voit}, G.~M., {Donahue}, M., {et~al.} 2009, \apj, 693, 1142

\bibitem[{{Tasse} {et~al.}(2021){Tasse}, {Shimwell}, {Hardcastle}, {O'Sullivan}, {van Weeren}, {Best}, {Bester}, {Hugo}, {Smirnov}, {Sabater}, {Calistro-Rivera}, {de Gasperin}, {Morabito}, {R{\"o}ttgering}, {Williams}, {Bonato}, {Bondi}, {Botteon}, {Br{\"u}ggen}, {Brunetti}, {Chy{\.z}y}, {Garrett}, {G{\"u}rkan}, {Jarvis}, {Kondapally}, {Mandal}, {Prandoni}, {Repetti}, {Retana-Montenegro}, {Schwarz}, {Shulevski}, \& {Wiaux}}]{Tasse21}
{Tasse}, C., {Shimwell}, T., {Hardcastle}, M.~J., {et~al.} 2021, \aap, 648, A1

\bibitem[{Taylor \& Babul(2005)}]{taylor05}
Taylor, J.~E. \& Babul, A. 2005, Monthly Notices of the Royal Astronomical Society, 364, 515

\bibitem[{{Tempel} {et~al.}(2017){Tempel}, {Tuvikene}, {Kipper}, \& {Libeskind}}]{tempel17}
{Tempel}, E., {Tuvikene}, T., {Kipper}, R., \& {Libeskind}, N.~I. 2017, \aap, 602, A100

\bibitem[{{Timmerman} {et~al.}(2022){Timmerman}, {van Weeren}, {Botteon}, {R{\"o}ttgering}, {McNamara}, {Sweijen}, {B{\^\i}rzan}, \& {Morabito}}]{timmerman22}
{Timmerman}, R., {van Weeren}, R.~J., {Botteon}, A., {et~al.} 2022, \aap, 668, A65

\bibitem[{{van den Bosch} {et~al.}(2014){van den Bosch}, {Jiang}, {Hearin}, {Campbell}, {Watson}, \& {Padmanabhan}}]{vandenbosch14}
{van den Bosch}, F.~C., {Jiang}, F., {Hearin}, A., {et~al.} 2014, \mnras, 445, 1713

\bibitem[{{van Weeren} {et~al.}(2021){van Weeren}, {Shimwell}, {Botteon}, {Brunetti}, {Br{\"u}ggen}, {Boxelaar}, {Cassano}, {Di Gennaro}, {Andrade-Santos}, {Bonnassieux}, {Bonafede}, {Cuciti}, {Dallacasa}, {de Gasperin}, {Gastaldello}, {Hardcastle}, {Hoeft}, {Kraft}, {Mandal}, {Rossetti}, {R{\"o}ttgering}, {Tasse}, \& {Wilber}}]{vweeren21}
{van Weeren}, R.~J., {Shimwell}, T.~W., {Botteon}, A., {et~al.} 2021, \aap, 651, A115

\bibitem[{{van Weeren} {et~al.}(2016){van Weeren}, {Williams}, {Hardcastle}, {Shimwell}, {Rafferty}, {Sabater}, {Heald}, {Sridhar}, {Dijkema}, {Brunetti}, {Br{\"u}ggen}, {Andrade-Santos}, {Ogrean}, {R{\"o}ttgering}, {Dawson}, {Forman}, {de Gasperin}, {Jones}, {Miley}, {Rudnick}, {Sarazin}, {Bonafede}, {Best}, {B{\^\i}rzan}, {Cassano}, {Chy{\.z}y}, {Croston}, {Ensslin}, {Ferrari}, {Hoeft}, {Horellou}, {Jarvis}, {Kraft}, {Mevius}, {Intema}, {Murray}, {Orr{\'u}}, {Pizzo}, {Simionescu}, {Stroe}, {van der Tol}, \& {White}}]{vweeren16}
{van Weeren}, R.~J., {Williams}, W.~L., {Hardcastle}, M.~J., {et~al.} 2016, \apjs, 223, 2

\bibitem[{{Von Der Linden} {et~al.}(2007){Von Der Linden}, {Best}, {Kauffmann}, \& {White}}]{vonderlinden07}
{Von Der Linden}, A., {Best}, P.~N., {Kauffmann}, G., \& {White}, S. D.~M. 2007, \mnras, 379, 867

\bibitem[{{Williams} {et~al.}(2016){Williams}, {van Weeren}, {R{\"o}ttgering}, {Best}, {Dijkema}, {de Gasperin}, {Hardcastle}, {Heald}, {Prandoni}, {Sabater}, {Shimwell}, {Tasse}, {van Bemmel}, {Br{\"u}ggen}, {Brunetti}, {Conway}, {En{\ss}lin}, {Engels}, {Falcke}, {Ferrari}, {Haverkorn}, {Jackson}, {Jarvis}, {Kapi{\'n}ska}, {Mahony}, {Miley}, {Morabito}, {Morganti}, {Orr{\'u}}, {Retana-Montenegro}, {Sridhar}, {Toribio}, {White}, {Wise}, \& {Zwart}}]{william16}
{Williams}, W.~L., {van Weeren}, R.~J., {R{\"o}ttgering}, H.~J.~A., {et~al.} 2016, \mnras, 460, 2385

\bibitem[{{W{\'o}jtowicz} {et~al.}(2023){W{\'o}jtowicz}, {Stawarz}, {Cheung}, {Werner}, \& {Rudka}}]{wojtowicz23}
{W{\'o}jtowicz}, A., {Stawarz}, {\L}., {Cheung}, C.~C., {Werner}, N., \& {Rudka}, D. 2023, \apj, 944, 195

\end{thebibliography}
\bibliographystyle{aa}

\begin{appendix}

\section{Spectral index error maps}\label{app:a}

Here we have shown the spectral index error maps corresponding to the Figure~\ref{spec-map}.

\begin{figure*}
    \centering
    \includegraphics[width=0.34\textwidth]{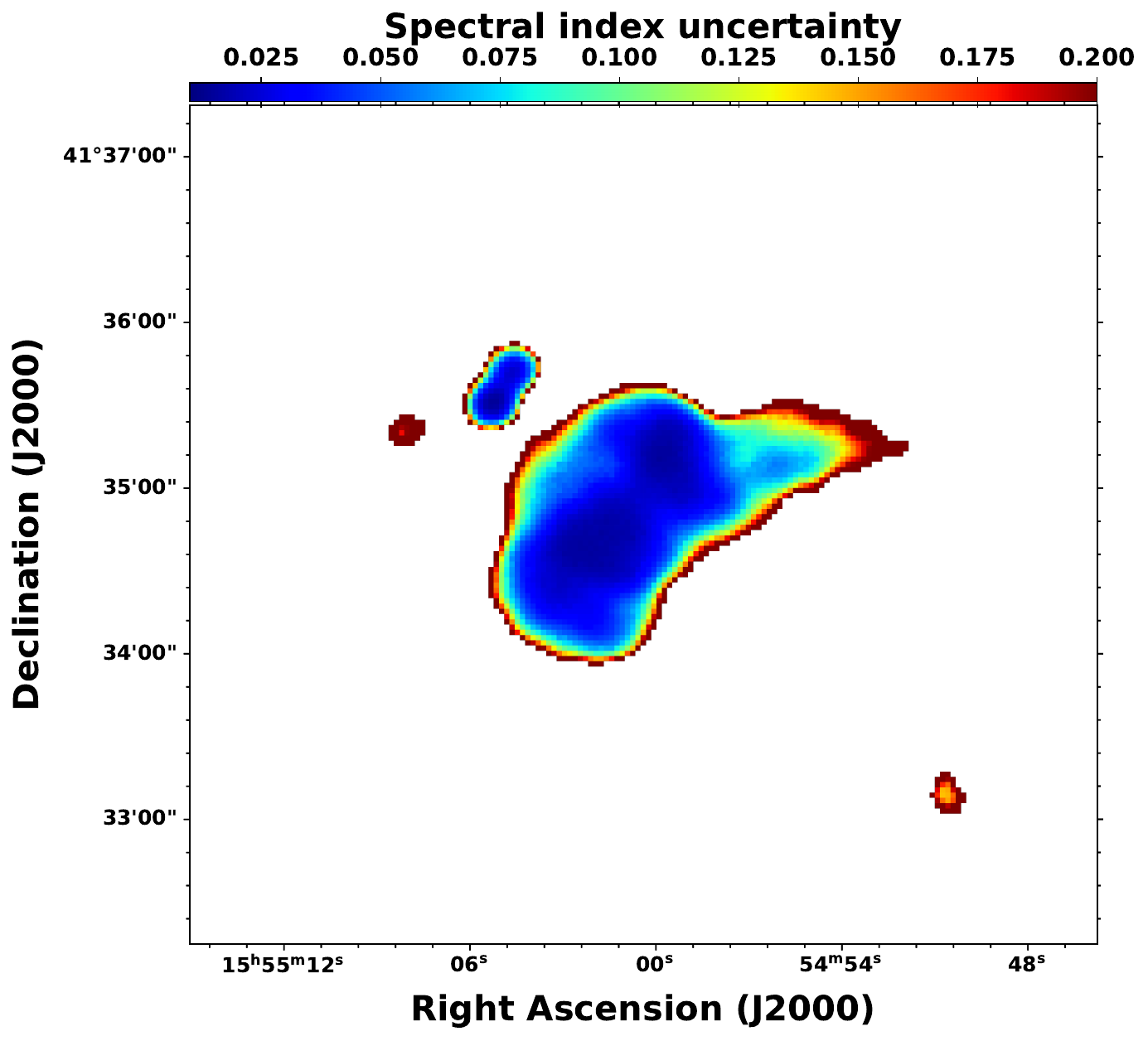} \includegraphics[width=0.29\textwidth]{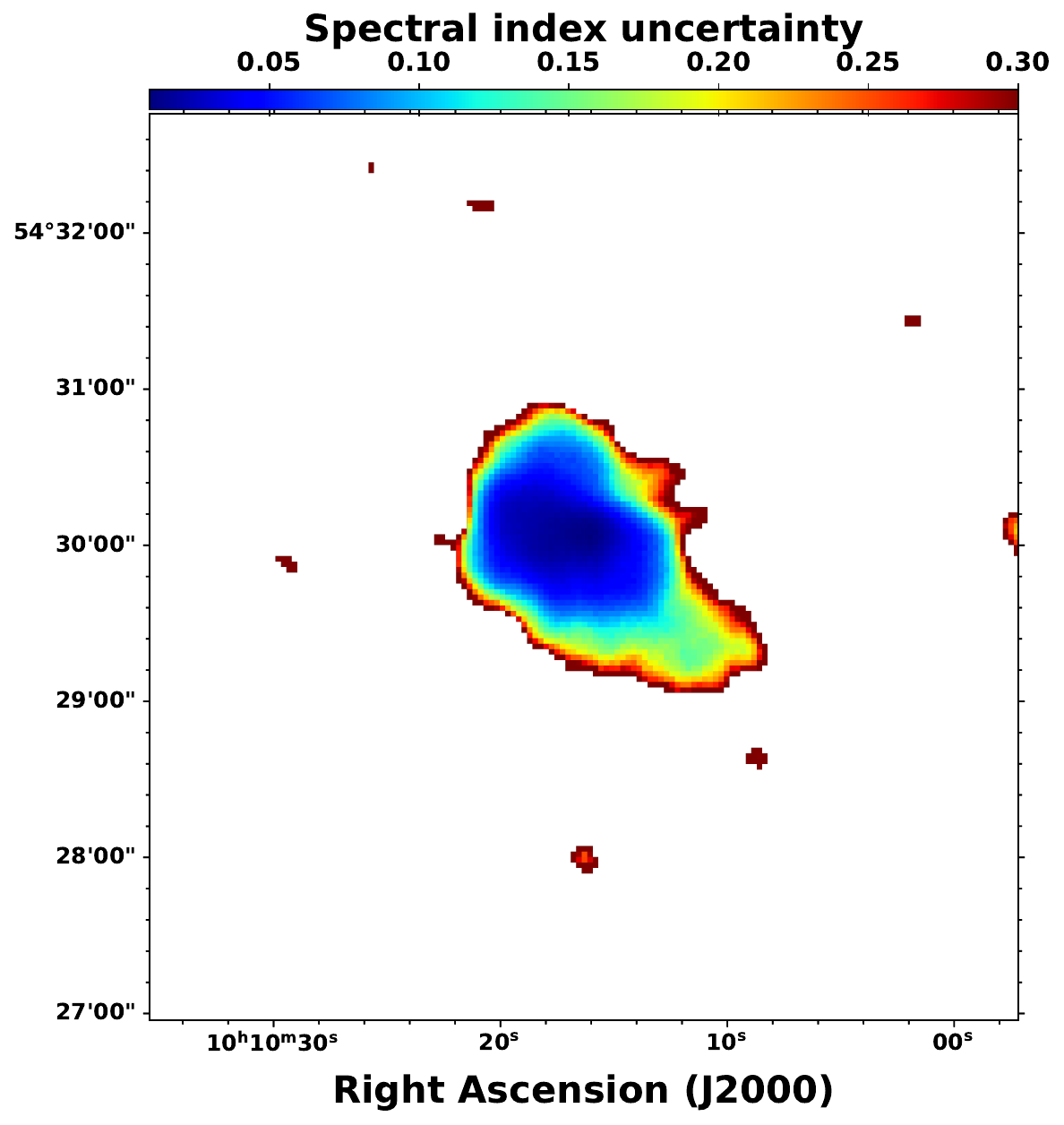} \includegraphics[width=0.30\textwidth]{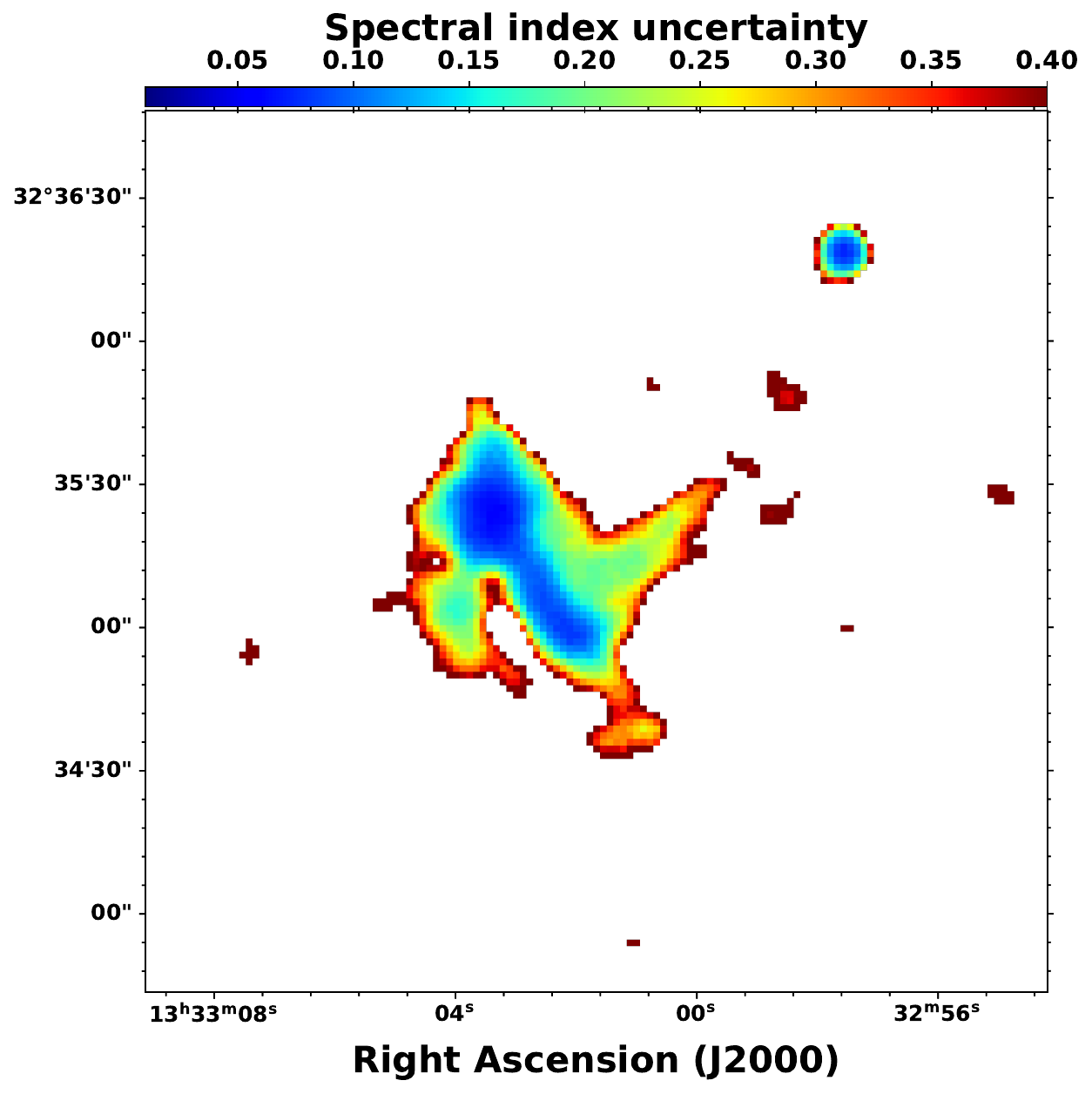}

    \caption{\textit{Left:} The spectral index error map between 144 and 400 MHz is shown for SDSSTG8102. \textit{middle:} The same is shown for the SDSSTG16393. \textit{Right:} The spectral error map is shown for SDSSTG28674 in a colour scale.}
    \label{spec-map-err}
\end{figure*}

\section{Image for the SDSSTG16393} \label{diff-regions}

The X-ray surface brightness map for the full FoV of the group SDSSTG16393. 
\begin{figure} [h!] 
\centering
      \includegraphics[height=7.6cm]{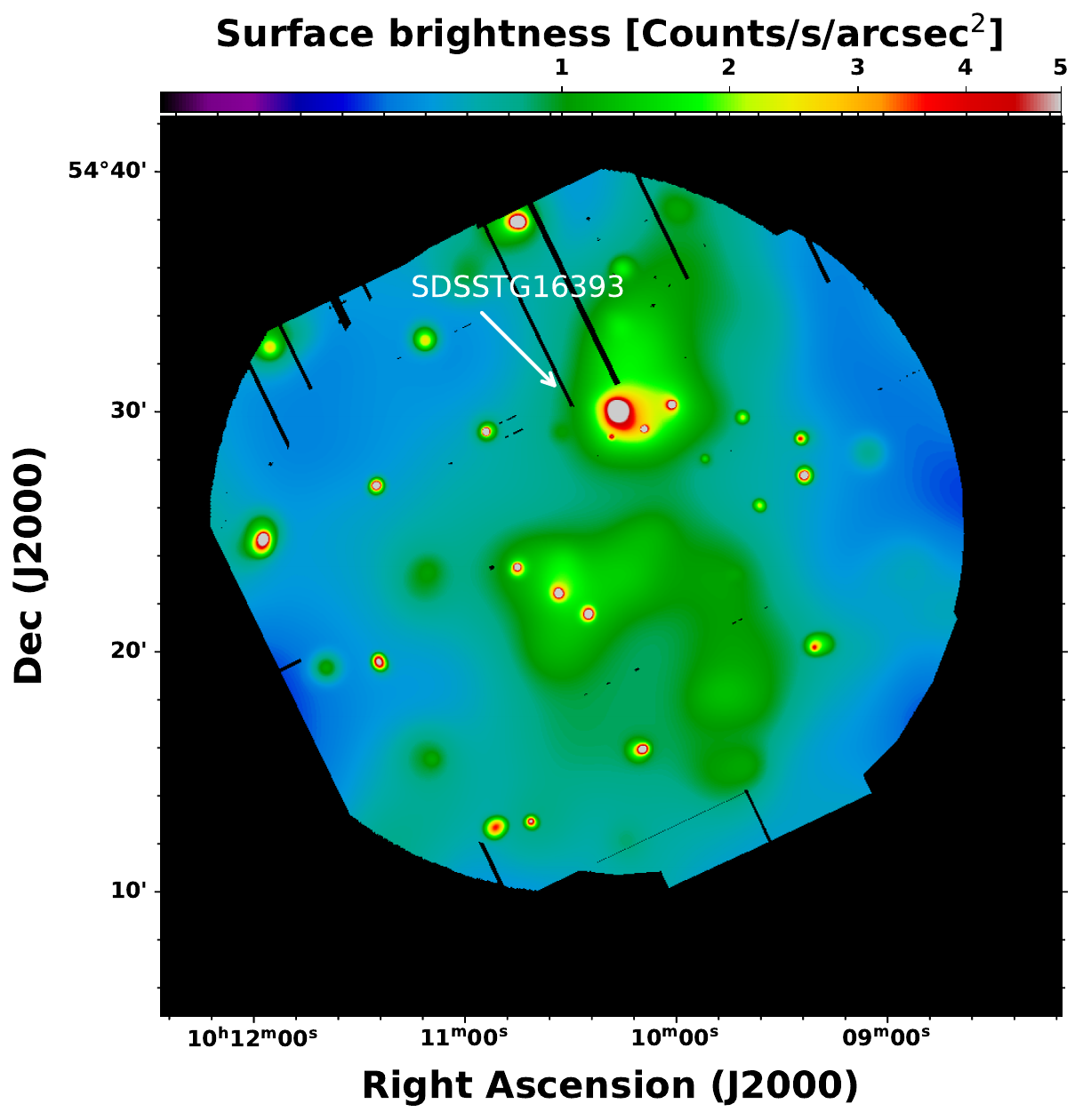}    
      \caption{Adaptively smoothed 0.7 $-$ 1.2 keV image of the full FoV is shown in colour for SDSSTG16393}.
    \label{xgap_16393_full_fov_img}
\end{figure}

\end{appendix}

\end{document}